\documentclass[fleqn,10pt]{wlscirep}
\usepackage[utf8]{inputenc}
\usepackage[T1]{fontenc}
\usepackage{empheq} 
\usepackage{amsmath,amssymb} 
\usepackage{threeparttable} 
\usepackage{mdframed} 
\usepackage{wrapfig} 
\usepackage{boxedminipage} 
\usepackage{bm} 
\usepackage{placeins} 
\usepackage{algorithm}
\usepackage{algpseudocode}

\title{Network dynamical stability analysis reveals key ``mallostatic'' natural variables that erode homeostasis and drive age-related decline of health} 

\author[1,$\dagger$]{Glen Pridham}
\author[1,$\ddagger$,*]{Andrew D. Rutenberg}
\affil[1]{Department of Physics and Atmospheric Science, Dalhousie University, Halifax, Nova Scotia, Canada, B3H 4R2}

\affil[$\dagger$]{glen.pridham@dal.ca}
\affil[$\ddagger$]{adr@dal.ca}
\affil[*]{Corresponding author}


\begin{abstract}
Using longitudinal study data, we dynamically model how aging affects homeostasis in both mice and humans. We operationalize homeostasis as a multivariate mean-reverting stochastic process. We hypothesize that biomarkers have stable equilibrium values, but that deviations from equilibrium of each biomarker affects other biomarkers through an interaction network --- this precludes univariate analysis. We therefore looked for age-related changes to homeostasis using dynamic network stability analysis, which transforms observed biomarker data into independent ``natural'' variables and determines their associated recovery rates. Most natural variables remained near equilibrium and were essentially constant in time. A small number of natural variables were unable to equilibrate due to a gradual drift with age in their homeostatic equilibrium, i.e. allostasis. This drift caused them to accumulate over the lifespan course and makes them natural aging variables. Their rate of accumulation was correlated with risk of adverse outcomes: death or dementia onset. We call this tendency for aging organisms to drift towards an equilibrium position of ever-worsening health ``mallostasis''. We demonstrate that the effects of mallostasis on observed biomarkers are spread out through the interaction network. This could provide a redundancy mechanism to preserve functioning until multi-system dysfunction emerges at advanced ages.
\end{abstract}
\begin{document}

\flushbottom
\maketitle
\thispagestyle{empty}

\vspace{-8mm}
\section*{Introduction}
Homeostasis is the self-regulating process that maintains internal stability \cite{Billman2020-lj}. Yet as individuals age, it is characteristic for biomarkers to drift away from healthy levels;  something about homeostasis is therefore “lost” during the aging process \cite{Schmauck-Medina2022-hy}. For example, loss of protein homeostasis is believed to cause the hallmark accumulation of unfolded, misfolded and aggregate proteins with age \cite{Lopez-Otin2013-pv}. Accumulation is observed at multiple biological scales, including oxidative damage \cite{Campisi2019-gp}, epigenetic age \cite{Li2020-hl}, senescent cells \cite{Karin2019-gf}, and regulatory T-cells \cite{Raynor2012-pc} at the cellular scale, and extending up to the whole organism scale where clinical deficits \cite{Mitnitski2015-ia}, including chronic diseases \cite{Fabbri2015-vd}, accumulate with age. Sehl and Yates performed univariate analysis of 445 health biomarkers and found that almost all of them accumulate negatively with age --- typically showing linear decline \cite{Sehl2001-ld}. Such accumulation of biomarker values in a particular direction appears to be a generic feature of aging. When biomarkers reach abnormal values, they are associated with dysfunction and poor health, independently of age \cite{Blodgett2017-jp, Cohen2013-hj}. A general mechanism of how accumulation and poor health emerge from homeostasis has, nevertheless, been missing.

Prior work suggests that accumulation may be a consequence of a drifting equilibrium position. Allostasis, literally ``homeostasis through change'' \cite{McEwen2000-fn}, describes a version of homeostasis in which the equilibrium position is mutable, adapting as necessary to environmental demands \cite{Juster2010-kw}. Over time, ``wear-and-tear'' of this adaptive stress-response causes a subclinical accumulation of dysfunction known as ``allostatic load'' \cite{Juster2010-kw}. We hypothesize that these allostatic changes may be asymmetric, causing a coherent, population-level drift in equilibrium biomarker values with age, and ultimately leading to accumulating biomarker values in particular directions.

Directly estimating an individual's allostatic load remains an open challenge \cite{Juster2010-kw}, owing to the confounding effects of the underlying interaction networks \cite{Cohen2013-hj}. Instead, most algorithms infer allostatic load by outlier detection \cite{Juster2010-kw, Cohen2013-hj} or other symmetric indicators, agnostic to any preferred biomarker accumulation direction \cite{Liu2021-zu, Yashin2007-py}.  These approaches have not been reconciled with generic, age-associated biomarker accumulation, which proceeds in preferred directions \cite{Sehl2001-ld}. It therefore remains unclear how allostatic load leads to worsening health. Other theories posit that outlying biomarker values indicate damage, which promotes further damage e.g.\ as quantified by the number of health deficits (``frailty index'', FI)\cite{Mitnitski2017-gn,Blodgett2017-jp, Stubbings2020-uc}. Needed is direct evidence of allostasis and how it is associated with worsening health.

Instability is another mechanism for accumulation. While linear accumulation is the norm \cite{Sehl2001-ld}, some biomarkers accumulate exponentially with age e.g.\ senescent cells \cite{Karin2019-gf} and the FI \cite{Mitnitski2015-ia}. Exponential growth indicates an instability \cite{Avchaciov2022-ws}. Nevertheless, a weak instability can appear linear until advanced ages.  As a result, it remains unclear whether age-related accumulation proceeds due to a shifting equilibrium, a weak instability, or some hybrid of the two.

Operationalizing and quantifying homeostatic changes is challenging \cite{Juster2010-kw} because homeostasis is a property of the whole system, not individual constituent parts \cite{Billman2020-lj,Cohen2013-hj}. In the language of complexity science \cite{Corning2002-jk}, homeostasis is an emergent property of a network of interacting variables.  Each variable measures a part of the system, but changes to one part can be balanced by other parts. For example, heart rate declines with age but can be compensated for by increased stroke volume \cite{Sehl2001-ld} in order to maintain arterial blood pressure \cite{Billman2020-lj}. The essential aspects of homeostasis are: (i) a multivariate interacting dynamical system, (ii) an equilibrium state, which may vary with age (allostasis), (iii) the system spontaneously returns to the equilibrium state (dynamical stability), and (iv) stresses (and interventions) provide random shocks to the system. Altogether, homeostasis can be operationalized as a multivariate, mean-reverting stochastic process \cite{Yashin2007-py}. 

Dynamical stability analysis uses eigen-decomposition to probe the stability of arbitrary systems \cite{Ledder2013-em, Ives1995-fs}. The system is first linearized around an equilibrium position \cite{Ledder2013-em}.  Orthogonal eigenvectors are then identified that decouple the interactions between variables. Eigenvectors are composite health measures that serve as \textit{natural variables} since they do not interact or compensate for each other, and so can be analyzed individually. Each such natural variable has an associated eigenvalue that determines its stability via a characteristic recovery rate or timescale ($-\text{eigenvalue}=\text{rate}=  \text{timescale}^{-1}$). A system is stable if and only if all recovery rates are positive \cite{Ledder2013-em}. Conversely, dynamical instability arises only if at least one recovery rate is negative.

We confront homeostasis with minimal assumptions. We seek generic changes to biomarker equilibrium and stability within aging organisms. We investigate multiple longitudinal datasets with multiple organisms (mice and humans) and multiple outcomes (dementia and death). In contrast to earlier work by Sehl and Yates \cite{Sehl2001-ld}, our model is multivariate and generic so that we can model homeostasis without constraining its dynamical behaviour. We find that allostatic drift is consistent with the observed data. Importantly, we find that  a small set of natural variables drive mortality and can be used to characterize an individual's health state. We do not observe any dynamical instabilities.
%

\section*{Model}
To analyse stability for deterministic \cite{Ledder2013-em} or stochastic \cite{Ives1995-fs} dynamics, we use a linear approximation near a stable point, 
\begin{align}
    \vec{y}_{i n+1} &= \vec{y}_{i n} + \boldsymbol{W}\Delta t_{i n+1}(\vec{y}_{in} - \vec{\mu}_{in}) +\vec{\epsilon}_{i n+1}, \nonumber\\
    \vec{\epsilon}_{i n+1} &\sim \mathcal{N}(0,\boldsymbol{\Sigma}|\Delta t|_{i n+1} ) \nonumber \\
    \vec{\mu}_{in} &\equiv \vec{\mu}_{0}+\boldsymbol{\Lambda}\vec{x}_{in}+ \vec{\mu}_{age} t_{in}
    \label{eq:sf0}
\end{align}
where $i$ indexes the individual, $n$ indexes the timepoint, $t_n$ is the age, $\vec{y}$ is a vector of observed biomarkers,  and $\vec{x}$ is a vector of covariates that includes sex. $\boldsymbol{W}$, $\boldsymbol{\Sigma}$, and $\boldsymbol{\Lambda}$ are constant matrices, independent of $i$ and $n$. If we take the average over individuals (indicated by angled brackets) then we can obtain rates of average change
\begin{align}
    \frac{\langle \Delta y_{ij n+1} \rangle}{\langle \Delta t_{in+1} \rangle} &=   W_{jj}\langle y_{ijn} - \mu_{ijn} \rangle + \sum_{k\neq j} W_{jk}\langle y_{ikn} - \mu_{ikn} \rangle. \label{eq:dydt}
\end{align}
We see that changes to the mean of a particular biomarker, $y_j$, are due either to recovery of $y_j$ towards the equilibrium position, $\mu_j$, or because of interactions with a compensating variable, $y_{k\neq j}$ through off-diagonal elements of $\boldsymbol{W}$.  This provides both a mechanism for biological redundancy --- if the organism can actively influence some of the $y_k$ then it can use them to steer others --- and a mechanism for mutual dysfunction --- since $\boldsymbol{W}$ couples dysregulation of $y_{k\neq j}$  to that of $y_j$. In Supplemental Section~S8 we show that equation~(\ref{eq:sf0}) approximates general nonlinear dynamics \cite{Ledder2013-em}. We also specifically show that it approximates the stochastic process model \cite{Yashin2007-py}, a framework for aging biomarker dynamics. Note that the model permits unequally-spaced sampling of individuals through $\Delta t_{in+1}$, which is the time between measurements of individual $i$ at time $t_n$ and $t_{n+1}$.

The stability of the model depends on the eigenvalues of $\boldsymbol{W}$. We can decouple variable means with the eigenvector transformation matrix $\boldsymbol{P}$. We obtain 
\begin{align}
    z_{ijn+1} &= z_{ijn} + \lambda_j\Delta t_{in+1}(z_{ijn} - \tilde{\mu}_{ijn}) +\tilde{\epsilon}_{ij}, \label{eq:z0}
\end{align}
where $\vec{z}_n\equiv \boldsymbol{P}^{-1}\vec{y}_n$, $\lambda_j \equiv P_{j\cdot}^{-1} \boldsymbol{W} P_{\cdot j}$, $\tilde{\vec{\mu}}_{n} \equiv \boldsymbol{P}^{-1}\mu_{n}$ and $\tilde{\vec{\epsilon}} \equiv \boldsymbol{P}^{-1}\vec{\epsilon}$. We refer to $\vec{z}$ as \textit{natural variables}. The natural variables build correlations only through the noise term, $\tilde{\epsilon}$ --- in addition to any correlated initial conditions. The system is mutually-diagonal if $\tilde{\epsilon}$ is uncorrelated. While our dynamics are discrete, it is also helpful to consider continuous dynamics corresponding to the limit $\Delta t \rightarrow 0$; see Figure~\ref{fig:1} and Box~1 (also Supplemental Section~S8 for more details). 

The parameters $\boldsymbol{W}$, $\vec{\mu}_0$, $\vec{\mu}_{age}$ and $\boldsymbol{\Lambda}$ are estimated from the data ($\boldsymbol{\Sigma}$ can also be).  The stochastic term, $\vec{\epsilon}$, is assumed to be normally distributed and independent across timepoints. See Supplemental Section~S6 for details. (For the remainder of the paper, we simplify notation by dropping the tilde and suppressing the individual $i$ and timepoint $n$ indices.) Optimal parameter values are selected by maximizing the likelihood. For uncorrelated noise this reduces to weighted linear regression.

If a mutually-diagonal system reaches steady-state --- having run long enough to forget initial conditions --- then the natural variables, $\vec{z}$, are the principal components, ranked by stability (Supplemental equation~(S49)). We used principal component analysis (PCA) as a preprocessing step.  If $\boldsymbol{P}$ is orthogonal (which it is from PCA) then Parseval's theorem states that $\langle \sum_j y_{jn}^2 \rangle = \langle \sum_j z_{jn}^2 \rangle = \sum_j  ( \text{Var}(z_j) + \langle z_j \rangle^2)$: this means that a single $z_{k}$ with large mean and variance can dominate that of the $\vec{y}$.

\begin{mdframed}[frametitle={Box 1. Ordinary differential equation (ODE) behaviour}]
        \begin{wrapfigure}{r}{0.45\textwidth}
            \begin{center}
            \includegraphics[width=.45\textwidth]{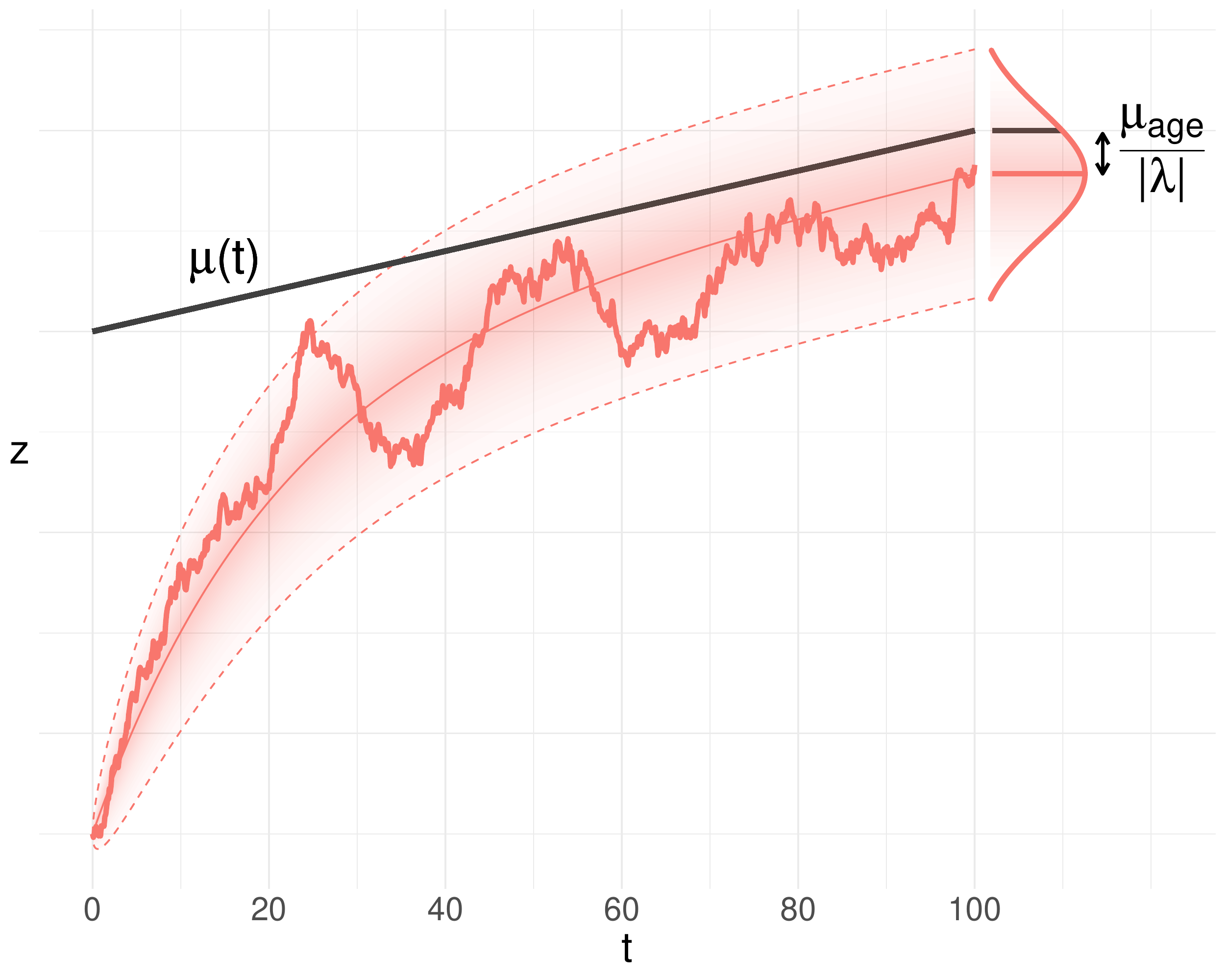} 
            \end{center}
            \caption{Simulation example of a stable system, with $\lambda < 0$. Initial conditions can differ from $\mu(t)$. A stable system is attracted to $\mu(t)$ (black line), but will be offset by $-\mu_{age}/|\lambda|$ in the steady-state.  ODE solutions are super imposed for mean and variance (dotted lines are 95\% interval). Fill density is proportional to probability density. Observing an ensemble at any time will yield Gaussian statistics.} \label{fig:1}
            \vspace{-30pt} 
        \end{wrapfigure}
        
        Consider a 1-dimensional space, $z$. If we take the limit $\Delta t\to 0$ then equation~(\ref{eq:z0}) is a modified Ornstein-Uhlenbeck process. The mean and variance are solutions to ordinary differential equations. The mean is described by
        \begin{align}
            \frac{d}{dt} \langle z \rangle = \lambda(\langle z \rangle - \mu(t)) = \lambda\langle z \rangle - \lambda \mu_0 - \lambda \mu_{age} t \label{eq:meanode}
        \end{align}
        where $\mu_{age} t$ is the time-dependent part of $\mu_n$ and $\mu_0$ is the remaining part. The general solution of equation~(\ref{eq:meanode}) is
        \begin{align}
            \langle z \rangle(t) &= \big(\langle z_0 \rangle  - \frac{\mu_{age}}{\lambda} - \mu_0\big)e^{\lambda t} + \frac{\mu_{age}}{\lambda} + \mu_0 + \mu_{age} t \label{eq:meant}
        \end{align}
        where $\langle z_0 \rangle$ is the initially observed mean at $t=0$. The exponential factors dampen or aggregate the mean depending on the sign of $\lambda$. If $\lambda < 0$ the system is stable and once $|\lambda| t \gg 1$ a dynamical steady-state (ss) is reached,
         \begin{align}
            \langle z \rangle_{ss}(t) &= \frac{\mu_{age}}{\lambda} + \mu_0 + \mu_{age} t =   \mu(t) - \frac{\mu_{age}}{|\lambda|}. \label{eq:meanss}
        \end{align}
        The steady-state is equivalent to the system forgetting its initial conditions. This steady-state behaviour can explain the drift observed by Sehl and Yates \cite{Sehl2001-ld} (Supplemental Section~S8.6). In the steady-state, the mean drifts at a constant rate,
         \begin{align}
            \frac{d}{dt}\langle z \rangle_{ss}(t) &= \mu_{age}, \label{eq:ddtmeanss}
        \end{align}   
        but there is a constant lag of $\langle z - \mu \rangle_{ss} = \mu_{age}/\lambda$; Figure~\ref{fig:1} illustrates. Only when $\mu_{age}=0$ (no drift) is $\mu(t)$ the steady-state position. Outside of steady-state, the mean is displaced by
    \begin{align}
        \langle z - \mu \rangle(t) &=  \langle z - \mu \rangle(t_0) e^{\lambda(t-t_0)} + \frac{\mu_{age}}{\lambda}(1-e^{\lambda (t-t_0)}) = \text{Memory} + \text{Drift} \label{eq:dz} 
    \end{align}           
        for reference time $t_0$. The first term encodes the system's initial conditions, whereas the last term encodes long-time drifting behaviour. Systems near steady-state exhibit $\text{Memory} \ll \text{Drift}$.
        
        If $\lambda=0$ the system is marginally stable and preserves its initial conditions. If $\lambda>0$ the system is unstable and the initial conditions grow exponentially over time. In either case the steady-state is never reached.
        In contrast to the mean, for $\lambda < 0$ the variance eventually equilibrates, reaching a constant value. The variance is described by
        \begin{align}
            \frac{d}{dt} \text{Var}(z) = 2\lambda\text{Var}(z) + \sigma^2 \label{eq:varode}
        \end{align}        
        where $\sigma^2$ is the noise strength. The general solution is given by
        \begin{align}
         \text{Var}(z)(t) = \text{Var}(z_0)e^{2\lambda t} - \frac{\sigma^2}{2\lambda}( 1 - e^{2\lambda t} ) \xrightarrow[\lambda < 0,~t \to \infty]{\text{steady state}} \frac{\sigma^2}{2|\lambda|}. \label{eq:vart}
        \end{align}
        Approaching instability, with $\lambda\to 0$, the system accumulates noise. 
       
\end{mdframed}

\section*{Results}
We analysed four datasets originating from three studies: two human and two mouse. Our analysis focused on the key properties of homeostasis: stability and equilibrium position. We used model selection to compare our model to the null hypothesis and to pick an optimal model form (Supplemental Section~S5). 

We observed that both an interaction network, $\boldsymbol{W}$, and an equilibrium term, $\vec{\mu}$, were needed to optimally predict future biomarker values. We saw no evidence of nonlinear terms in the dynamics. We found that fitting equation~(\ref{eq:z0}) using principal components (PCs) yielded equivalent performance to the model with full flexibility, equation~(\ref{eq:sf0}), but was already in diagonal form. Hence for each dataset we analysed a set of decoupled, one-dimensional equations in $z_j$ (with $j$ sorted by stability, so that $j=1$ is the least stable in each study).

For covariates, we generally found non-significant improvements in prediction --- though we kept them to improve interpretability (to reduce confounding). The exception was the age covariate, $\mu_{age}$, which significantly improved the fit of the SLAM Het3 mice (SLAM C57/BL6 were almost significant). The presence of $\mu_{age}$ indicates allostasis in the form of a time-dependent homeostasis.

\begin{figure*}[!ht] 
     \centering
        \includegraphics[width=.9\textwidth]{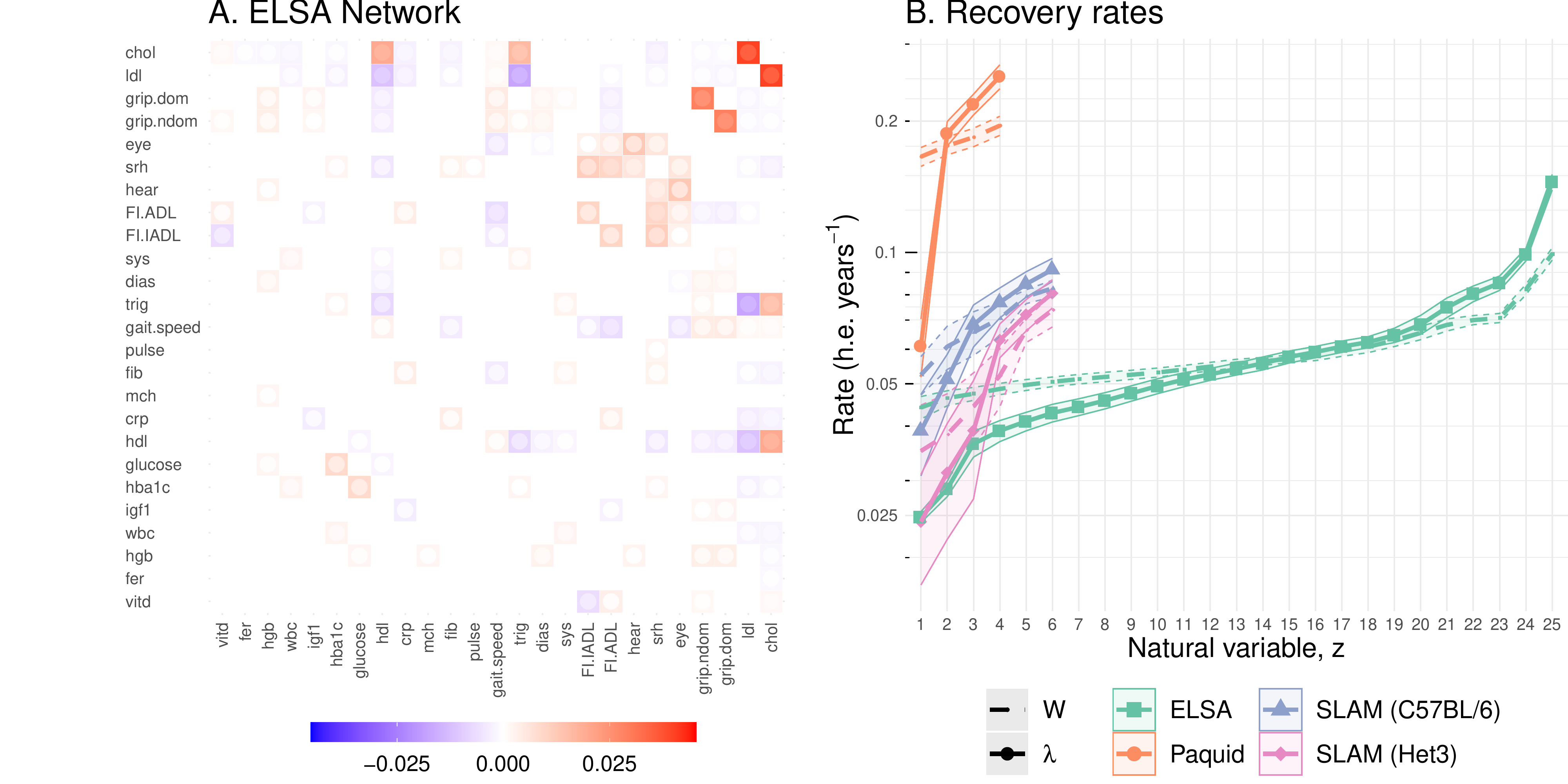}  
    \caption{\textbf{A.} ELSA interaction network. Tile colour indicates interaction strength (saturation) and direction (colour) of the interaction from the y-axis variable to the x-axis variable. Inner dot colour indicates the limit of the 95\% confidence interval (CI) closest to zero (more visible point indicates lower significance). Non-significant interactions have been whited-out. Diagonal has been suppressed for visualization (see dotted lines in B). The matrix is real and symmetric because the data were diagonalized by an orthogonal matrix (PCA).  Variables are sorted by diagonal strength in both A.\ and B.\ (increasing rate). 
    \textbf{B.} Recovery rates in human-equivalent (h.e.) years i.e.\ negative eigenvalues ($-\lambda$). The smallest recovery rates determine system stability \cite{Ledder2013-em}.  A recovery rate of $0.025$ implies $1-e^{-1}=63\%$ recovery after $-\lambda^{-1}=40$~years (95\% recovery after 120~years). The survival data all have similar minimum rates near $0.025$, whereas the dementia data was faster (Paquid). The dotted lines are network diagonals ($-W_{jj}$); the solid lines are rates ($-\lambda_j$). }
    \label{fig:networks0}
\end{figure*}

The interaction networks between variables can be represented by the respective weight matrices, e.g.\ Figure~\ref{fig:networks0}A. For ELSA we see expected relationships, e.g.\ total/HDL/LDL cholesterol, and non-dominant/dominant grip strength. ELSA also shows a block of like-variables including the FI-ADL, FI-IADL, self-reported health (srh), and gait speed, which could relate to frailty \cite{Pridham2023-yj}. See Supplemental Figure~S9 for networks from the other datasets.

The interactions between observed biomarkers prevents us from assessing stability directly. However, we can eigen-decompose the networks to yield an equivalent non-interacting network of natural variables. Each natural variable has a characteristic recovery rate, Figure~\ref{fig:networks0}B. All natural variables were stable, with $\lambda<0$. Faster recovery rates indicate higher stability (resilience) \cite{Ives1995-fs}. It takes 3 timescales for the system mean to recover 95\% of the way to equilibrium. For each mortality dataset the recovery timescale of its slowest natural variable was comparable to the organism's lifespan, $\approx 40~$human-equivalent years; only the mental acuity dataset (Paquid) was faster ($\approx 20$~years). In all datasets the rates for the natural variables extended to higher and lower values than the diagonal elements of the observed biomarkers (compare solid to dotted lines) --- indicating that network interactions play an important role in recovery dynamics. 

\begin{figure*}[!ht] 
     \centering
        \includegraphics[width=.9\textwidth]{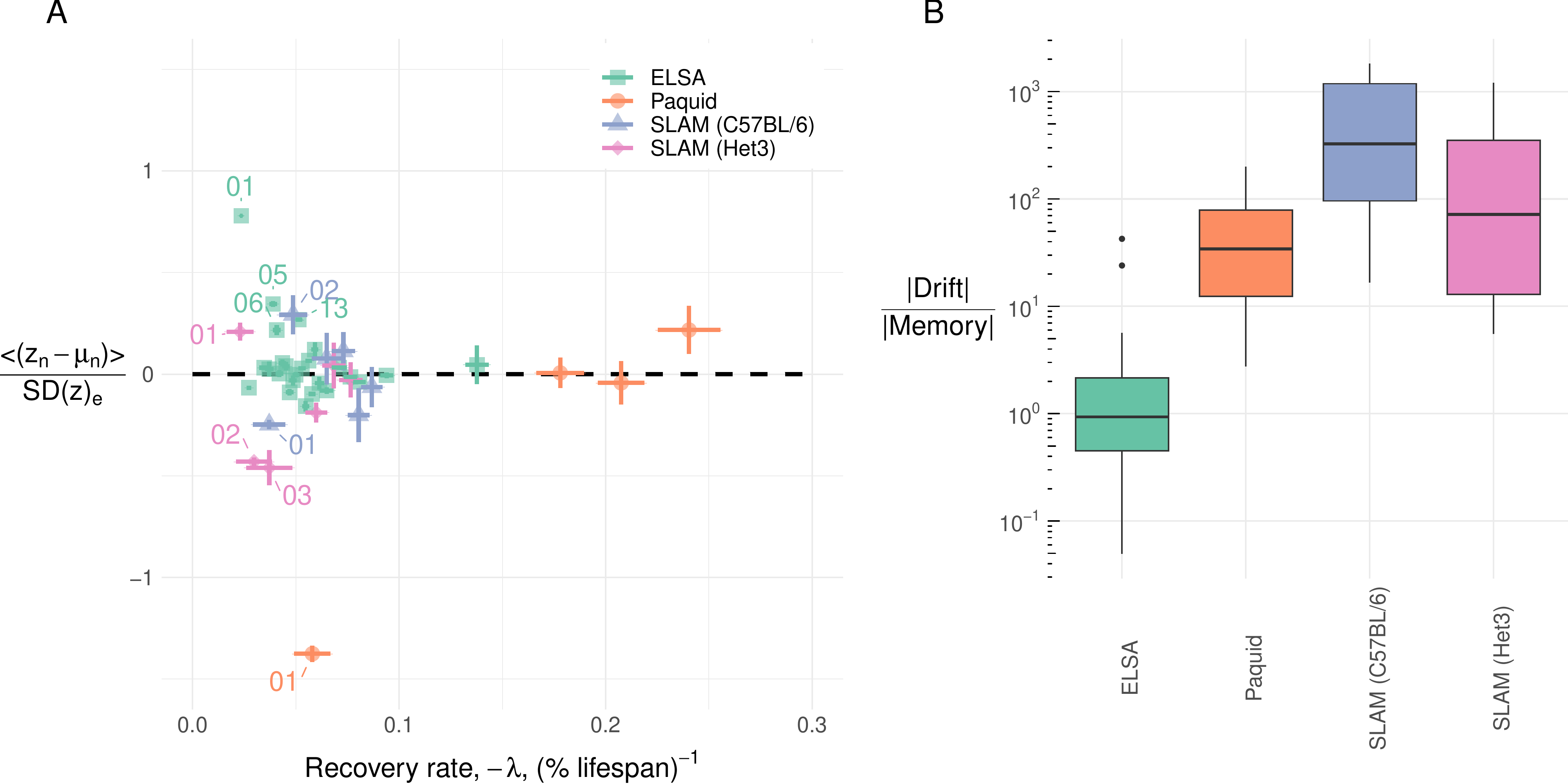}
    \caption{\textbf{A.}  Position relative to equilibrium vs recover rate. Most natural variables were homeostatic (near equilibrium at $0$). Some (labeled) variables were observed to be far from equilibrium; variables are labelled by rank e.g.\ $01 \equiv z_{01}$ has the fastest recovery (furthest left). \textbf{B.} Characterization of natural variable deviations from equilibrium using equation~(\ref{eq:dz}). Observe that ELSA is the only dataset where memory may dominate the system behaviour (ratio $ \lesssim 1 = 10^0$),  indicating that the followup period may have been too short to reach a steady-state. In both figures only mouse (SLAM) data points over age 80~weeks were used since biomarkers had a u-shaped curve over the lifespan \cite{Palliyaguru2021-jh}.}
    \label{fig:allo}
\end{figure*}

We summarize homeostasis in Figure~\ref{fig:allo}A, using the population means. If variables are in homeostasis then the mean should be close to $\mu_n$, where the scale is determined by the native dispersion. Each dataset had most natural variables near $0$ with a few outliers, such as $z_1$ for all datasets. (In contrast, the majority of observed biomarkers had large deviations from equilibrium --- see Supplemental Figure~S10). We characterized the natural variable dynamics using equation~(\ref{eq:dz}) in Figure~\ref{fig:allo}B. Excluding ELSA, most data points were in a steady-state, as indicated by their small memory term (relative to drift). The steady-state mean includes a drift caused by $\mu_{age}$. Across variables, the deviations from equilibrium observed in Figure~\ref{fig:allo}A, $\langle z_n  - \mu_n \rangle$, were very strongly correlated with $\mu_{age}$, with correlation coefficients: -0.988 ($p=2\cdot10^{-4}$, SLAM BL/6), -0.947 ($p=10^{-3}$, SLAM Het3), -0.989 ($p=0.01$, Paquid), and -0.302 ($p=0.14$, ELSA). This is consistent with equation~(\ref{eq:meanss}), and supports our use of an allostatic model with equilibrium drifts given by $\mu_{age}$. The smaller correlations observed with the ELSA dataset are consistent with the strong memory effect seen in Figure~\ref{fig:allo}B --- violating the steady-state assumption of equation~(\ref{eq:meanss}). ELSA may have failed to reach steady-state due to the limited followup period, which was the shortest of all datasets by a factor of $2$, or could indicate the confounding effects of medical interventions, which are not relevant for the other datasets. 

\begin{figure*}[!ht] 
     \centering
    \includegraphics[width=.9\textwidth]{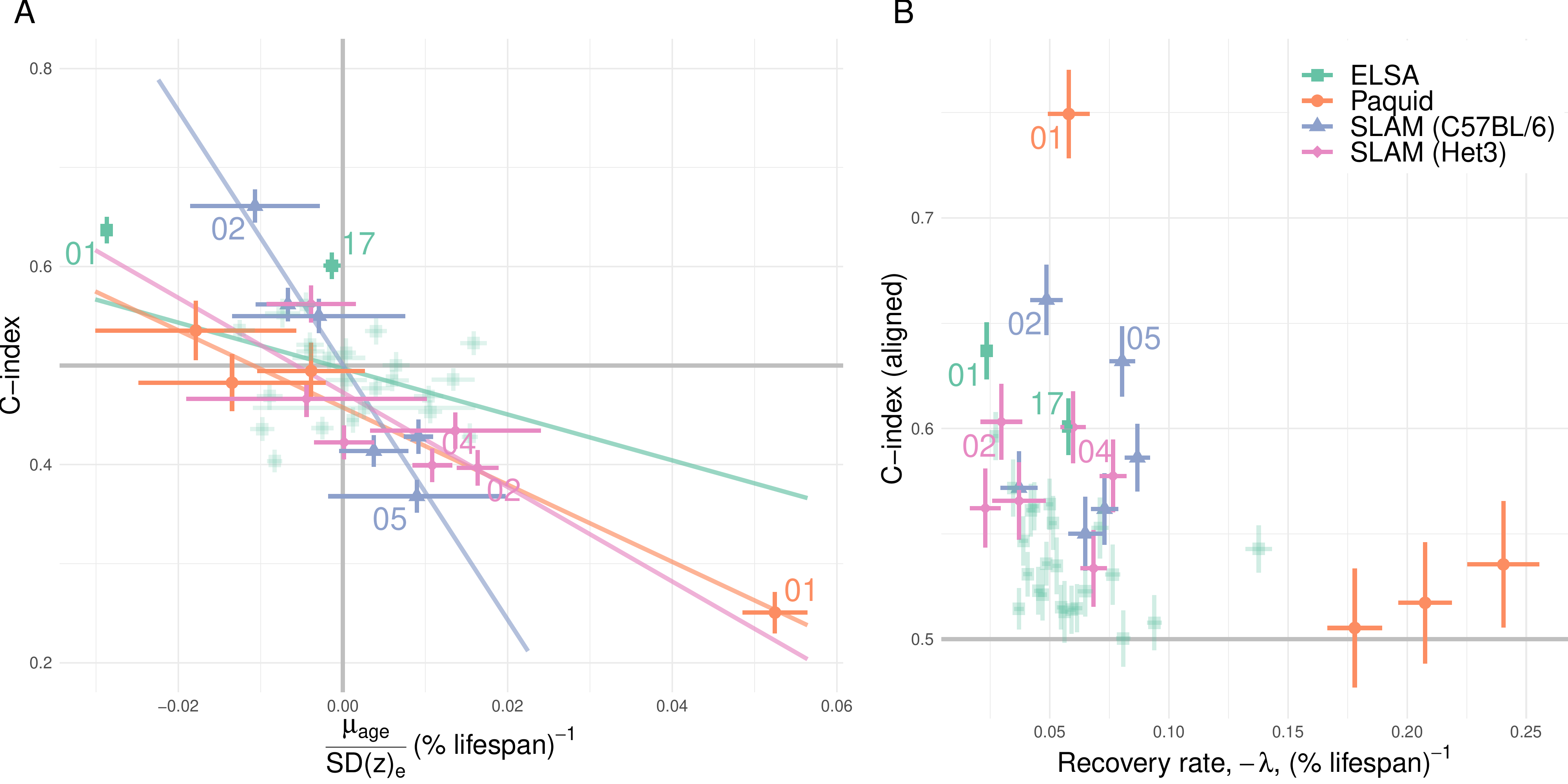}
    \caption{Survival effects. \textbf{A.} Allostasis drifts towards the risk direction, ``mallostasis''. The relationship appears to be linear (lines), with strong correlations: -0.96 (SLAM BL/6), -0.71 (SLAM Het3), -0.99 (Paquid), and -0.53 (ELSA). The equilibrium dispersion provides a native scale for each variable. High risk natural variables for each dataset have been labelled by eigenvalue rank (e.g.\ $z_1 \equiv 01$ has the smallest eigenvalue, $z_2 \equiv 02$ the second smallest, etc). \textbf{B.} Recovery rate (-eigenvalue), $-\lambda$, has an ambiguous relationship with survival. Smaller eigenvalues appear to be important survival dimensions (e.g. $01$ for ELSA and Paquid), but the overall correlation is weak ($\rho=-0.254$, $p=0.1$). The C-index measures the relative risk for pairs of individuals based on the value of $z_j$ (C-index of 0.5 indicates no risk; C-index larger than $0.5$ means small values are bad).}
    \label{fig:allo_survival0}
\end{figure*}

Most natural variables have small drift and are effectively homeostatic --- with only a few strongly drifting allostatic natural variables. The steady-state drift rate of natural variables, $\mu_{age}$, was correlated with the survival risk for each dimension: Figure~\ref{fig:allo_survival0}A. The correlations were typically strong: -0.958 ($p=0.002$, SLAM BL/6), -0.713 ($p=0.1$ SLAM Het3), -0.987 ($p=0.01$, Paquid), and -0.534 ($p=0.006$ ELSA); overall: -0.742 ($p=3\cdot10^{-8}$). The correlation was weakest for ELSA, which had not reached steady-state. The Cox proportional hazards coefficients, conditioned on age and sex, showed a similarly strong correlation with $\mu_{age}$, $0.70$ ($p=10^{-7}$, all data) (Supplemental Figure~S14). Furthermore, we see that the drift direction, $\text{sign}(\mu_{age})$, is the same as the risk direction ($p=0.0003$, Fisher test). Hence, not only does homeostasis drift with age, the direction of the drift is \textit{towards ill-health}. The primary risk directions were $z_1$ for ELSA and Paquid and $z_2$ for SLAM. Interestingly, $z_2$ of the Het3 mice is nearly identical to $z_1$ of the C57BL/6 mice in terms of covariates and survival effect --- hence $z_1$ of the C57BL/6 is also likely a key risk direction (Supplemental Figures~S11 and S18). Regardless, $z_1$ or $z_2$ exhibited the strongest survival effect for their each dataset (Figure~\ref{fig:allo_survival0}A). These variables also both had small eigenvalues ($z_1$ is rank 1 and $z_2$ is rank 2). However, this relationship between survival and eigenvalue magnitude does not appear to generalize, see  Figure~\ref{fig:allo_survival0}B. 

\begin{figure*}[!ht] 
     \centering
        \includegraphics[width=.9\textwidth]{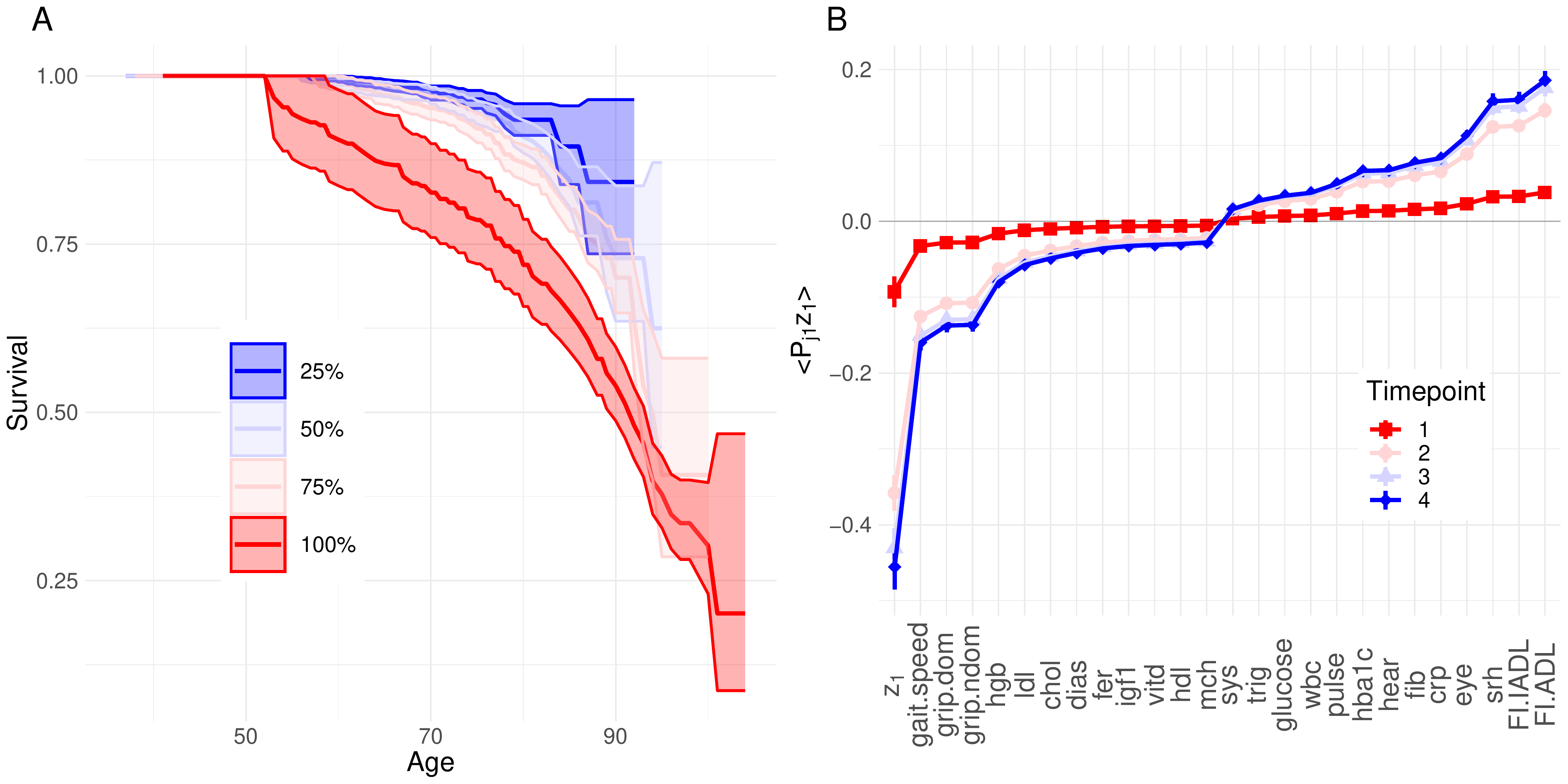}
    \caption{\textbf{A.} Composite health measure of survival $b\equiv (\vec{\mu}_{age}^T\vec{z})$, stratified by quartile (ELSA). Separation is excellent, indicating a strong survival predictor.  Fill is 95\% confidence interval. See Supplemental Figure~S15 for the other datasets. \textbf{B.} Natural variables can drive changes in observable biomarkers. The $z_1$ mean is accumulating in the negative direction. This accumulation is mapped into observable variables with $\langle P_{j1} z_1 \rangle$ for indicated timepoints each separated by approximately 4 years. The drift direction is overwhelmingly unhealthy: increased disability measures (srh, eye, hear, FI.ADL and FI.IADL --- high is bad), decreased physical ability scores (gait and grip), increased inflammation (crp), increased glucose, etc. The effect of the drift is concentrated in $z_1$ but dilute across its covariates, which could make the effect of unhealthy $z_1$ subclinical in the observed biomarkers.  All variables are on standardized scale. Similar effects were observed for the other datasets (Supplemental Figure~S13).}
    \label{fig:ba_survival0}\label{fig:zdrift}
\end{figure*}

As an illustration of the utility of the correlation between survival and $\mu_{age}$, we consider a simple summary health measure. The Cox proportional hazards model assumes the hazard scales as $\exp{(\vec{\beta}^T\vec{z})}$, where the $j$th coefficient, $\beta_j$, is the log--hazard ratio per unit increase of $z_j$. As mentioned in the previous paragraph, $\vec{\beta} \sim \vec{\mu}_{age}$ were correlated; the relative hazard can therefore be approximated by $\exp{(\vec{\mu}_{age}^T\vec{z})}$. Indeed, we observed that $b\equiv \vec{\mu}_{age}^T\vec{z}$ is an excellent predictor of survival, see Figure~\ref{fig:ba_survival0}A and Supplemental Figure~S15.

The natural variables with large $|\mu_{age}|$ will eventually experience the largest drift, according to equation~(\ref{eq:ddtmeanss}). $z_1$ in Figure~\ref{fig:zdrift}B is an example of such an accumulating variable. The other variables with large $|\mu_{age}|$ experienced a similar accumulation (Supplemental Figure~S13). For an orthogonal transformation such as $\boldsymbol{P}^{-1}$, the sum of the variance and squared mean is conserved (Parseval's theorem). Natural variables with large means and variances will therefore disproportionately affect the means and variances of observed biomarkers. The effect is demonstrated for ELSA in Figure~\ref{fig:zdrift}B. As the dominant natural variable drifts it influences observable biomarkers to drift as well. 

Slower recovery rates (eigenvalues) take longer to forget  perturbations, causing the associated natural variables to accumulate variance due to noise. Recall that the slowest recovery rates were on the order of a lifespan (Figure~\ref{fig:networks0}B). The Pearson correlations between the estimated variance and rate (-eigenvalue) were strong: -0.852 ($p=0.03$, SLAM BL/6), -0.802 ($p=0.05$ SLAM Het3), -0.998 ($p=0.002$, Paquid), and -0.764 ($p=9\cdot 10^{-6}$ ELSA) (log-log scale; see Supplemental Figure~S16). Hence the variances we observe at old ages will be dominated by the variables with the smallest eigenvalues, $\lambda$ (e.g.\ $z_1$ and $z_2$). As we have seen before, these variables are often --- but not always --- strongly associated with adverse effects, depending on the drift rate $\mu_{age}$. This suggests that most of the age-related changes to health were concentrated in a few $z_k$ which drive both biomarker drift (mean) and dispersion (variance). Growing variance along these dimensions may capture individual accumulation of stochastic damage, such as genetic damage or disease.

\section*{Discussion}

We fit a homeostasis model of equilibrium and stability to four longitudinal aging datasets (two mouse and two human) using generic health biomarkers. Our model is lightweight, can be estimated using standard statistical algorithms, and is sufficient to capture essential information about the aging process. Health biomarkers have an equilibrium position, $\vec{\mu}$. Their corresponding stochastic term, which has covariance $\boldsymbol{\Sigma}$,  represents random stresses that drive individuals away from equilibrium; as well as residual effects such as individual variability and nonlinearities. An interaction network, $\boldsymbol{W}$,  pulls individuals towards equilibrium either through recovery (diagonal terms) or by one variable compensating for another (off-diagonal terms). By eigen-decomposing $\boldsymbol{W}$ we transformed the dataset into non-interacting, natural variables --- which are linear combinations of the input biomarkers. This increases interpretability and simplifies analysis. The stability of the system is described by the recovery rates of the natural variables --- which are the corresponding eigenvalues with flipped signs, $-\lambda$.

We modelled homeostasis as stability around an equilibrium mean value. Stability can only be assessed using the natural variables because the original biomarkers have interactions between them. Homeostasis was violated by some natural variables (Figure~\ref{fig:allo}A). Although most natural variables had average values near the homeostatic equilibrium --- indicative of homeostasis --- several were far away. We determined that this latter group were out of homeostasis because they were chasing drifting equilibrium positions from behind. This equilibrium drift with age represents allostasis, i.e. a changeable equilibrium position.

Allostatic variables accumulated over the course of the study period as they chased after the drifting equilibrium, systematically increasing or decreasing. This was facilitated by an age-dependent equilibrium position, governed by $\mu_{age}$, and was typically accompanied by a small eigenvalue. The gap between the population and allostatic equilibrium position is governed by $\mu_{age}/\lambda$, so a small $\lambda$ enables a large gap such that the entire population drifts coherently towards the moving equilibrium --- causing population-level accumulation. This makes the linear drift term, $\mu_{age}$, the primary culprit for causing biomarkers to drift with age. Presumably $\mu_{age}$ arises from either (a) the effects of unknown biomarkers/mechanisms not included in the model (i.e.\ $\mu_{age} t \approx \sum_{k\neq j} W_{jk}\langle y_{ikn} - \mu_{ikn} \rangle$ in equation~(\ref{eq:dydt}) for a set of unknown $y_{ikn}$), or (b) asymmetric stressors, which cannot be captured by our symmetric stochastic term (e.g.\ there is no such thing as negative damage so health deficits skew positive \cite{Mitnitski2015-ia}).

The transformation to natural variables effectively compressed the drifting (accumulating) mean of many variables into a small number of natural variables. The natural variables can be thought of as the underlying cause of the observed biomarker drift (Figure~\ref{fig:zdrift}B). In this manner, the widely observed age-related decline in biomarkers\cite{Sehl2001-ld} are governed by a few  natural variables --- which are not directly observed. The effect is spread out by the transformation, potentially hiding the observed biomarker decline below diagnostic thresholds. This may be a redundancy mechanism: the network permits the biological system to spread out the age-related decline to keep biomarkers in healthy ranges for longer. The trade-off may be that many biomarkers would reach unhealthy ranges concurrently, leading to multisystem dysfunction. For example, the effects of chronic kidney disease are mild and non-specific until the patient nears kidney failure --- at which point multisystem failure is imminent, typically leading to death via cardiovascular disease \cite{Mizdrak2022-jl}. This tradeoff assumes that diagnostic thresholds represent critical values beyond which deterioration of a biological system accelerates. Univariate dynamical modelling of senescent cell count in mice \cite{Karin2019-gf} and \emph{E.\ coli} membrane integrity \cite{Yang2023-xj} supports the existence of such critical values, where repair mechanisms saturate and decline accelerates. From this perspective, an individual's robustness would depend on their buffer space available to absorb new insults, which could be quantified by the natural variable scores together with the stressor effect strength which should be proportional to the noise $\sigma$. 
Consistent with this perspective\cite{Karin2019-gf}, the allostatic drift rate, $\mu_{age}$, strongly correlated with the mortality/dementia risk associated with each natural variable (Figure~\ref{fig:allo_survival0}A and Supplemental Figure~S14). Since $\mu_{age}$ is the steady-state drift rate of the mean, the steady-state behaviour is continually worsening health due to the drifting mean. Prior work on operationalizing allostasis has neglected the existence of a preferred risk direction, instead using the absolute distance from allostasis as a mortality factor\cite{Yashin2007-py, Cohen2013-hj, Liu2021-zu}, irrespective of whether biomarkers are high or low. In contrast, our results indicate that numerous natural variables do, in fact, have preferred risk directions. Aging researchers should be aware of this symmetry breaking. This means that the adaptive changes due to allostasis at best mitigate declining health and, at worst, lead to a further decline in health. We refer to this phenomenon as ``mallostasis'': the tendency of an aging biological system towards an ever-worsening equilibrium position. We have used this phenomenon both to identify important survival variables and to generate a novel composite health measure.

Our key quantitative results coincide with three key qualitative predictions made by allostatic load theory: (i) a shifting equilibrium position for biomarkers indicative of adaptive changes (allostasis, Figure~\ref{fig:allo}A), (ii) the shift is associated with adverse outcomes (mallostasis, Figure~\ref{fig:allo_survival0}A), and (iii) the shift is subclinical due to compensating mechanisms between biomarkers (transformation, Figure~\ref{fig:zdrift}B) \cite{Juster2010-kw}. This is compelling evidence that allostasis is a generic aging phenomenon, rather than being specific to neuroendocrinology. Our proposed composite health measure is therefore a novel estimator of allostatic load. In contrast to conventional estimators \cite{Juster2010-kw}, we were able to estimate allostatic drift directly as $\mu_{age}$. Our results rely on using natural variables, which are canonical coordinates that greatly simplify analysis.

Allostatic load is believed to arise from the long-term costs of short-term protection against stressors \cite{McEwen2000-fn}, making it an example of antagonistic pleiotropy \cite{Gavrilov2002-pb}. Alternatively, long-term costs could reflect imperfect repair. Regardless, long-term costs that accumulate in a given direction would lead to the allostatic drift which we have observed and characterized. Furthermore, we observed slow dynamical rates for the dominant mortality-risk natural variables (Figure~\ref{fig:networks0}B). Accordingly, the dynamical timescale of these effects are comparable to the organismal lifetime --- consistent with long-term costs. 

Interestingly, we did not observe any instabilities or nonlinearities. We had expected that ``allostatic overload'' --- the final state of allostatic load theory \cite{Juster2010-kw} --- would operationalize as an instability.  However, instabilities, and their associated exponential growth, are rare among health biomarkers \cite{Sehl2001-ld}; although they are observed in summary measures of health such as the FI (frailty index) \cite{Mitnitski2015-ia}. Other unstable, FI-like variables can be extracted from generic biomarkers using nonlinear techniques such as a deep neural network \cite{Avchaciov2022-ws}, diagnostic thresholds \cite{Blodgett2017-jp}, or quantile-based preprocessing \cite{Stubbings2020-uc, Stubbings2021-ug}. Since we did not see evidence of instabilities or other nonlinearities in the natural variables, the nonlinear embedding or discretization should be considered as a possible cause for observed FI-like instabilities. It may be that biological systems naturally suppress nonlinear effects of aging --- obscuring the effects --- or, conversely, that aging is primarily a linear phenomenon that slowly pushes individuals towards nonlinear tolerance thresholds for dysfunction/damage, e.g. saturation of repair \cite{Karin2019-gf,Yang2023-xj} and/or emergence of chronic disease. A non-trivial issue is that exponential growth often appears linear, for example the FI in mice and younger humans ($\lesssim 85$~years old) \cite{Rockwood2017-mg}. Nonlinear effects in biomarker dynamics may require special populations, such as the ill or exceptionally old, to be observed. 
The key model variables, $z_1$ and $z_2$, dominate the aging process. These natural variables with smaller $\lambda$ carried the majority of the variance and become the dominant principal components in the steady-state model (Supplemental Figure~S16 and equation~(S49), respectively). Applying Parseval's theorem, these variables will control the variance of directly observed biomarkers. Since they also dominate the means via allostatic drift, they will determine the aging phenotype that we observe. Both effects get stronger with age. This means that the empirically observed age-related changes in the mean and variance of biomarkers will be predominantly caused by only a few key natural variables. Hence the nearly-universal linear decline in health biomarkers observed by Sehl and Yates \cite{Sehl2001-ld} may simply be a few declining natural variables spread across the observed biomarkers (Supplemental Section~S8.6). Furthermore, this implies that a single dimensional decline can drive many observed biomarkers, which is the foundational assumption of ``biological age'' estimators \cite{Klemera2006-mw, Jylhava2017-wc}. Our results provide much needed support for such low dimensional representations of aging --- which should become increasingly accurate with advancing age since the means of the key natural variables grow fastest, and their variances grow largest.
The natural variables, $z$, should be good choices for targeting and monitoring interventions. They are prospective biomarkers with the convenient property that if you can intervene on one it will not affect the others. In contrast, we know from the network of interactions that intervening on any single biomarker is likely to affect many other biomarkers. In the steady-state, the mallostatic drift rate, controlled by $\mu_{age}$, is a proxy for the hazard and therefore identifies the most important targets of intervention. The coefficients of the transformation, $\boldsymbol{P}$, provides both hints at what mechanisms each $z_j$ is capturing as well as a map for which biomarkers will be affected by interventions on $z_j$. For example, $z_1$ of ELSA shares many features with frailty: strong age dependence, large effect in gait, weakness (grip strength), disability and self-reported health, and large survival effect \cite{Pridham2023-yj}. $z_1$ is thus a prospective biomarker of frailty and can be used both to monitor an individual's frailty and to engineer interventions. The strong signals we see in Figure~\ref{fig:zdrift}B for gait, grip strength and activities of daily living are hints that loss of physical fitness is one mechanism by which frailty proceeds and therefore one mechanism by which we can intervene, consistent with a meta-analysis which has shown that physical activity can reduce frailty in humans \cite{Negm2019-rt}. Remarkably, we observed other prospective targets in addition to $z_1$ of ELSA. Given an organism and set of biomarkers, each $z_j$ with substantial drift should be considered a prospective intervention target, with the faster drifting being the most important. 

We note a few limitations of our study. We assumed linear, time-invariant interactions through the network, $\boldsymbol{W}$ --- following previous work that suggested that interactions are linear and time-invariant \cite{Farrell2022-lw} (as are the principal components \cite{Pridham2023-yj}). The networks we extracted were symmetric and hence acausal due to our use of PCA as a preprocessing step, although we did estimate more general networks and found they performed no better. This could be a consequence of the data which were entirely observational, obfuscating causality. Understanding interventions using our results is similarly subject to the caveat that biomarkers behave the same whether they are observed or intervened upon \cite{Dawid2010-mv}. This seems plausible since observational studies include everyday interventions such as disease, medicine, or lifestyle changes. Finally, our model is at the population-level and hence we cannot resolve homeostatic changes at the individual level.

We see exciting opportunities for future work. Our observation that principal components could be effectively used as independent variables  suggests that more complex statistical models could also be applied. For example, individual-level model parameter estimates via mixed-effects modelling would help to determine whether individual health changes are gradual or sudden (or possibly critical \cite{Scheffer2012-sx}). Changes in our model parameters due to age, chronic or acute illness, or medical interventions is particularly interesting, but will require specialized datasets to assess. Fortunately, small datasets are tractable with our linearized model. The generic nature of the model and its ability to find accumulating natural variables could also be applied to other biological or temporal scales. Others have postulated that damage aggregates due to dysfunction in regulatory systems or other intermediate scales \cite{Lopez-Otin2013-pv}, which could be tested. Composite health measures, including biological age \cite{Jylhava2017-wc}, are also interesting to explore using our approach. Applying our approach using multiple biological ages as biomarkers \cite{Li2020-hl} will naturally extract salient information regarding stability and mallostasis, as well as a smaller set of essential natural variables. New datasets will open up new opportunities for this analysis pipeline. It is interesting to consider leveraging the effect of the natural variables to intervene and observe in clever ways. For example, $z_1$ appears to be a biomarker of frailty, which affects both mental and physical health \cite{Negm2019-rt}, hence we could potentially intervene based on a physical mechanism but monitor using cognitive changes.

We have developed and applied a lightweight network model that includes the salient features of homeostasis: equilibrium values and recovery rates. Equilibrium values are allowed to drift, to accommodate allostatic changes. Across datasets and species we consistently observed that the linear decline of biomarkers with age was governed by a small set of accumulating natural aging variables. This accumulation can be described as mallostasis: homeostatic dysfunction and associated declining health. These variables appear to be important measures of age-related decline, including health and mortality. Their effects are spread out by a network of interactions, driving drift in the observed biomarkers, and potentially diluting and obfuscating the effects of age.  We find that generic biomarkers spontaneously move towards an equilibrium position which is itself continuously drifting towards ill-health. Mallostasis is a generic feature of the aging process.

\section*{Methods}
\subsection*{Materials}
We used 4 longitudinal datasets originating from 3 studies (organism, primary outcome): Paquid (human, dementia) \cite{Proust-Lima2017-uz}, SLAM (mouse, death) \cite{Palliyaguru2021-jh,Palliyaguru2021-hj} and ELSA (human, death) \cite{Banks2021-jt}. We directly modelled biomarkers, $\vec{y}$, and included covariates, $\vec{x}$, in the homeostatic term, $\vec{\mu}$, using equation~(\ref{eq:sf0}).

The Paquid dataset is a random subset of 500 humans (212 males and 288 females) from the Paquid prospective cohort study, enriched in dementia prevalence \cite{Proust-Lima2017-uz}. Age range: 66-95~years-old. Individuals were measured on average every 3.2~years for a maximum of 9 timepoints.  We modelled four ordinal variables, including three measures of mental acuity: mini-mental state examination (MMSE), Benton visual retention test (BVRT) and Isaacs set test (IST), along with a self-reported depression score (CESD). We considered for covariates: sex, age and education level (completed primary vs not).

The Study of Longitudinal Aging in Mice (SLAM) includes two datasets, one for each mouse strain. Both include body composition measures and glucose serum at 12~week intervals starting at 7~weeks of age and continuing for the lifespan of each mouse \cite{Palliyaguru2021-hj}. Body composition and serum measurements were staggered and had to be imputed. Covariates included age and sex. We dropped 538/66138 measurements that were recorded after death, ostensibly these were coding errors. After preprocessing, the first dataset included 608 C57BL/6 mice (303 male and 305 female) measured on average every 6.2~weeks for a maximum of 20 timepoints (every 4.9~human-equivalent years). C57BL/6 mice are genetically similar (inbred) and prone to lymphoma and metabolic dysfunction \cite{Mitchell2015-pg}. The second included 611 Het3 mice (304 male and 307 female) measured on average every 4.2~weeks for a maximum of 27 timepoints (every 3.6~human-equivalent years). Het3 mice are a genetically heterogeneous cross of four inbred mice (including C57BL/6) \cite{Mitchell2015-pg}. We converted to human-equivalent years using the ratio of median survival times of each strain to ELSA. Full details of the study are described elsewhere \cite{Palliyaguru2021-hj,Palliyaguru2021-jh}.

The English Longitudinal Study of Ageing (ELSA) is a representative sample of English people aged 50 and over (with some younger) \cite{Banks2021-jt}. We used physical functioning questionnaire data and blood tests for 9330 humans (4063 males and 5267 females), reported at 4 timepoints, each separated by approximately 4 years. Our choice of 25 variables includes frailty measures, cardiometabolic biomarkers, and immune biomarkers (Supplemental Table~S1). We considered waves 2, 4, 6 and 8, since only these contained the full suite of biomarkers. Covariates included age and sex. We considered only individuals whom were present both in wave 2 and in subsequent waves, thus excluding new recruits.  Despite the large number of individuals, ELSA appeared to have the worst quality data due to high individual heterogeneity and low number of timepoints.

\subsection{Data handling}
All missing data were imputed. Dead individuals were also imputed, as it reduced bias due to mortality in simulated data (Supplemental Section~S4.1). We compared several imputation strategies, including carry forward/back, multivariate imputation using chained equations (MICE) \cite{Van_Buuren2010-oy}, and using our model to impute the model mean. Ultimately, we used carry forward/back followed by the model mean, except for ELSA which used each individual's mean biomarker value followed by the population multivariate normal mean then model mean. See Supplemental Section~S4 for details.

Estimation of our model, equation~(\ref{eq:sf0}), used Supplemental Algorithm~S1, which iteratively ($\times 5$) applied the maximum likelihood estimator:
\begin{align}
    \boldsymbol{\hat{\Lambda}} &=\langle |\Delta t_{in+1} |\vec{y}_{in}\vec{x}_{in}^T \rangle_{i,n} \big( \langle |\Delta t_{in+1} |\vec{x}_{in}\vec{x}_{in}^T \rangle_{i,n} \big)^{-1} -  \boldsymbol{W}^{-1}\langle \text{sign}(\Delta t_{in+1})(\vec{y}_{in+1}-\vec{y}_{in})\vec{x}_{in}^T \rangle_{i,n} \big( \langle |\Delta t_{in+1} |\vec{x}_{in}\vec{x}_{in}^T \rangle_{i,n} \big)^{-1}
\end{align}
for $\boldsymbol{\Lambda}$ (which includes $\mu_0$ through $x_0 = 1$), and
\begin{align}
    \boldsymbol{\hat{W}} &= \langle \text{sign}(\Delta t_{in+1})(\vec{y}_{in+1}-\vec{y}_{in})(\vec{y}_{in}-\vec{\mu}_{in})^T \rangle_{i,n} \big( \langle |\Delta t_{in+1}|(\vec{y}_{in}-\vec{\mu}_{in})(\vec{y}_{in}-\vec{\mu}_{in})^T \rangle_{i,n} \big)^{-1}  \label{eq:west}
\end{align}
for $\boldsymbol{W}$, where the expectation values are to be taken over times, $n$, and individuals, $i$. For the diagonal models we instead used weighted linear regression. Missing values were imputed with the model prediction after each iteration (except ELSA). Estimators are described and validated in Supplemental Sections~S6 and S7, respectively. We used a time-dependent Cox model to assess survival. We assumed stepwise constant covariates via start-stop formatting \cite{Moore2016-rh}. All correlations are Pearson. All errorbars are standard errors unless stated otherwise.

\subsection{Model assessment}
We simultaneously estimated both parameter uncertainty and model performance using the standard deviation from $100 \times$ repeat bootstrap resampling. We compared model performance using the root-mean squared error (RMSE) and mean absolute error (MAE). Both were estimated using out-of-sample bootstrap \cite{Hastie2017-wl}. In validation tests we found that a simple 632 estimator i.e., $\text{RMSE}_{632} \equiv 0.632 \cdot \text{RMSE}_{\text{test}} + 0.368 \cdot \text{RMSE}_{\text{train}}$, provided a good estimate for the true values of both performance metrics (Supplemental Figure~S7). 0.632 is the expected fraction of unique individuals in each bootstrap \cite{Hastie2017-wl}.

\section*{Acknowledgements}
ELSA is funded by the National Institute on Aging (R01AG017644), and by UK Government Departments coordinated by the National Institute for Health and Care Research (NIHR). A.R.\ thanks the Natural Sciences and Engineering Research Council of Canada (NSERC) for operating Grant RGPIN-2019-05888.

\section*{Author contributions statement}
A.R.\ conceived the project,  G.P.\ built the model and analysed the data.  All authors reviewed the manuscript. 

\section*{Additional information}
The authors declare no competing interests.

\section*{Data availability}
All data used are publicly available. The SLAM datasets are available from a previous publication\cite{Palliyaguru2021-jh}. Paquid is available from a software package\cite{Proust-Lima2017-uz}. ELSA \cite{Banks2021-jt} is available from the UK Data Service \url{https://ukdataservice.ac.uk/}. Software for fitting and simulating our model is available at \url{https://github.com/GlenPr/stochastic_finite_model}.


\renewcommand{\theequation}{S\arabic{equation}}
\renewcommand{\thefigure}{S\arabic{figure}}
\renewcommand{\thetable}{S\arabic{table}}
\renewcommand{\thesection}{S\arabic{section}}
\renewcommand{\thealgorithm}{S\arabic{algorithm}}

\setcounter{equation}{0}
\setcounter{figure}{0}
\setcounter{table}{0}
\setcounter{section}{0}
\setcounter{algorithm}{0}

\renewcommand*{\thefootnote}{\fnsymbol{footnote}}

\section{Supplemental Information}
\begin{figure*}[!h] 
        \centering
        \includegraphics[width=\textwidth]{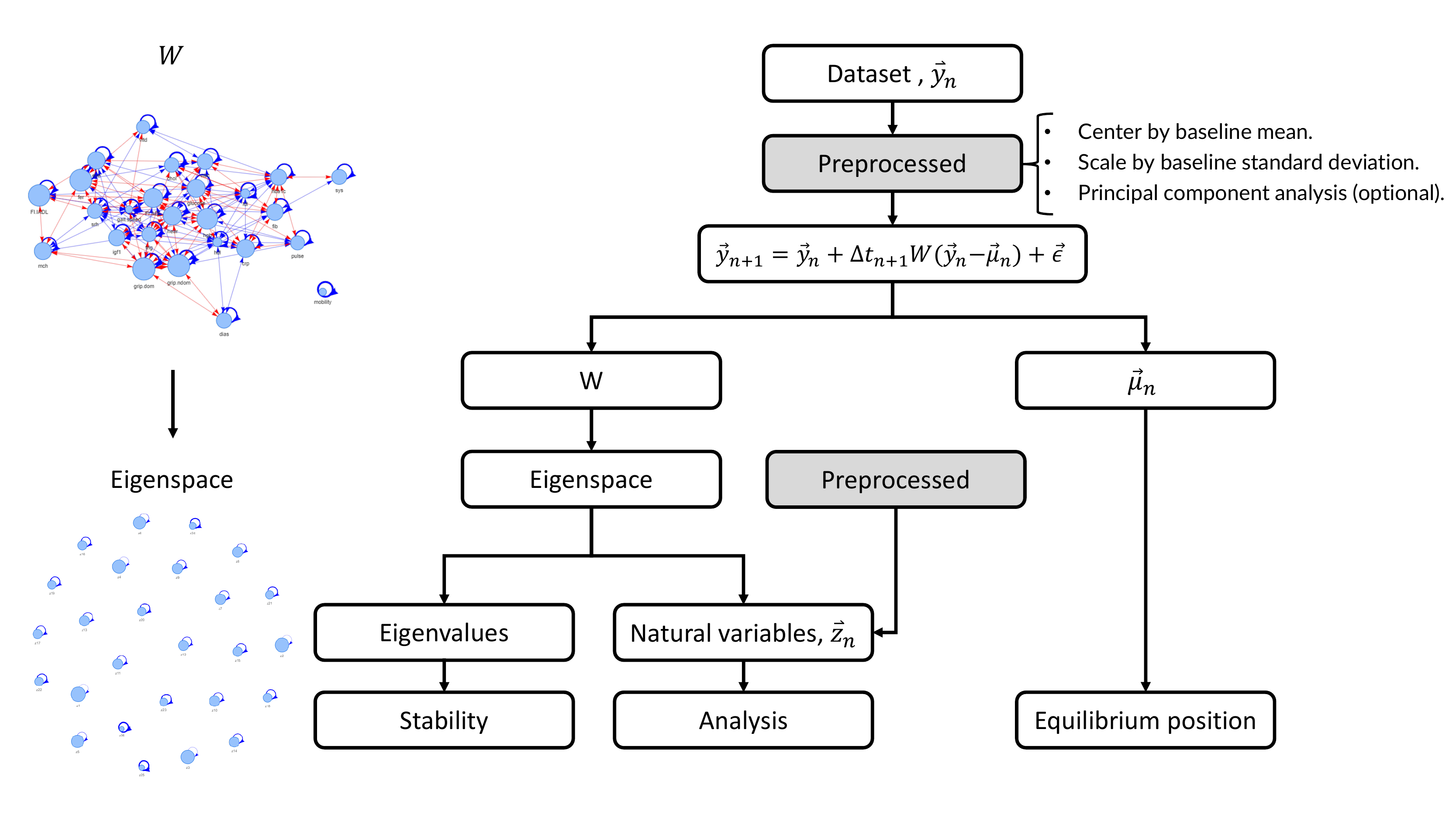}
    \caption{Study pipeline. We analysed four datasets using our proposed model. We model the dynamics of biomarkers, $\vec{y}_n$, over time using equation~(\ref{eq:sf}). Our model extracts an interaction network, $\boldsymbol{W}$, and equilibrium positions $\vec{\mu}_n$, where the latter are allowed to depend on covariates (e.g.\ age and sex). The estimated network, $\boldsymbol{W}$, captures arbitrary linear interactions between biomarkers which can be removed by working with the natural variables, $\vec{z}_n$. Natural variables are defined by a linear mapping into the eigenspace of $\boldsymbol{W}$. The natural variables allowed us to analyse stability. We were also able to infer changes to the mean and variance of the observed variables based on changes in the natural variables.}
    \label{fig:pipeline}
\end{figure*}

We modelled generic health biomarker data as a mean-reverting stochastic process. Our study pipeline is summarized in Figure~\ref{fig:pipeline}. Our model describes generic dynamics near an equilibrium solution (Section~\ref{sec:gen}). In this supplemental we provide additional information to support and validate both our methods and our conclusions. We provide a complete description of the data in Section~\ref{sec:materials} and how we preprocessed it in Section~\ref{sec:preprocess}. We performed a number of consistency checks on missing data which were imputed according to Section~\ref{sec:missing}. We consider variations of our model in Section~\ref{sec:select} which demonstrates that our final model best describes the data. We provide the mathematics necessary to estimate model parameters, along with an iterative estimation algorithm in Section~\ref{sec:estimation}. We then validate our algorithm using synthetic data in Section~\ref{sec:validation}. Additional mathematics useful for understanding our model and its connection to the literature are described in Section~\ref{sec:math}. Finally, we include additional results in Section~\ref{sec:results} which support our conclusions.

\begin{table}[!h]
\caption{Dataset Summary --- Biomarkers} \label{tab:datasets}
    \centering
\begin{threeparttable}
    \begin{tabular}{l l l l l}
    \hline
        Dataset & Species & Primary Outcome & Variable & Description \\ \hline
        SLAM (C57BL/6) & Mouse, C57BL/6 & Death & bw & Body weight \\ 
        SLAM (C57BL/6) & Mouse, C57BL/6 & Death & fat & Total fat mass \\ 
        SLAM (C57BL/6) & Mouse, C57BL/6 & Death & lean & Total lean mass \\ 
        SLAM (C57BL/6) & Mouse, C57BL/6 & Death & fluid & Total fluid mass \\ 
        SLAM (C57BL/6) & Mouse, C57BL/6 & Death & glucose & Blood glucose (fasting) \\ 
        SLAM (C57BL/6) & Mouse, C57BL/6 & Death & lactate & ~ \\ 
        SLAM (Het3) & Mouse, Het 3 & Death & bw & Body weight \\ 
        SLAM (Het3) & Mouse, Het 3 & Death & fat & Total fat mass \\ 
        SLAM (Het3) & Mouse, Het 3 & Death & lean & Total lean mass \\ 
        SLAM (Het3) & Mouse, Het 3 & Death & fluid & Total fluid mass \\ 
        SLAM (Het3) & Mouse, Het 3 & Death & glucose & Blood glucose (fasting) \\ 
        SLAM (Het3) & Mouse, Het 3 & Death & lactate & ~ \\ 
        Paquid & Human & Dementia & MMSE & Mini-mental state exam\tnote{\dag} \\ 
        Paquid & Human & Dementia & BVRT & Benton visual retention test \\ 
        Paquid & Human & Dementia & IST & Isaacs set test \\ 
        Paquid & Human & Dementia & CESD & Center for Epidemiological Studies depression scale\tnote{\ddag}  \\ 
        ELSA & Human & Death & vitd & Vitamin d \\ 
        ELSA & Human & Death & srh & Self-reported health (higher: worse) \\ 
        ELSA & Human & Death & eye & Self-reported (corrected) eyesight (higher: worse) \\ 
        ELSA & Human & Death & hear & Self-reported (corrected) hearing (higher: worse) \\ 
        ELSA & Human & Death & FI.ADL & Activities of Daily Living (ADL\cite{Edemekong2021-gf}) FI\tnote{*} \\ 
        ELSA & Human & Death & FI.IADL & Instrumental ADL FI\tnote{*} \\ 
        ELSA & Human & Death & gait.speed & Time to walk 8~feet (2.44~m)\\ 
        ELSA & Human & Death & grip.ndom & Grip strength, non-dominant hand \\ 
        ELSA & Human & Death & grip.dom & Grip strength, dominant hand \\ 
        ELSA & Human & Death & crp & C-reactive protein\tnote{$\star$} \\ 
        ELSA & Human & Death & hba1c & Glycohaemoglobin \\ 
        ELSA & Human & Death & glucose & Glucose \\ 
        ELSA & Human & Death & hgb & Haemoglobin \\ 
        ELSA & Human & Death & mch & Mean corpuscular haemoglobin \\ 
        ELSA & Human & Death & fer & Ferritin\tnote{$\star$} \\ 
        ELSA & Human & Death & chol & Cholesterol \\ 
        ELSA & Human & Death & ldl & Low density lipoprotein \\ 
        ELSA & Human & Death & hdl & High density lipoprotein \\ 
        ELSA & Human & Death & trig & Triglycerides\tnote{$\star$} \\ 
        ELSA & Human & Death & sys & Systolic blood pressure \\ 
        ELSA & Human & Death & dias & Diastolic blood pressure \\ 
        ELSA & Human & Death & pulse & Pulse \\ 
        ELSA & Human & Death & fib & Fibrogen \\ 
        ELSA & Human & Death & igf1 & Insulin-like growth factor-1 \\ 
        ELSA & Human & Death & wbc & White blood cell count\tnote{$\star$} \\ \hline
    \end{tabular} 
\begin{tablenotes}
\item[*]  FI: frailty index; defined as average number of health deficits\cite{Searle2008-xi}.
\item[\dag] Transformed as $-\sqrt{\max(\text{MMSE})-\text{MMSE}}$.
\item[\ddag] Square root transformed for normality.
\item[$\star$] Log transformed for normality.
\end{tablenotes}
\end{threeparttable}
\end{table}
\clearpage

\section{Materials} \label{sec:materials}
We analysed 4 datasets derived from 3 longitudinal studies. The datasets and predictors (``biomarkers'') used are summarized in Table~\ref{tab:datasets}. All predictor variables were continuous or, in the case of Paquid, ordinal with many scales ($> 15$).

We included covariates to reduce confounding effects and to look for allostasis, which depends on age. We included age (continuous) and binary variables. The covariates used are summarized in Table~\ref{tab:cov}.
\begin{table}[!h]
\caption{Covariate Summary} \label{tab:cov}
    \centering
    \begin{tabular}{l l l}
    \hline
        Dataset & Covariate & Description \\ \hline
        SLAM (C57BL/6) & Age & Chronological age in weeks \\
        SLAM (C57BL/6) & Sex & 0: male, 1: female \\
        SLAM (Het3) & Age & Chronological age in weeks \\
        SLAM (Het3) & Sex & 0: male, 1: female \\
        Paquid & Age & Chronological age in years  \\
        Paquid & Sex & 0: male, 1: female  \\
        Paquid & Education & 0: did not complete primary school, 1: did  \\
        ELSA & Age & Chronological age in years \\
        ELSA & Sex & 0: male, 1: female \\ \hline
    \end{tabular}
\end{table}

\section{Preprocessing} \label{sec:preprocess}
The data we analyzed were longitudinal with regular sampling rates. For this reason, data were conveniently stored as 3-dimensional arrays (individuals, biomarkers, time points), meaning that each individual had the same number of variables and measurements (although many of them missing). This means that some timepoints for some individuals had to be `invented' (instantiated as NA) based on the sampling rate of the study in question.

The Study of Longitudinal Aging in Mice (SLAM) datasets were both processed using the same criteria. The initial data were downloaded and processed using the analysis script of another publication \cite{Palliyaguru2021-jh}. We then applied additional preprocessing as follows. Biomarkers were visually investigated for normality and deemed adequate. The sex-specific mean and standard deviation of the first measurement of each biomarker was used to center and scale all timepoints. Mice with less than 2 timepoints were excluded from analysis (about 1\% of mice). Any observations made past the reported death age of each mouse were excluded from analysis (about 1\% of observations). We excluded the first two timepoints from analysis because after encoding we found that approximately half of individuals had not yet had a body composition measurement (imputed values looked unrealistic). Missing timepoints --- which occurred due to staggered data collection --- were instantiated using a piecewise linear model between known observations. Data from SLAM and the other datasets were stored in 3-dimensional arrays, with missing values imputed according to Section~\ref{sec:missing}. The final arrays were size: (608, 6, 22) for SLAM C57/BL6, and (611, 6, 29) for SLAM Het3 (individuals, biomarkers, time points).

The Paquid dataset we used is available as part of a software package \cite{Proust-Lima2017-uz}. Biomarkers were visually investigated for normality. To improve normality we transformed CESD by the square-root and MMSE by $-\sqrt{30-\text{MMSE}}$ where $30$ is the maximum allowed score for the MMSE. All biomarkers were centered and scaled by their respective mean and standard deviation from the first timepoint. Missing timepoints were instantiated using a piecewise linear model between known observations. The final array was size (500, 4, 9); (individuals, biomarkers, time points).

The English Longitudinal Study of Ageing (ELSA) dataset is available from the UK data service (\url{https://ukdataservice.ac.uk/}). We analysed all of the waves which included lab work: 2, 4, 6 and 8 (i.e. the ``nurse'' waves). We included only individuals present in wave 2, thus excluding later recruits. Biomarkers were visually investigated for normality. We found that the log transformation improved normality for C-reactive protein, ferritin, triglycerides, and white blood-cell count. All biomarkers were centered and scaled by their respective mean and standard deviation from the first timepoint. Skipped timepoints were instantiated using linear interpolation of the available timepoints. Censored (or died) timepoints were instantiated using the mean followup time (which was uniform due to the study design). The final array was size (9330, 23, 4); (individuals, biomarkers, time points).

\section{Missing data} \label{sec:missing}
We were presented with two forms of missing data for an individual at a particular timepoint. The entire timepoint could be missing or some subset of values could be missing. In either case the missingness could be informative; for example an individual may have temporarily left the study due to poor health and their biomarkers could have had abnormal values reflecting their poor health. In this way the population may appear abnormally healthy as it ages. Under such circumstances, failure to impute can lead to biased study conclusions \cite{Sterne2009-jq}, such as parameter estimates (Section~\ref{sec:censor}).

We considered three imputation approaches and selected the approach which gave the most reasonable values. First (``simplest''), we imputed
 a single value using either the individual's mean biomarker value, carry forward the previous value (and then carry back any skipped values), the conditional population mean (assuming multivariate Gaussian statistics), or the individual mean followed by the conditional mean for individuals whom did not have that variable reported. Second (``model mean''), we considered an iterative approach after applying one of the first methods wherein values were imputed according to the model mean (i.e. model prediction), equation~(\ref{eq:ynimp}) and equation~(\ref{eq:ynp1imp}). Third (``MICE''), we considered multivariate imputation using chained equations (MICE) \cite{Van_Buuren2010-oy}. MICE is a multiple imputation technique that uses a Gibbs' sampler along with a predictive model. We considered MICE using both classification and regression trees (CART) and 2-level modelling (normal for continuous variables and logistic for binary).

When imputing the model mean, at each iteration we estimated the model parameters then imputed the conditional mean for each missing value (Algorithm~\ref{alg:est}). Let $\vec{y}$ denote the biomarker, $\vec{u}$ denote all unobserved $y$ and $\vec{o}$ denote all observed $y$. The statistics are Gaussian (equation~(\ref{eq:sf})), so we can use the factorization theorem \cite{Brandt2012-iz} to compute the expectation value. 

If $\vec{y}_{n+1}$ is known but a set of $\vec{y}_n$ are unknown then
\begin{align}
    E(\vec{u}_{n}|\vec{o}_{n},\vec{y}_{n+1}) &= \langle \vec{y}_{u n} \rangle + \boldsymbol{\Sigma}_{uo}\boldsymbol{\Sigma}_{oo}^{-1}(\vec{o}_{n}-\langle \vec{y}_{o n} \rangle )~~~~\text{where,}  \nonumber \\
    \langle \vec{y}_{u n} \rangle &= (\boldsymbol{I}+\Delta t_{n+1} \boldsymbol{W})^{-1}_{u\cdot}\big(\vec{y}_{n+1} + \boldsymbol{W}\Delta t_{n+1} \vec{\mu}_n\big), \nonumber \\
    \langle \vec{y}_{o n} \rangle &= (\boldsymbol{I}+\Delta t_{n+1} \boldsymbol{W})^{-1}_{o\cdot}\big(\vec{y}_{n+1} + \boldsymbol{W}\Delta t_{n+1} \vec{\mu}_n\big), \nonumber \\
    \Sigma_{uo} &= ((\boldsymbol{I}+\Delta t_{n+1} \boldsymbol{W})^T\boldsymbol{Q}(\boldsymbol{I}+\Delta t_{n+1} \boldsymbol{W}))^{-1}_{uo}, \nonumber \\
    \Sigma_{oo}^{-1} &= \bigg(((\boldsymbol{I}+\Delta t_{n+1} \boldsymbol{W})^T\boldsymbol{Q}(\boldsymbol{I}+\Delta t_{n+1} \boldsymbol{W}))^{-1}_{oo} \bigg)^{-1}, \label{eq:ynimp}
\end{align}
where $E(x|y)$ denotes expectation value of $x$ conditional on $y$, $\boldsymbol{Q}$ is the precision matrix defined below, and $\boldsymbol{I}$ is the identity matrix. Note that
\begin{gather}
\boldsymbol{\Sigma} \equiv \boldsymbol{Q}^{-1} =
\begin{bmatrix} 
\boldsymbol{\Sigma}_{oo} & \boldsymbol{\Sigma}_{ou} \\
\boldsymbol{\Sigma}_{uo} & \boldsymbol{\Sigma}_{uu}
\end{bmatrix} 
\end{gather}
is the block-decomposition of the noise covariance rearranged for observed ($o$) and unobserved ($u$) variables.
If instead $\vec{y}_{n}$ is known but a set of $\vec{y}_{n+1}$ are unknown then
\begin{align}
    E(\vec{u}_{n+1}|\vec{o}_{n+1},\vec{y}_{n}) 
    &= \vec{y}_{un} + W_{u\cdot}\Delta t_{n+1}(\vec{y}_{n}-\vec{\mu}_n) + \boldsymbol{\Sigma}_{uo}\boldsymbol{\Sigma}_{oo}^{-1}(\vec{o}_{n+1} - \vec{y}_{on} - W_{o\cdot}\Delta t_{n+1}(\vec{y}_{n}-\vec{\mu}_n)). \label{eq:ynp1imp}
\end{align}
We used the simplest approach as an initial imputation (e.g.\ carry forward/back). We then imputed $\vec{y}_1$ first using equation~(\ref{eq:ynimp}) then each subsequent timepoint using equation~(\ref{eq:ynp1imp}). 

We compared the imputation quality and found that the model mean was both straightforward and effective, and therefore elected to use it for both SLAM datasets and Paquid. We initialized imputation with carry forward/back: that is, forward carrying previous values until the last timepoint was reached then backwards carrying to fill any values still missing. In rare cases a few data points were missing after imputation, these were simply ignored (we used all available case data). The ELSA dataset was sensitive to the model mean --- perhaps due to the limited number of data points --- hence we used a single imputation which combined first imputing the individual-specific variable mean followed by the conditional mean, assuming multivariate Gaussian statistics at each timepoint. Note that since we elected to use bootstrapping, we imputed each bootstrap replicate and then averaged to get an estimate for each missing value along with a standard error. 

\begin{figure*}[!h]
     \centering
        \includegraphics[width=.8\textwidth]{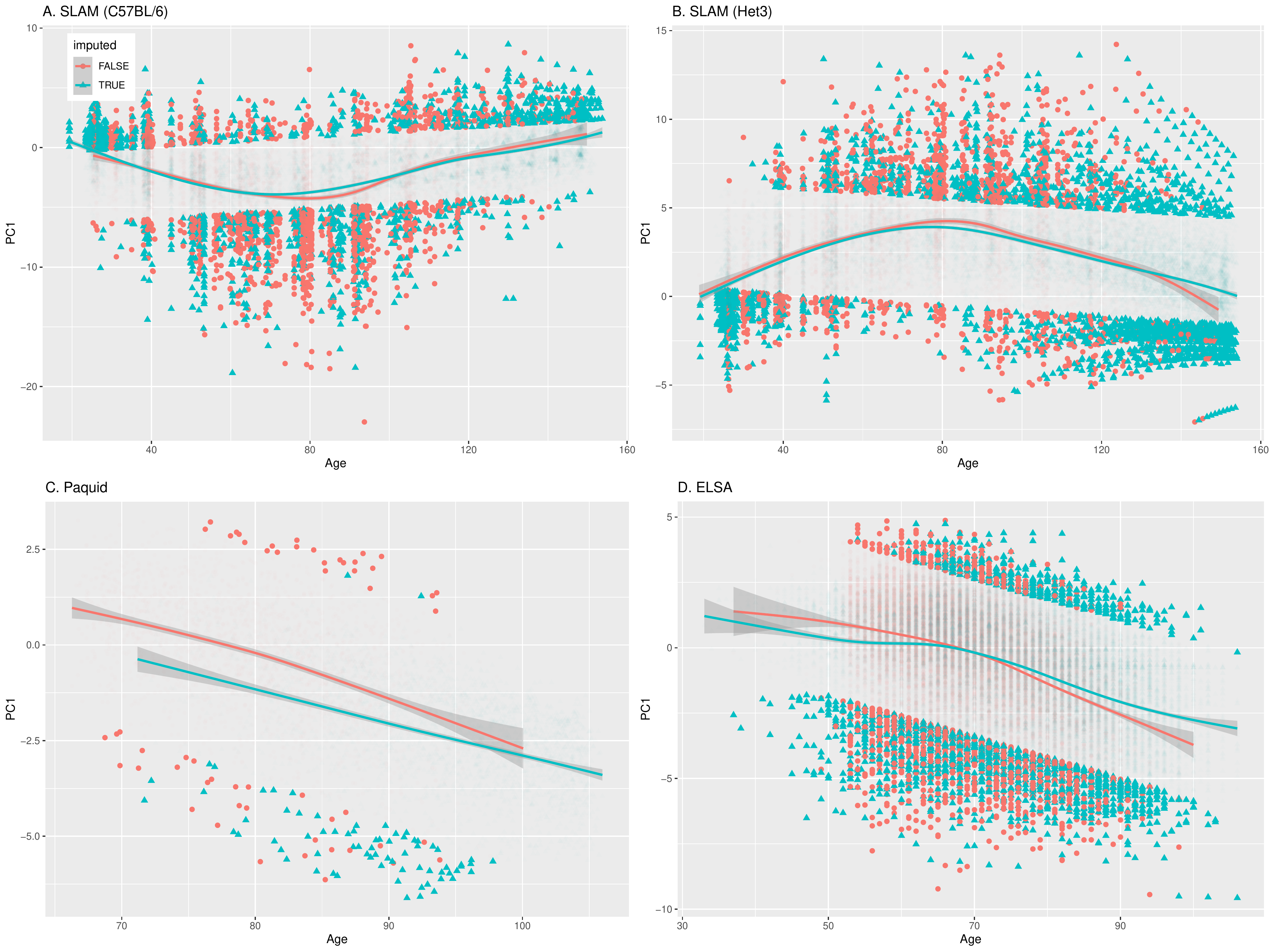}  
    \caption{Final imputation quality check, visualized using principal component 1. \textbf{A.} C57BL/6 mice (SLAM). \textbf{B.} Het3 mice (SLAM). \textbf{C.} Paquid (human, dementia). \textbf{D.} ELSA (human). Imputed values appear to be reasonable for each dataset. Principal component analysis (PCA) was applied to each dataset in the entirety, flatted across timepoints. Good quality imputation (blue triangles) should show the same trend and dispersion as the observed data (red points). Censored individuals likely have worse health, so imputed values may look a little `older' than the observed. Age-dependence is indicated by the solid lines with confidence intervals (cubic spline; \texttt{geom\_smooth} with defaults \cite{Wickham2016-kw}). Outlying points are highlighted ($l \pm 3$ where $l$ is the ordinary linear regression model). Data points were labelled as imputed (blue triangles) if the preponderance of the rotation weights were missing: $\sum_{i=missing} |U_{i1}|/(\sum_j |U_{j1}|) > 0.5$; where $\boldsymbol{U}$ is the PCA rotation matrix.}
    \label{fig:imp_pc1}
\end{figure*}

The final imputation was assessed for quality, Figure~\ref{fig:imp_pc1} --- and looked reasonable. When inspecting imputation quality we are looking for the same age-dependent pattern for both the imputed and observed values, both in terms of mean and dispersion. Informative censorship is possible, so for variables with survival effects we can expect that missing values should be at higher risk because they include individuals whom were censored due to poor health (or death). Risk can be inferred by the direction of drift with respect to age: data points which look `older' are likely higher risk. Hence imputed values may look a little `older' than observed values.

\subsection{Informative censorship} \label{sec:censor}
Is it better to impute dropped individuals (dead, censored) or not? Dropped individuals may have abnormal biomarker values leading to their exclusion, i.e.\ informative censorship. There is ``substantial'' evidence that dropped individuals in longitudinal studies have worse health \cite{Hardy2009-nl} and their health biomarkers will reflect this, leading to a potential survivorship bias. We used simulated data to test for potential bias and observed that --- if done well --- imputing values for dropped individuals can reduce this bias.

We simulated data from our model equation~(\ref{eq:sf}) using randomly generated parameters then imposed informative censorship. We simulated $100$ times with $100$ individuals in each simulation. Each simulation included $2$ biomarkers. Parameters and biomarkers were draw from normal random variables. The diagonal of $\boldsymbol{W}$ was mean $-1/4$, the off-diagonals were mean $0$ and the overall standard deviation was $0.1$. The mean $\mu_0$ was $0$ and standard deviation was $0.1$. No covariates were simulated. The noise was diagonal, $\boldsymbol{\Sigma} = 0.5 \boldsymbol{I}$ (also used for instantiating the population). We censored using Gompertz statistics with proportional hazards for biomarker values \cite{Bender2005-ag} (shape: $\alpha=0.1$, scale: $\lambda=10^{-5}$). The proportional hazards coefficients were randomly sampled from a normal distribution with mean $1$ and standard deviation $0.1$, this ensured that large values of the biomarkers were preferentially censored.

\begin{figure*}[!h]
     \centering
        \includegraphics[width=.8\textwidth]{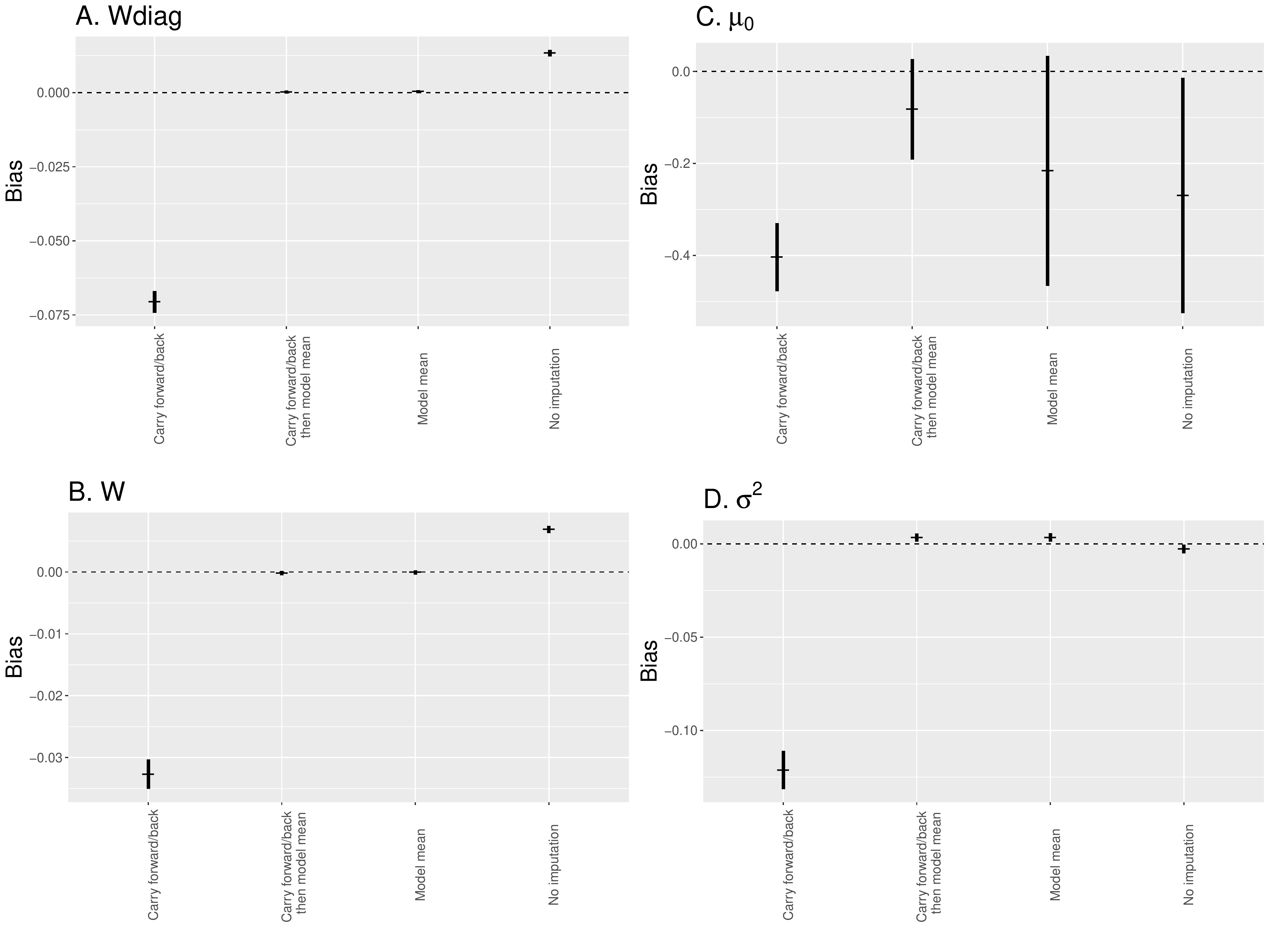}  
    \caption{Imputation of dropped individuals can reduce bias. We simulated informative censorship and here compare estimates from different missing data handling strategies. Observe that both the diagonal elements of $\boldsymbol{W}$ (A.) and all elements of $\boldsymbol{W}$ (B.) were biased high when data were not imputed. However, if we imputed using the model mean, the bias was greatly reduced. For $\mu_0$ (C.) we also reduced the bias with the combined imputation strategy, which was the strategy employed on the real data. Imputation did introduce a small bias in the noise estimate (D.). The bias was largest if we used only the carry forward/back method.}
    \label{fig:zombie_imp}
\end{figure*}

The results of the simulation are shown in Figure~\ref{fig:zombie_imp} for various imputation strategies, with the horizontal dashed line indicating unbiased results. We observed that a significant bias existed in both the diagonal and off-diagonal elements of $\boldsymbol{W}$, which were systematically over-estimated if the data were not imputed. Conversely, if we used only the simple carry forward/back imputation, an even worse bias ensued in the opposing direction. If we used the model mean imputation, however, we reduced the bias in $\boldsymbol{W}$ to nearly $0$ without significantly increasing bias in the other parameters. For this reason, we imputed all dropped individuals: censored and dead.

\section{Model selection} \label{sec:select}
Our goal with model selection was two fold: (i) to find the optimal model(s) that best fit the data, and (ii) to test which model parameters were essential for fitting the data. This allows us to infer the existence of which model parameters are robustly supported by the data. Fit quality was measured using the root-mean squared error (RMSE) and mean absolute error (MAE). We used the 632 estimator for errors, which is a linear combination of 63.2\% test error and 36.8\% training error \cite{Hastie2017-wl}. Test error was estimated via out-of-sample bootstrap replication, with 100 resamples. Each bootstrap selected a new dataset of the same size as the original by resampling individuals with replacement. The out-of-sample individuals are those whom were not selected. Estimation algorithms are reported in Section~\ref{sec:estimation}.

\begin{figure*}[!h]
     \centering
         \includegraphics[width=\textwidth]{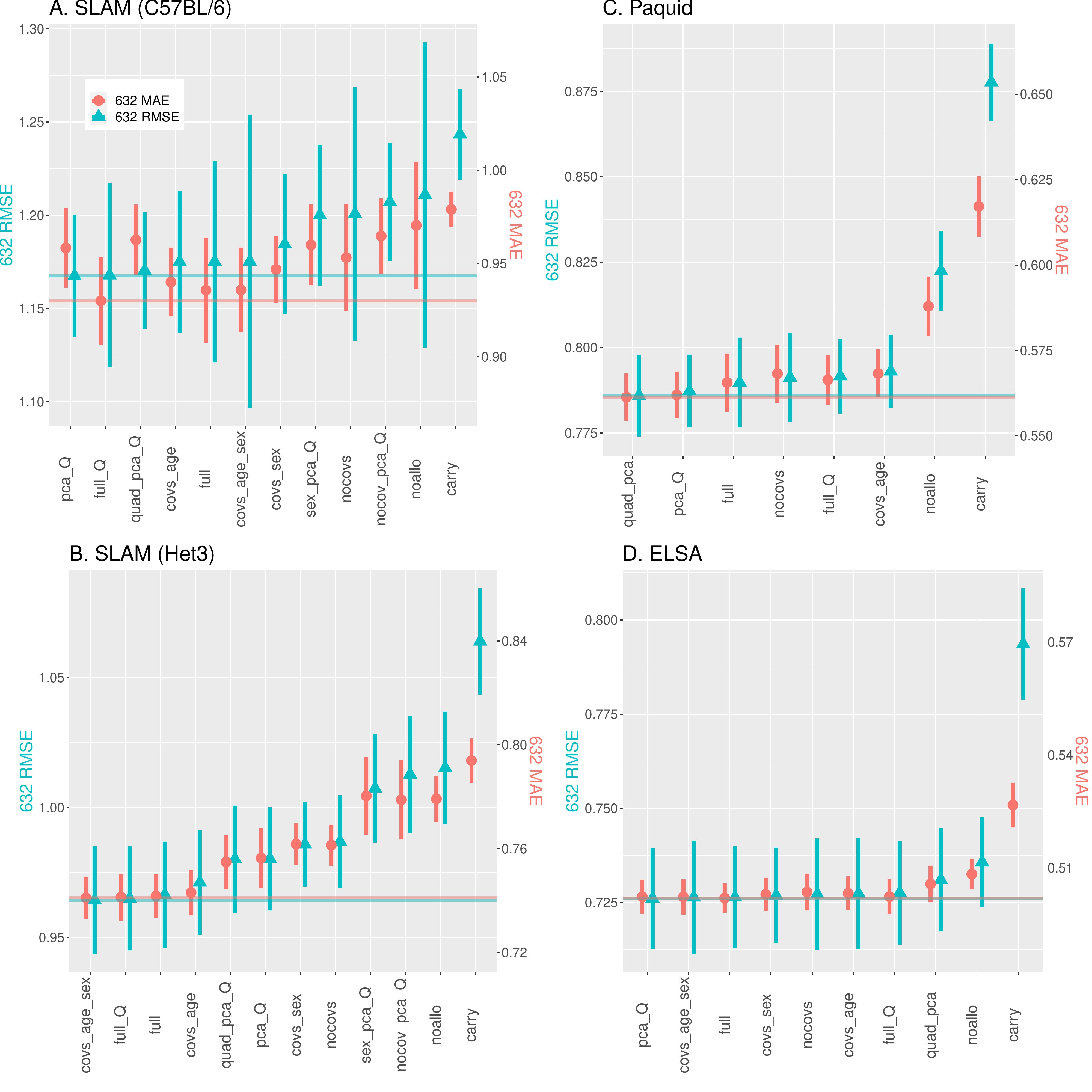}
    \caption{Model selection. \textbf{A.} C57BL/6 mice (SLAM). \textbf{B.} Het3 mice (SLAM). \textbf{C.} Paquid (human, dementia). \textbf{D.} ELSA (human). Lower error is better. y-axis is 632-RMSE on left and 632-MAE on right. Horizontal lines indicate the best performing model. We are looking for the simplest model that consistently hits those lines across datasets. We considered models significantly worse if they do not have an error interval overlapping this line; prioritizing RMSE. Models: carry: equation~(\ref{eq:carry}); fast: equation~(\ref{eq:fast}); noallo: equation~(\ref{eq:noallo}); quad: equation~(\ref{eq:quad}); full: equation~(\ref{eq:sf}). Additional parameters: pca: equation~(\ref{eq:z}) with PCA preprocessing and diagonal noise; $Q$: the noise was estimated; covs: prefix, after which included covariates are listed; nocovs: no covariates were used. For example, \texttt{sex\_pca\_Q} included sex as a covariate (\texttt{sex}), used PCA as a preprocessing step and assumed diagonal $\boldsymbol{W}$ and $\boldsymbol{Q}$, and fit equation~(\ref{eq:z}) (\texttt{pca}), and estimated $Q$ from the data (\texttt{Q}). The fast model, equation~(\ref{eq:fast}), performed much worse for all datasets (points above plot region), 632-RMSE: 0.91(2) (Paquid), 0.92(1) (ELSA), 2.03(6) (SLAM C57) and 2.21(7) (SLAM HET3); 632-MAE: 0.68(1) (Paquid), 0.702(5) (ELSA), 1.32(3) (SLAM C57) and 1.47(3) (SLAM HET3).}
    \label{fig:models}
\end{figure*}

We compare model performance in Figure~\ref{fig:models}; we explain the model labels here. The general form of our model is (`full')
\begin{align}
    \vec{y}_{n+1} &= \vec{y}_{n} + \boldsymbol{W} \Delta t_{n+1}(\vec{y}_{n} - \vec{\mu}_n) +\vec{\epsilon}, \nonumber\\
    \vec{\epsilon} &\sim \mathcal{N}(0,\Sigma|\Delta t|) \nonumber \\
    \vec{\mu}_{n} &\equiv \vec{\mu}_{0}+\boldsymbol{\Lambda} \vec{x}_{n}+ \vec{\mu}_{age} t,
    \label{eq:sf}
\end{align}
where $t$ is the age.
The error can be expressed in terms of the precision matrix,
\begin{align}
    \boldsymbol{Q} \equiv \boldsymbol{\Sigma}^{-1}.
\end{align}

We considered both using $\boldsymbol{Q}=\boldsymbol{I}$, the identity matrix (default), and estimating $\boldsymbol{Q}$ from the data using the log-likelihood (`Q').
Transforming into natural variables --- wherein $\boldsymbol{W}$ is diagonal --- we have 
\begin{align}
    z_{jn+1} &= z_{jn} + \lambda_j\Delta t_{n+1}(z_{jn} - \tilde{\mu}_{jn}) + \tilde{\epsilon}, \label{eq:z}
\end{align}
where $\vec{z}_n\equiv \boldsymbol{P}^{-1}\vec{y}_n$, $\lambda_i \equiv P_{i\cdot}^{-1}\boldsymbol{W} P_{\cdot i}$, $\tilde{\vec{\mu}}_{n} \equiv \boldsymbol{P}^{-1}\mu_{n}$ and $\tilde{\vec{\epsilon}} \equiv \boldsymbol{P}^{-1}\vec{\epsilon}$. 

We considered (`pca') the possibility that principal component analysis (PCA) could be used as a preprocessing step to decouple the biomarkers such that we could fit equation~(\ref{eq:z}), assuming independent noise between the $z_j$. Equation~(\ref{eq:ypca}) states that in the steady-state the principal components are equivalent to the eigenvectors of $\boldsymbol{W}$, this self-consistency motivates using PCA. Prior work has also suggested that principal components don't change much during the aging process \cite{Pridham2023-yj}.

We considered that $\mu_n$ may be time-dependent and may also depend on other covariates (`covs'), which is discussed in Section~\ref{sec:materials}.

We considered simpler, nested forms of equation~(\ref{eq:sf}). Recall that the data were standard-deviation-scaled and mean-centered by the baseline value, which justifies some of the simplifications. Simplified forms allowed us to test whether $\boldsymbol{W}$ and $\vec{\mu}$ were necessary to fit the data. Removing these parameters leads to special cases of the model. The simplest model for the data is to simply carry forward the previous value. If recovery is small, $W\Delta t \to 0$ then we have (`carry'),
\begin{align}
    \vec{y}_{n+1} &= \vec{y}_{n} + \tilde{\epsilon}  \label{eq:carry}
\end{align} 
which corresponds to carrying the previous value forward ($\langle \vec{y}_{n+1} \rangle = \langle \vec{y}_{n} \rangle$). This model does not require any parameters to make predictions; it was used for the initial imputation of the Paquid and SLAM datasets (Section~\ref{sec:missing}).

If recovery is complete between each timepoint then $\boldsymbol{W} \Delta t \to -\boldsymbol{I}$ and we instead have the second simplest model (`fast'),
\begin{align}
    \vec{y}_{n+1} &= \vec{\mu}_{n} +\vec{\epsilon}  \label{eq:fast}
\end{align} 
which corresponds to biomarkers being randomly distributed about some mean value which depends on covariates ($\langle \vec{y}_{n+1} \rangle = \langle \vec{\mu}_{n} \rangle$). Alternatively, we could have $\vec{\mu}_n\equiv0$ in which case we have (`noallo'),
\begin{align}
    \vec{y}_{n+1} &= \vec{y}_{n} + W\Delta t_{n+1}\vec{y}_{n} +\vec{\epsilon}  \label{eq:noallo}
\end{align} 
which we refer to as the no allostasis model (it also implicitly sets homeostatic equilibrium to $0$). In 1-dimension, equation~(\ref{eq:noallo}) is simple exponential growth/decay in the mean (for small $\Delta t$).

While nonlinear behaviour can be captured by our model (Section~\ref{sec:math}), we also directly investigated nonlinear behaviour by including a quadratic term (`quad'),
\begin{align}
    z_{jn+1} &= z_{jn} + \lambda_j\Delta t_{n+1}(z_{jn} - \tilde{\mu}_{jn}) + \gamma \Delta t_{n+1}^2 z_{jn}^2 + \tilde{\epsilon}.  \label{eq:quad}
\end{align} 
We only considered a quadratic term for the diagonal model, equation~(\ref{eq:z}) with PCA preprocessing.

We compare model performance in Figure~\ref{fig:models}. We found that the fast model, equation~(\ref{eq:fast}), fit very poorly, having error so large that it did not fit in the plot region. Within the plot region, the carry-forward model performed the worst, equation~(\ref{eq:carry}) (`carry'). Note the implication: biomarker recovery towards equilibrium is much closer to none (`carry') than complete (`fast'). Next worse was excluding $\mu$, equation~(\ref{eq:noallo}) (`noallo'). The remaining models typically performed similarly-well. The SLAM datasets both saw a noteworthy improvement in fit when age was included as a covariate (in $\mu_n$). We observed no improvement with inclusion of a quadratic term, equation~(\ref{eq:quad}) (`quad'). Finally, and importantly, we found that a diagonal fit on principal components (PCs) yielded equivalent performance to the full model. This permitted a greatly simplified methodology since we were able estimate using weighted linear regression (Section~\ref{sec:wlm}).


\section{Estimation} \label{sec:estimation}
We provide useful results for fitting equation~(\ref{eq:sf}) and its simplified form, equation~(\ref{eq:z}). The latter can be solved using weighted linear regression.

\subsection{(Weighted) Linear Regression} \label{sec:wlm}
If the noise term is diagonal then the equations decouple and we have a set of linear equations which can be independently solved using linear regression. In the present study we used PCA (principal component analysis) as a preprocessing step prior to fitting a diagonal model. That is, we assumed the PCs do not interact with each other. We can rewrite equation~(\ref{eq:z}) as
\begin{align}
    z_{ijn+1} - z_{ijn} &= \lambda_j(\Delta t_{in+1} z_{ijn}) + \vec{\beta_j}^T (\Delta t_{in+1} \vec{x}_{in}) + \epsilon_{ij},~~~~\text{where} \nonumber \\
    \beta_{j0} &= \lambda_j \mu_0, \nonumber \\
    \beta_{j age} &= \lambda_j\mu_{j age}, \nonumber \\
    \beta_{jk} &= \lambda_j \Lambda_{k \cdot},~~~~\text{and} \nonumber \\
    \epsilon_{ij} &\sim \mathcal{N}(0,\sigma_j^2|\Delta t_{in+1}|). \label{eq:wlr}
\end{align}
This is a weighted linear regression problem\cite{Xiaogang2009-hi} where the predictors are $\Delta t_{in+1} z_{ijn}$, and $\Delta t_{in+1} x_{ijn}$; the weights are $|\Delta t_{in+1}|^{-1}$. During model selection, Section~\ref{sec:select}, we found that equation~(\ref{eq:wlr}) fit the data as well as the full model, equation~(\ref{eq:sf}).

\subsection{Maximum likelihood estimators (MLEs)} \label{sec:mle}
We derive the MLEs for equation~(\ref{eq:sf}) in full generality. For convenience define
\begin{align}
    \hat{y}_{ibn+1} \equiv y_{ibn} + \Delta t_{in+1}\sum_k W_{bk}(y_{ikn}- \mu_{ikn}) = y_{ibn} + \Delta t_{in+1}\sum_k W_{bk}(y_{ikn}-\sum_j \Lambda_{kj}x_{ijn})
\end{align}
where we have $p$ variables, $N$ individuals and $T+1$ timepoints. We index the $N$ individuals with $i$ and the $T$ timepoint-pairs with $n$. For convenience we drop $\mu_0$ and define the equivalent $\vec{\mu}_{in} \equiv \boldsymbol{\Lambda}\vec{x}_{in}$; where we use $x_{i0n}\equiv 1$ to recover $\mu_0$. Estimators are denoted with a hat e.g.\ $\hat{\boldsymbol{W}}$ estimates $\boldsymbol{W}$.

The log-likelihood is,
\begin{align}
    l &= -\frac{1}{2}\sum_{i,n}\ln{(\det\big|2\pi \boldsymbol{Q}^{-1}|\Delta t_{in+1}|\big|)} - \frac{1}{2}\sum_{i,n}(\vec{y}_{in+1}-\hat{\vec{y}}_{in+1})^T\frac{\boldsymbol{Q}}{|\Delta t_{in+1}|}(\vec{y}_{in+1}-\hat{\vec{y}}_{in+1}) \nonumber \\
    &= \frac{1}{2}NT\ln{(\det\big|\boldsymbol{Q}\big|)} - \frac{p}{2}\sum_{i,n} \ln{(2\pi|\Delta t_{in+1}|)} - \frac{1}{2}\sum_{i,n}(\vec{y}_{in+1}-\hat{\vec{y}}_{in+1})^T\frac{\boldsymbol{Q}}{|\Delta t_{in+1}|}(\vec{y}_{in+1}-\hat{\vec{y}}_{in+1}).
\end{align}

We derive analytical forms for the MLEs as well as providing derivatives for gradient-based optimization algorithms. We also report the curvature since this is used to estimate the asymptotic error via the inverse Fisher matrix\cite{Held2014-kf}. We found that the asymptotic errors are well-calibrated for $\boldsymbol{W}$, but tend to be too small for $\vec{\mu}_n$ (Section~\ref{sec:validation}). In the present study, we report bootstrap errors.

Note that we find it useful to express the estimators in terms of the uncentered (cross)covariance,
\begin{align}
    \text{Cov}_2 (\vec{x}_{in}) &\equiv \langle \vec{x}\vec{x}^T \rangle_{i,n}, ~\text{and} \nonumber \\
    \text{Cov}_2 (\vec{x}_{in},\vec{y}_{in}) &\equiv \langle \vec{x}\vec{y}^T \rangle_{i,n}
\end{align}
where the expectation value is taken over individuals, $i$, and timepoints, $n$. In general, $\langle f(x_{in}) \rangle_{i,n}$ denotes the average of $f(x_{in})$ over individuals, $i$, and timepoints $n$.

We start by considering $\boldsymbol{W}$. The derivatives are
\begin{align}
    \frac{\partial l}{\partial W_{\alpha\beta}} &= \sum_{i,n,a}\text{sign}(\Delta t_{in+1})Q_{a\alpha}(y_{ian+1}-y_{ian}-\Delta t_{in+1} \sum_k W_{ak}(y_{ikn}-\mu_{ikn}))(y_{i\beta n}-\mu_{i\beta n}) \nonumber \\ 
    \nabla_{W} l &= \sum_{i,n}\text{sign}(\Delta t_{in+1})\boldsymbol{Q}(\vec{y}_{in+1}-\vec{y}_{in}-\Delta t_{in+1} \boldsymbol{W}(\vec{y}_{in}-\vec{\mu}_{in}))(\vec{y}_{in}-\vec{\mu}_{i n})^T
    \label{eq:sfWgrad}
\end{align}
where $\nabla_W$ denotes the gradient with respect to (vectorized) $\text{vec}(\boldsymbol{W})$.
The MLE is thus
\begin{align}
    \hat{\boldsymbol{W}} \langle |\Delta t_{in+1}|(\vec{y}_{in}-\vec{\mu}_{in})(\vec{y}_{in}-\vec{\mu}_{in})^T \rangle_{i,n} &= \langle \text{sign}(\Delta t_{in+1})(\vec{y}_{in+1}-\vec{y}_{in})(\vec{y}_{in}-\vec{\mu}_{in})^T \rangle_{i,n} \nonumber \\
    \text{Cov}_2(\sqrt{|\Delta t_{in+1}|}(\vec{y}_{in}-\vec{\mu}_{in}))\hat{\boldsymbol{W}}^T &= \text{Cov}_2(\text{sign}(\Delta t_{in+1})(\vec{y}_{in}-\vec{\mu}_{in}),\vec{y}_{in+1}-\vec{y}_{in}).     \label{eq:west}
\end{align}
The latter equation is useful for linear algebra software packages. Alternatively, we can invert the uncentered covariance $\langle |\Delta t_{in+1}|(\vec{y}_{in}-\vec{\mu}_{in})(\vec{y}_{in}-\vec{\mu}_{in})^T \rangle_{i,n}$ which yields equation~(12).

The curvature of $\boldsymbol{W}$ is
\begin{align}
    \frac{\partial^2 l}{\partial W_{\gamma\delta}W_{\alpha\beta}} &= -NT Q_{\gamma\alpha}\langle |\Delta t_{in+1}| (y_{i\beta n}-\mu_{i\beta n})(y_{i\delta n}-\mu_{i\delta n}) \rangle_{i,n}
\end{align}
the Fisher information is the negative of this. The covariance of the MLE is given by the inverse Fisher information,
\begin{align}
    I^{-1}_{\alpha\beta\gamma\delta} &= \frac{1}{NT}Q^{-1}_{\alpha \gamma} \langle |\Delta t_{in+1} | (\vec{y}_{in}-\vec{\mu}_{in})(\vec{y}_{in}-\vec{\mu}_{in}) \rangle^{-1}_{\beta\delta}
\end{align}
the standard errors are the square-roots of the diagonal elements,
\begin{align}
    \delta W_{\alpha\beta}^2 &= \frac{Q^{-1}_{\alpha\alpha}}{NT}\langle |\Delta t_{in+1} | (\vec{y}_{in}-\vec{\mu}_{in})(\vec{y}_{in}-\vec{\mu}_{in}) \rangle^{-1}_{\beta\beta}.
\end{align}

Next we consider $\vec{\mu}_n$. We condense all relevant parameters into $\boldsymbol{\Lambda}$ which has MLE,
\begin{align}
    \frac{\partial l}{\partial \Lambda_{\alpha\beta}} &= -\sum_{i,n,a,b} \frac{Q_{ab}}{|\Delta t_{in+1}|} (y_{ian+1}-\hat{y}_{ian+1})(\Delta t_{in+1} W_{b\alpha}x_{i\beta n}) \nonumber \\
    &= -NTW_{\alpha\cdot}^T\boldsymbol{Q}\langle (\vec{y}_{in+1}-\vec{y}_n)x_{i\beta n}\text{sign}(\Delta t_{in+1}) \rangle + NTW^T_{\alpha\cdot}\boldsymbol{Q}\boldsymbol{W}\langle |\Delta t_{in+1}|\vec{y}_{in}x_{i\beta n} \rangle \nonumber \\
    &- W_{\alpha\cdot}^T\boldsymbol{Q}\boldsymbol{W}\boldsymbol{\Lambda}\langle|\Delta t_{in+1}| \vec{x}_{in}x_{i\beta n} \rangle \nonumber \\
\implies \frac{1}{NT}\nabla_{\Lambda} l &= -\boldsymbol{W}^T\boldsymbol{Q}\langle (\vec{y}_{in+1}-\vec{y}_n)\vec{x}_{in}^T\text{sign}(\Delta t_{in+1}) \rangle + \boldsymbol{W}^T\boldsymbol{Q}\boldsymbol{W}\langle |\Delta t_{in+1}|\vec{y}_{in}\vec{x}_{in}^T \rangle \nonumber \\
    &- \boldsymbol{W}^T\boldsymbol{Q}\boldsymbol{W}\boldsymbol{\Lambda}\langle|\Delta t_{in+1}| \vec{x}_{in}\vec{x}_{in}^T \rangle \label{eq:sfLgrad}
\end{align}
This implies
\begin{align}
    \boldsymbol{W}^T\boldsymbol{Q}\boldsymbol{W}\hat{\boldsymbol{\Lambda}}\langle |\Delta t_{in+1} |\vec{x}_{in}\vec{x}_{in}^T \rangle &= \boldsymbol{W}^T\boldsymbol{Q}\boldsymbol{W}\langle |\Delta t_{in+1} |\vec{y}_{in}\vec{x}_{in}^T \rangle -  \boldsymbol{W}^T\boldsymbol{Q}\langle \text{sign}(\Delta t_{in+1})(\vec{y}_{in+1}-\vec{y}_{in})\vec{x}_{in}^T \rangle
\end{align}
which we can write
\begin{align}
\boldsymbol{W}^T\boldsymbol{Q}\boldsymbol{W}\hat{\boldsymbol{\Lambda}} &= \boldsymbol{W}^T\boldsymbol{Q}\boldsymbol{W}\text{Cov}_2(|\Delta t_{in+1} |\vec{y}_{in},\vec{x}_{in})\big(\text{Cov}_2(\sqrt{|\Delta t_{in+1} |}\vec{x}_{in})^{-1}\big) \nonumber \\
    &-\boldsymbol{W}^T\boldsymbol{Q}\text{Cov}_2(\text{sign}(\Delta t_{in+1})(\vec{y}_{in+1}-\vec{y}_{in}),\vec{x}_{in})\big(\text{Cov}_2(\sqrt{|\Delta t_{in+1} |}\vec{x}_{in})^{-1}\big). \label{eq:lambdaest}
\end{align}
We estimate from the general form, but note that equation~(\ref{eq:lambdaest}) can be greatly simplified when $\boldsymbol{W}$ is invertible, which is expected because it empirically has strong diagonal elements. For invertible $\boldsymbol{W}$ we get equation~(11). 

The curvature is
\begin{align}
    \frac{\partial^2 l}{\partial \Lambda_{\gamma\delta}\partial\Lambda_{\alpha\beta}} &= -(\boldsymbol{W}^T\boldsymbol{Q}\boldsymbol{W})_{\alpha\gamma}TN\langle |\Delta t_{in+1}| \vec{x}_{in}\vec{x}_{in}^T \rangle_{\beta\delta}
\end{align}
where the expectation is over times and individuals. The Fisher information is used to estimate the asymptotic error,
\begin{align}
    I^{-1}_{\alpha\beta\gamma\delta} &= \frac{1}{NT}(\boldsymbol{W}^T\boldsymbol{Q}\boldsymbol{W})^{-1}_{\alpha \gamma}\langle |\Delta t_{in+1}|\vec{x}_{in}\vec{x}_{in}^T \rangle^{-1}_{\beta\delta}
\end{align}
the fit error is the square-root of the diagonal,
\begin{align}
    \big(\delta \Lambda_{\alpha\beta}\big)^2 &= \frac{1}{NT}(\boldsymbol{W}^T\boldsymbol{Q}\boldsymbol{W})^{-1}_{\alpha \alpha}\langle |\Delta t_{in+1}|\vec{x}_{in}\vec{x}_{in}^T \rangle^{-1}_{\beta\beta}.
\end{align}

Finally, observe the equilibrium case where $\langle \vec{y}_{n+1} \rangle = \langle \vec{y}_n \rangle = \vec{\mu}_n$ and $\text{Cor}(\vec{y}_{n+1}-\vec{y}_{n},\vec{x})=\text{Cor}(\vec{y}_{n}-\vec{\mu}_{n},\vec{x})=0$ (i.e.\ the fluctuations are random) then equation~(\ref{eq:sfLgrad}) becomes
\begin{align}
\frac{1}{NT}\nabla_{\Lambda_{eq}} l &= \boldsymbol{W}^T\boldsymbol{Q}\boldsymbol{W}\langle |\Delta t_{in+1}|\vec{y}_{in}\vec{x}_{in}^T \rangle - \boldsymbol{W}^T\boldsymbol{Q}\boldsymbol{W}\boldsymbol{\Lambda}\langle|\Delta t_{in+1}| \vec{x}_{in}\vec{x}_{in}^T \rangle 
\end{align}
and we have
\begin{align}
    \hat{\boldsymbol{\Lambda}}_{eq} &= \bigg\langle |\Delta t_{in+1}|\vec{y}_{in+1}\vec{x}_{in}^T \bigg\rangle \bigg( \bigg\langle |\Delta t_{in+1}| \vec{x}_{in}\vec{x}_{in}^T \bigg\rangle \bigg)^{-1},\nonumber \\
    &=\text{Cov}_2(|\Delta t_{in+1}|\vec{y}_{in+1},\vec{x}_{in})\big( \text{Cov}_2(\sqrt{|\Delta t_{in+1}|} \vec{x}_{in})^{-1} \big) .\label{eq:lambdaesteq}
\end{align}
equation~(\ref{eq:lambdaesteq}) is useful for an initial $\boldsymbol{\Lambda}$ estimate as it does not depend on $\boldsymbol{W}$.

\subsection{Noise estimator}
We used a simple estimator for the noise, $\boldsymbol{\Sigma}$. For our model, equation~(\ref{eq:sf}), a simple estimator is derived by observing
\begin{align}
    \vec{y}_{n+1} - \langle \vec{y}_{n+1} \rangle &= \vec{\epsilon}
\end{align}
which implies that
\begin{align}
    \langle (\vec{y}_{n+1} - \langle \vec{y}_{n+1} \rangle)(\vec{y}_{n+1} - \langle \vec{y}_{n+1} \rangle)^T \rangle &= \langle \vec{\epsilon}\vec{\epsilon}^T \rangle = \boldsymbol{\Sigma} |\Delta t|
\end{align}
we conclude that
\begin{align}
     \hat{\boldsymbol{\Sigma}} &= \bigg\langle \frac{1}{|\Delta t_{in+1}|}(\vec{y}_{i n+1} - \langle \vec{y}_{i n+1} \rangle)(\vec{y}_{i n+1} - \langle \vec{y}_{i n+1} \rangle)^T \bigg\rangle_{i,n}. \label{eq:sigmaest}
\end{align}
Where the expectation must be taken over individuals, $i$, and timepoints, $n$. Note that $\vec{y}_{n+1} - \langle \vec{y}_{n+1} \rangle$ is the model residual, which is easily computed after the model has been fit.

\subsection{Iterative estimation}
The maximum likelihood estimators from Section~\ref{sec:mle} imply that the parameters cannot, in general, be simultaneously estimated. We found that a simple iterative scheme of alternating estimators was able to correctly recovery true parameter values in a simulation study, Section~\ref{sec:validation}. The scheme proceeds according to Algorithm~\ref{alg:est} (see below). We defaulted to $\text{numIter} = 5$ iterations. Note that in the special, diagonal case, of equation~(\ref{eq:z}) we simultaneously estimated $\boldsymbol{\Lambda}$ and $\boldsymbol{W}$ using weighted linear regression, Section~\ref{sec:wlm} (``PCA'' case).

While we did estimate the asymptotic error, we found that the bootstrapped error had lower bias and hence we only report the latter (see Section~\ref{sec:validation}). To estimate the errors in parameters and prediction we bootstrapped Algorithm~\ref{alg:est} and took the standard deviation as the error estimate (100 resamples).

\begin{algorithm}
\caption{Iterative estimator}\label{alg:est}
\begin{algorithmic}
\If{imputeFirst}
    \State Impute missing $\vec{y}$ using simple algorithm (e.g.\ carry forward/back).
\EndIf
\If{doPCA}
    \State Estimate PCA rotation, $\boldsymbol{U}$, on first timepoint, $\vec{y}_1$, then apply to all timepoints.
\EndIf
\State Estimate $\boldsymbol{\Lambda}$ using equation~(\ref{eq:lambdaesteq}) which assumes $\langle \vec{y}_{n+1} \rangle = \langle \vec{y}_n \rangle = \vec{\mu}_n$ and $\text{Cor}(\vec{y}_{n}-\vec{\mu}_{n},\vec{x})=0$.
\State Estimate $\boldsymbol{W}$ using equation~(\ref{eq:west}).
\For{i in 1 to numIter}
\If{estimateNoise}
    \State Estimate $\boldsymbol{\Sigma}$ using equation~(\ref{eq:sigmaest}) and $\boldsymbol{Q} = \boldsymbol{\Sigma}^{-1}$.
\EndIf
\If{imputeMean}
    \State Transform model parameters into observed space using $\boldsymbol{U}^{-1}=\boldsymbol{U}^T$.
    \State Impute $\vec{y}_1$ with the model mean using equation~(\ref{eq:ynimp}).
    \For{n in 2 to numTimes}
        \State Impute $\vec{y}_n$ with the model mean using equation~(\ref{eq:ynp1imp}).
    \EndFor
\EndIf
\State Estimate $\boldsymbol{\Lambda}$ using equation~(\ref{eq:lambdaest}).
\State Estimate $\boldsymbol{W}$ using equation~(\ref{eq:west}).
\EndFor
\If{doPCA}
    \State Transform imputed values and parameters into observed space using $\boldsymbol{U}^{-1}=\boldsymbol{U}^T$.
\EndIf
\State Estimate asymptotic errors.
\State Return $\boldsymbol{W}$, $\boldsymbol{\Lambda}$, $\boldsymbol{\Sigma}$ and imputed values, $\vec{y}_{imp}$.
\end{algorithmic}
\end{algorithm}
Where imputeFirst, doPCA, estimateNoise and imputeMean are Boolean user settings. numTimes is the number of observation timepoints in the dataset, $T+1$.

\section{Validation} \label{sec:validation}
We used synthetic (simulated) data to validate: (i) Algorithm~\ref{alg:est}, (ii) the parameter errorbars, and (iii) the prediction error estimator (RMSE). We used synthetic data based on the SLAM C57/BL6 dataset for validation. We fit the full model equation~(\ref{eq:sf}) to the dataset: all 6 predictors and 2 covariates (sex and age), as well as estimating the noise. We used the fit parameters to generate new data then tested to see if Algorithm~\ref{alg:est} recovered the true parameters and errors. Algorithm~\ref{alg:est} was bootstrapped 100 times, the prediction error was estimated using both in-sample (train) and out-of-sample (test). We simulated 1000 times for each synthetic dataset size: 10, 50, 100, 500 and 1000 individuals. Each dataset had 22 timepoints.

\begin{figure*}
     \centering
        \includegraphics[width=\textwidth]{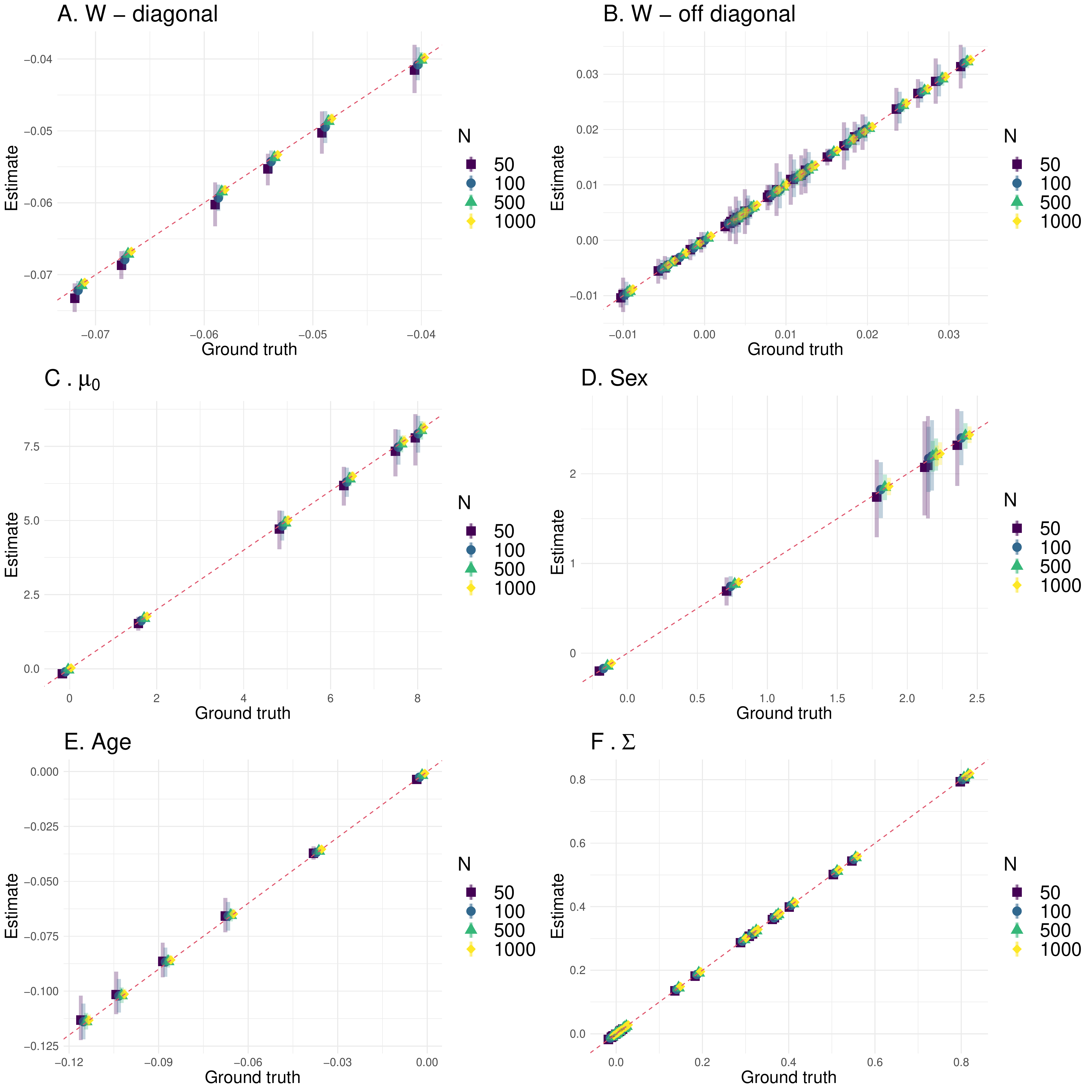}  
    \caption{Algorithm~\ref{alg:est} validation. For the indicated parameters in each measurement (A.-F.), the estimated value is plotted against the ground truth value for a variety of sample sizes (indicated by the legend). Points show mean; bands are the interquartile range (25th to 75th percentile). Bias is indicated by position of point relative to the red dashed line, $y=x$ (perfect estimator). Precision (and accuracy) are inferred by the dispersion (bands). As the number of individuals, $N$, is increased from $50$ to $1000$ we see the estimator becomes increasingly accurate and precise, with a small dispersion around the ground truth values for each parameter. Points are staggered for visualization. Note: $N=10$ had large errors and hence was excluded for better visualization.}
    \label{fig:fitvalidation}
\end{figure*}

We confirmed that Algorithm~\ref{alg:est} is able to accurately reproduce true model parameters. In Figure~\ref{fig:fitvalidation} we plot the parameter estimates versus the ground truth values. We see that the algorithm is accurate for $N\geq 50$. We see a bias-low for the diagonal elements of $\boldsymbol{W}$.

\subsection{Parameter error}
Here we test the calibration of our parameter errorbars. We compare both the bootstrap and asymptotic error estimates to the ground truth. Bootstrap errors were estimated using the standard deviation of bootstrap replicates. Asymptotic errors were estimated using the estimators in Section~\ref{sec:mle}. As we will demonstrate in this section, the asymptotic errorbars can be too small, whereas the bootstrap errors appeared to be correctly calibrated. For this reason, we always used the bootstrap estimates in the main text. The asymptotic error estimates are much faster to compute and are presented for posterity. 

\begin{figure*}
     \centering
        \includegraphics[width=.49\textwidth]{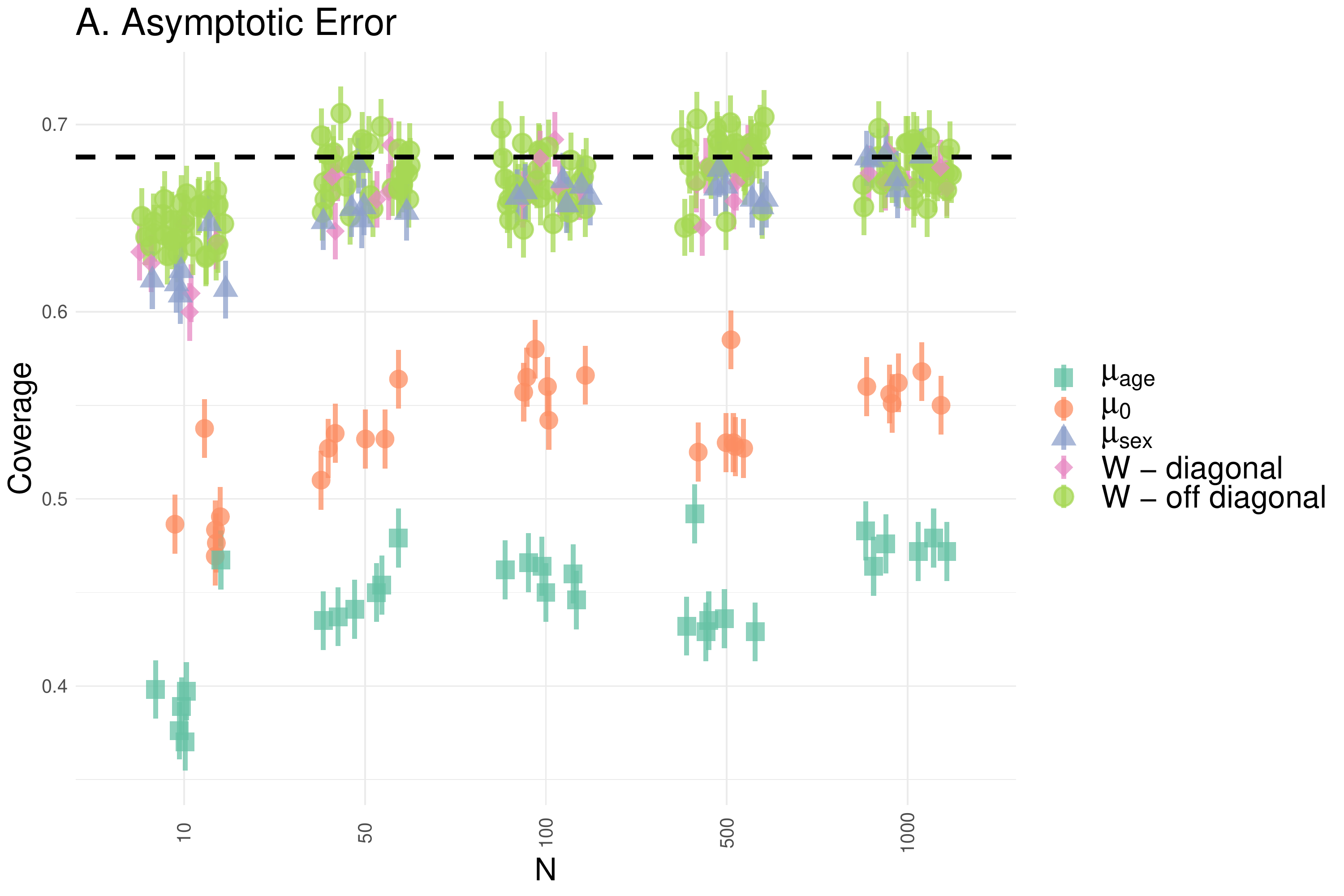}        \includegraphics[width=.49\textwidth]{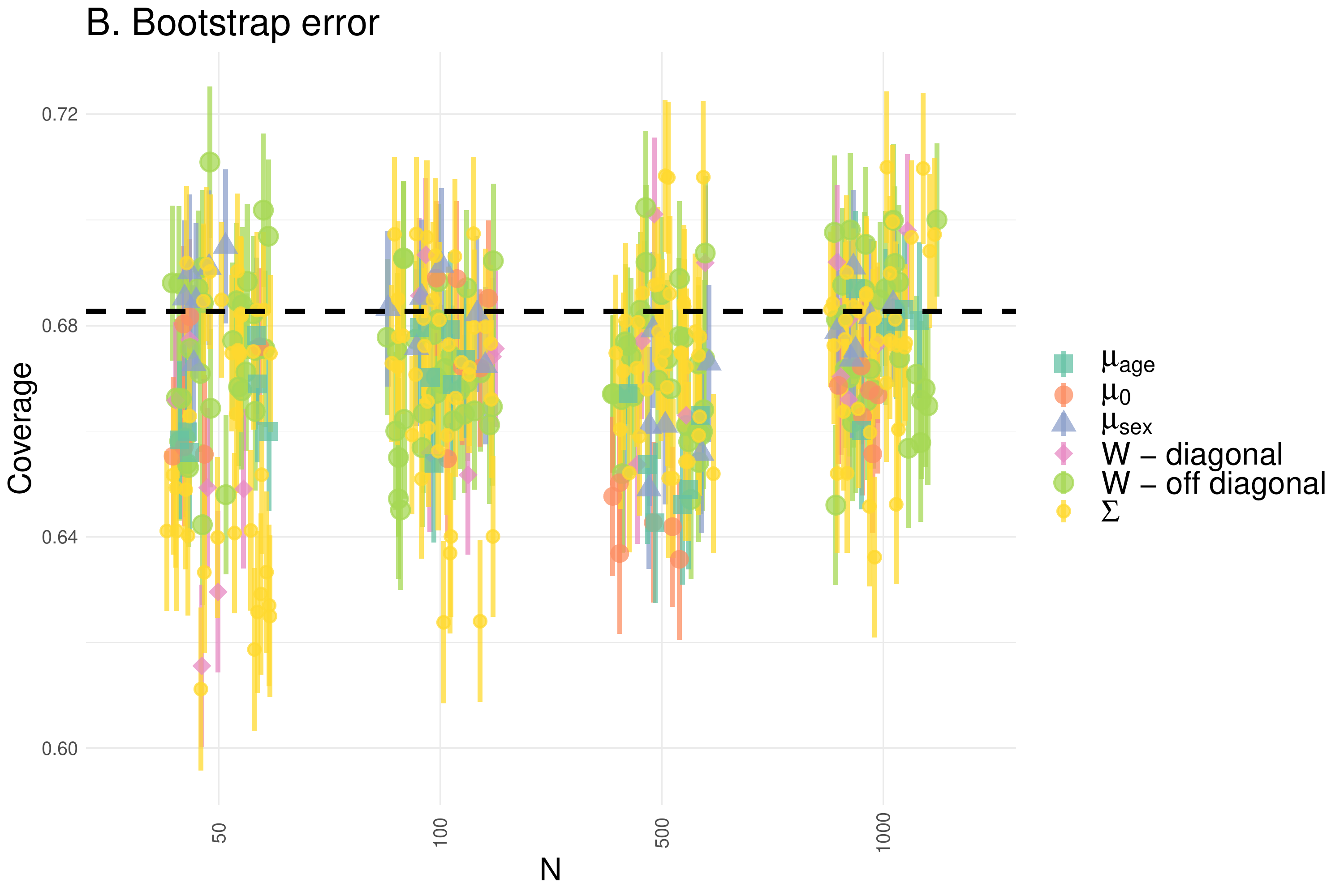}  
    \caption{Parameter errorbar validation (coverage). Asymptotic errorbars can be too small, whereas bootstrap errorbars appear to be valid. \textbf{A.} Asymptotic error clearly has abnormally low coverage for $\mu_0$ and $\mu_{age}$, perhaps due to strong correlations between the two parameters. Asymptotic error estimates for the other parameters look good. \textbf{B.} bootstrap error coverage looks good: parameters are close to the nominal rate (dashed line) and are (mostly) symmetrically distributed above and below. Note the scale. Errorbars are standard error in the mean. x-axis not to scale.}
    \label{fig:parcoverage}
\end{figure*}

In Figure~\ref{fig:parcoverage} we present the coverage of both error estimators. The coverage is the fraction of times that the true parameter value fell within the estimated error interval. The nominal coverage is 68.3\% for a normal random variable. In Figure~\ref{fig:parcoverage}A we present the coverage of the asymptotic error estimates and find that they are unsatisfactory for $\mu_{age}$ and $\mu_0$ (errorbars were too small). These may be due to strong correlations between the two parameters, for example the parameters for body weight correlated across simulations at $\text{cor}(\mu_{age},\mu_0)= -0.886$, which could make the asymptotic errors inaccurate. In Figure~\ref{fig:parcoverage}B we observe that all of the bootstrap error parameter coverages were close to the nominal rate (dashed line), and were symmetrically distributed above and below. This indicates that our bootstrap parameter errorbars were properly calibrated.

\subsection{Prediction error}
\begin{figure*}
     \centering
        \includegraphics[width=\textwidth]{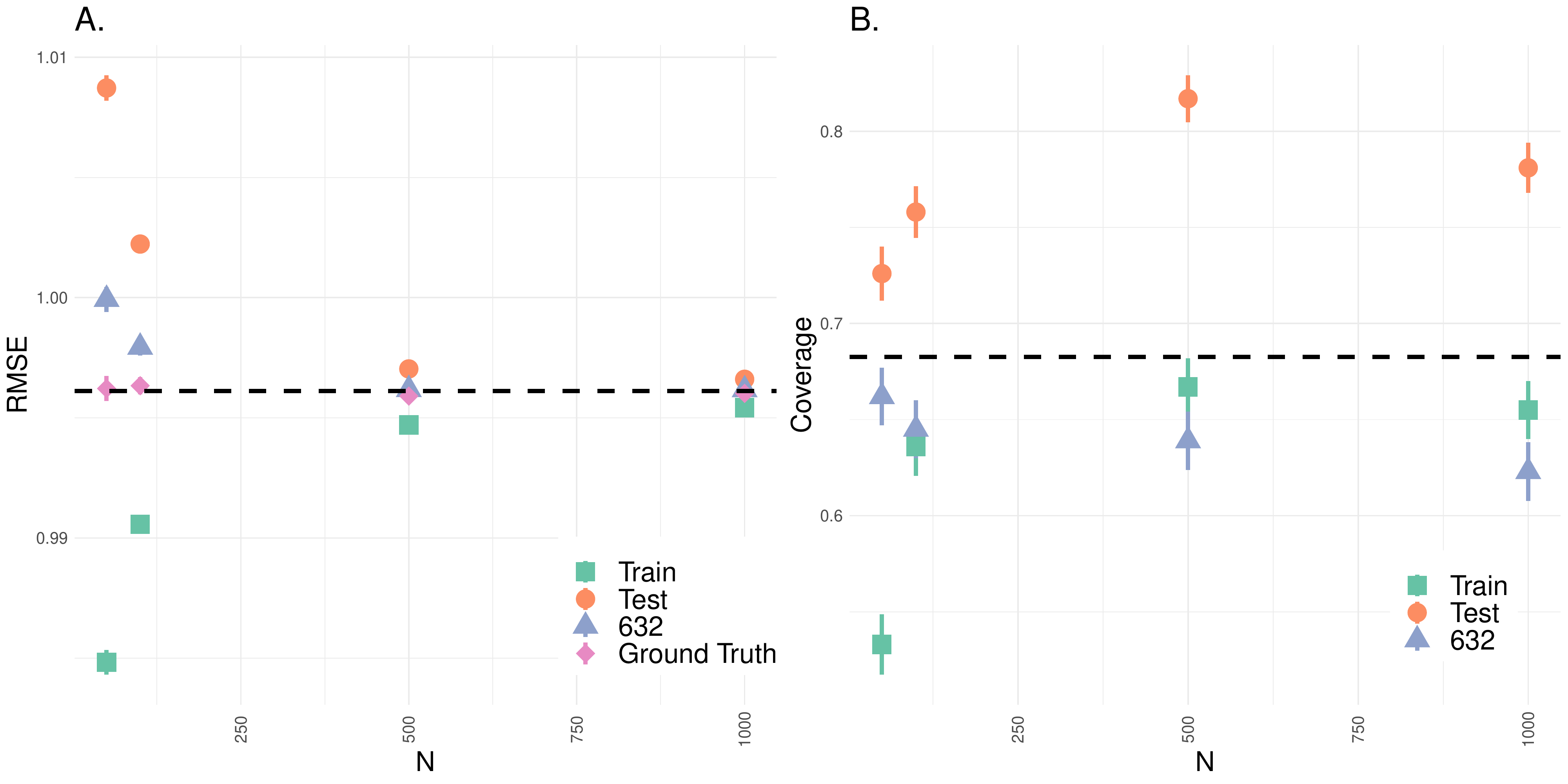}  
    \caption{Bootstrap error calibration. 632 error is a satisfactory estimator of the true error. \textbf{A.} Test error (out-of-sample) was biased high, training error (in-sample) was biased low, whereas 632 error was nearly unbiased relative to the ground truth. \textbf{B.} The coverage of the train and 632 error were close to the nominal rate, 68.3\% (dashed line). The test error clearly had abnormally high coverage, indicating the errorbars on the test error are too large. Note: the true (stochastic) error is difficult to precisely estimate due to non-uniform sampling, so we used the average ground truth to estimate the true error. Errorbars are standard error.}
    \label{fig:bootsfrmse}
\end{figure*}

Our primary measure of prediction error was the root-mean-squared error (RMSE). It is important that our measure is properly calibrated such that it estimates the correct RMSE, i.e. in a simulation study where the true error is known. We compare three RMSE estimators to the ground truth: (i) the testing error, which is the out-of-sample bootstrap error, (ii) the training error, which is the in-sample bootstrap error, and (iii) the 632 error which is a linear combination of 63.2\% testing error and 36.8\% training error \cite{Hastie2017-wl}. The ground truth error is the error of the sample given the correct parameter values: this is the error of a single sample, not the distribution of possible values. The average ground truth error should be an unbiased estimate of the true, distribution error. In Figure~\ref{fig:bootsfrmse}A we demonstrate that the 632 error is close to the ground truth error. In Figure~\ref{fig:bootsfrmse}B we present the coverage of each estimator and find they are all close to the nominal rate. We conclude that the 632 error is a satisfactory estimator of the true model error.

\FloatBarrier
\section{Math} \label{sec:math}
In this section we include supporting information for the model moments along with mappings to related approaches, i.e.\ other researcher's models.

\subsection{Ordinary differential equation} \label{sec:ode}
We consider only 1-dimension since we found that we could transform our multivariate biomarkers into a set of decoupled, 1-dimensional equations using the $\boldsymbol{W}$-diagonalizing matrix, $\boldsymbol{P}$. That is,
\begin{align}
    \vec{z} \equiv \boldsymbol{P}^{-1}\vec{y} \label{eq:zPinvy}
\end{align}
decouples the $z_j$ into a set of independent 1-dimensional equations. As needed, we can map back into $\vec{y}$ --- which we do in Section~\ref{sec:linear}.

In the limit of small $\Delta t$ our 1-dimensional model equation~(\ref{eq:z}) becomes a modified Ornstein-Uhlenbeck process as follows,
\begin{align}
    \lim_{\Delta t \to 0} z_{jn+1} &= z_{jn} + \lambda_j(z_{jn} - \tilde{\mu}_{jn})dt + \lim_{\Delta t \to 0} \tilde{\epsilon},
\end{align}
with
\begin{align}
    \tilde{\epsilon} \sim \mathcal{N}(0,\tilde{\sigma}_j^2|\Delta t|).
\end{align}
A Wiener process, $d\xi$, has three criteria\cite{Henderson2006-jg}: (i)  independence,  (ii) stationarity  (statistics doesn't change over time),  and  (iii) $\mathcal{N}(0,|\Delta t|)$--distributed. These criteria are satisfied by $\tilde{\epsilon}$ once we scale out $\tilde{\sigma}_j$. Substituting $t$ for timepoint $n$ and $t+dt$ for timepoint $n+1$ we have
\begin{align}
    z_{j}(t+dt) &= z_j(t) + \lambda_j(z_{j}(t) - \tilde{\mu}_{j}(t))dt + \tilde{\sigma}_jd\xi(t),
\end{align}
which can be rewritten as
\begin{align}
    dz_{j}(t) &= \lambda_j(z_{j}(t) - \tilde{\mu}_{j}(t))dt + \tilde{\sigma}_jd\xi(t), \label{eq:zou}
\end{align}
which is an Ornstein-Uhlenbeck process with a non-constant equilibrium, $\tilde{\mu}_{j}(t)$, which depends on time through $\tilde{\mu}_{j,age} t$ \cite{Henderson2006-jg}. Note that equation~(\ref{eq:zou}) holds for all $\tilde{\mu}_j(t)$ that can be Taylor expanded, since the nonlinear corrections go as $\mathcal{O}(\Delta t^2)$.

We rewrite equation~(\ref{eq:zou}) in a stripped down form as
\begin{align}
    dz &= \lambda(z-\mu(t))dt + \sigma d\xi \label{eq:out}.
\end{align}
The solution is then
\begin{align}
    z(t) = z_0e^{\lambda t}-\lambda e^{\lambda t}\int_0^t
    \mu(s) e^{-\lambda s}ds+\sigma e^{\lambda t}\int_0^t e^{-\lambda s}d\xi(s). \label{eq:ouint}
\end{align}
The integral is stochastic (Ito) and cannot be solved analytically, however its moments can be computed using two standard results \cite{Henderson2006-jg}. The mean is
\begin{align}
    \langle \int_0^t f(s,\xi)d\xi \rangle &= 0, 
\end{align}
and the two-point correlations are
\begin{align}
    \langle \int_0^t f(s,\xi)d\xi(s)\int_0^{t'} g(u,\xi)d\xi(u) \rangle &= \int_0^{\min(t,t')} \langle f(s,\xi)g(s,\xi) \rangle ds.
\end{align}
The statistics are Gaussian so all moments can be rewritten in terms of the mean and two-point correlations. 

\subsubsection{Moments}
Here we provide additional details supporting the math in Box 1.

Let
\begin{align}
    \mu(t) &= \mu_0 + \mu_{age} t,
\end{align}
that is, the only time-dependence is through the linear term $\mu_{age} t$ ($\mu_0$ can still depend on covariates, but they can't vary in time).

Starting from equation~(\ref{eq:ouint})
\begin{align}
    z(t) &= z_0e^{\lambda t}-\lambda e^{\lambda t}\int_0^t
    \mu(s) e^{-\lambda s}ds+\sigma e^{\lambda t}\int_0^t e^{-\lambda s}d\xi(s) \nonumber \\
    &= z_0e^{\lambda t}-\lambda e^{\lambda t}\bigg( -\frac{\mu_0}{\lambda}\big(e^{-\lambda t}-1\big) + \frac{\mu_{age}}{\lambda^2}\big( e^{-\lambda t}(-\lambda t-1) + 1\big) \bigg) +\sigma e^{\lambda t}\int_0^t e^{-\lambda s}d\xi(s).
\end{align}

The mean is easily computed as
\begin{align}
    \langle z(t) \rangle &= \langle z(0) \rangle e^{\lambda t}-\lambda e^{\lambda t}\bigg( -\frac{\mu_0}{\lambda}\big(e^{-\lambda t}-1\big) + \frac{\mu_{age}}{\lambda^2}\big( e^{-\lambda t}(-\lambda t-1) + 1\big) \bigg) \nonumber \\
    &= \langle z(0) \rangle e^{\lambda t}+ (\mu_0+\frac{\mu_{age}}{\lambda})(1-e^{\lambda t}) + \mu_{age} t. 
\end{align}
The auto-covariance doesn't depend on $\mu(t)$, it is simply,
\begin{align}
    \bigg\langle (z(t+\tau)-\langle z(t+\tau) \rangle)(z(t) - \langle z(t) \rangle) \bigg\rangle &= \sigma^2 e^{2\lambda t}e^{\lambda\tau} \bigg\langle \int_0^{t+\tau} e^{-\lambda s}d\xi(s) \int_0^t e^{-\lambda s}d\xi(s) \bigg\rangle \nonumber \\
    &= -\frac{\sigma^2}{2\lambda}\bigg(e^{\lambda|\tau|} - e^{2\lambda t}e^{\lambda\tau}\bigg).
\end{align}
The variance is the special case $\tau=0$,
\begin{align}
    \text{Var}(z(t)) &= -\frac{\sigma^2}{2\lambda}\bigg(1 - e^{2\lambda t}\bigg).
\end{align}
The mean and auto-covariance completely characterize Gaussian statistics, all other statistics can be calculated from them.

The above moments neglect the possibility that we may be unable to measure the system at $t=0$. It is therefore useful to define the moments relative to a reference time, $t_r$. Doing some algebra we have
\begin{align}
    \langle z(t) \rangle &= \langle z (t_r) \rangle e^{\lambda(t-t_r)} + (\mu_0+\frac{\mu_{age}}{\lambda} + \mu_{age}t_r)(1-e^{\lambda (t-t_r)}) + \mu_{age} (t - t_r) \nonumber \\
    &= \langle z (t_r) \rangle e^{\lambda(t-t_r)} + (\mu(t_r)+\frac{\mu_{age}}{\lambda})(1-e^{\lambda (t-t_r)}) + \mu_{age} (t - t_r)  
\end{align}
for the mean, where $\mu(t_r)\equiv \mu_0 + \mu_{age}t_r$. Note that it is convenient to write
\begin{align}
    \langle z(t) \rangle - \mu(t) &=  (\langle z (t_r) \rangle - \mu(t_r)) e^{\lambda(t-t_r)} + \frac{\mu_{age}}{\lambda}(1-e^{\lambda (t-t_r)}).
\end{align}

For the variance we have 
\begin{align}
    \text{Var}(z(t)) &= \text{Var}(z(t_r))e^{2\lambda (t-t_r)} - \frac{\sigma^2}{2\lambda}\bigg(1 - e^{2\lambda (t-t_r)}\bigg).
\end{align}
Note that if we wait a long time, $t-t_r \gg 1/\lambda$, we reach steady-state values (so long as $\lambda < 0$). For example, the steady-state  variance is
\begin{align}
    \text{Var}(z)_{ss} &= - \frac{\sigma^2}{2\lambda}. \label{eq:eqvar}
\end{align}

\subsection{Biomarker Principal Components}
The biomarkers, $\vec{y}$, are connected to the natural variables, $\vec{z}$, by the transformation, $\boldsymbol{P}^{-1}$, equation~(\ref{eq:zPinvy}). $\boldsymbol{P}^{-1}$ is the (linear) diagonalizing transformation of $\boldsymbol{W}$. We can use this to calculate the steady-state principal components of $\vec{y}$,
\begin{align}
    \text{Cov}(\vec{y}_{ss},\vec{y}_{ss}) &= \langle (\vec{y}_{ss}-\langle \vec{y}_{ss} \rangle) (\vec{y}_{ss}-\langle \vec{y}_{ss} \rangle) ^T \rangle \nonumber \\
    &= \boldsymbol{P}\langle (\vec{z}_{eq}-\langle \vec{z}_{ss} \rangle) (\vec{z}_{eq}-\langle \vec{z}_{ss} \rangle) ^T \rangle \boldsymbol{P}^{T} \nonumber \\
    &= \boldsymbol{P} \text{Cov}(\vec{z}_{ss},\vec{z}_{ss}) \boldsymbol{P}^{T}. \label{eq:ypca}
\end{align}
If $\boldsymbol{P}$ is a rotation/orthogonal ($\boldsymbol{P}^{-1}=\boldsymbol{P}^T$) then, by definition\cite{Byron1992-af}, $\boldsymbol{P}$ is the diagonalizing transformation of $\text{Cov}(\vec{y}_{ss},\vec{y}_{ss})$, with eigenvalues equal to the steady-state variance of the $z_j$. Note: $\boldsymbol{P}$ is orthogonal if $\boldsymbol{W}$ is real and symmetric\cite{Byron1992-af}. If we rank-order the $\text{Var}(z_j)$ then we have exactly the principal components of $\vec{y}$.\cite{Pridham2023-yj} Hence, in the steady-state the principal components are exactly the same as the natural variables, $\vec{z}$, sorted in order of decreasing variance, equation~(\ref{eq:eqvar}).

\subsection{Small Timesteps, $\Delta t$}
Our model, equation~(\ref{eq:sf}), approximates an ordinary differential equation in the limit $|\lambda\Delta t| \ll 1$ (Sections~\ref{sec:ode} and \ref{sec:linear}). Sehl and Yates \cite{Sehl2001-ld} found that most biomarkers decay linearly at a rate of $\lambda < 0.01~\text{year}^{-1}$ with the fastest being about $0.03~\text{year}^{-1}$. The frailty index --- the average number of health deficits an individual has --- accumulates at a similarly small rate of $0.025-0.04~\text{year}^{-1}$.\cite{Mitnitski2015-ia} We observed typical rates in the range $0.025-0.05~\text{human-equivalent year}^{-1}$ (Figure~2B), with sampling times $\Delta t$ of 4 years for ELSA, 3 years for Paquid, 4.9~human-equivalent years for SLAM C57/BL6 and 3.6~human-equivalent years for SLAM Het3. This implies that we can expect $|\lambda\Delta t| \ll 1$ and therefore the small $\Delta t$ approximation of equation~(\ref{eq:sf}) is likely fine. This means our model should behave similarly to an ordinary differential equation.

\subsection{General dynamics} \label{sec:gen}
Linear and nonlinear dynamical models alike can be analysed for stability near an equilibrium position using the eigenvalues\cite{Ledder2013-em}. The system is linearized as
\begin{align}
    \frac{d}{dt}\vec{y} &= \boldsymbol{W} \vec{y} + \vec{b}.
\end{align}
The system is stable if and only if the real parts of the eigenvalues are always negative (positive recovery).
Observe that the mean of our model equation~(\ref{eq:sf}) can be written as
\begin{align}
    \frac{\langle\vec{y}_{n+1}-\vec{y}_n\rangle}{\langle \Delta t_{n+1} \rangle} &= \boldsymbol{W} \langle \vec{y}_n \rangle - \boldsymbol{W} \vec{\mu}_{n} \nonumber \\
    &= \boldsymbol{W} \langle \vec{y}_n \rangle + \vec{b} 
\end{align}
for $\vec{b}\equiv - \boldsymbol{W} \vec{\mu}_n$. Hence for small $\Delta t$ we have (approximately)
\begin{align}
    \frac{d}{dt}\langle\vec{y}(t)\rangle &= \boldsymbol{W} \langle \vec{y}(t) \rangle + \vec{b} 
\end{align}
hence our approach probes the mean-stability of arbitrary linear or nonlinear dynamics.

\subsection{Stochastic process model (SPM) approximation} \label{sec:linear}
Our model can be used to analyse arbitrary dynamical systems near equilibrium, as discussed in Section~\ref{sec:gen}. Here we show how a specific dynamical model --- the stochastic process model (SPM) --- is approximated by our model. Our model was motivated in part by earlier works which have shown that biomarker data can be modelled as a stochastic differential equation \cite{Yashin2007-py,Farrell2022-lw}. The earlier work by Yashin \textit{et al.} proposed the SPM as a generic framework for longitudinal aging biomarker data \cite{Yashin2007-py} where an individual's collection of biomarkers, $\vec{y}$, evolves over time as
\begin{align}
    d\vec{y} &= \boldsymbol{A}(t)(\vec{y} - \vec{\mu}(t))dt + \boldsymbol{B}(t)d\xi_t \label{eq:spm}
\end{align}
where $\vec{\mu}$ is the unknown equilibrium term (``functional state'' of the organism), $\boldsymbol{A}$ is the interaction network and $d\xi_t$ is a Wiener noise term modified by the matrix $\boldsymbol{B}$. Subsequent work by Farrell \textit{et al.} demonstrated that a deep neural network could be used to fit SPM and further that a time-independent linear interaction model was sufficient to describe the interaction network, $\boldsymbol{A}$, for ELSA data \cite{Farrell2022-lw}. Our model, equation~(\ref{eq:sf}), is the appropriate approximation for equation~(\ref{eq:spm}) for small timesteps. 

\textbf{Proof:} In Section~\ref{sec:ode} we showed that our 1-dimensional model is equivalent to a Wiener process in the limit of $\Delta t \to 0$. Consider the SPM with constant regulation matrix, $\boldsymbol{A}$, and linear functional state, $\mu$,
\begin{align}
    d\vec{y} &= \boldsymbol{A}(\vec{y}-\vec{\mu}(t))dt + \boldsymbol{B}d\xi
\end{align}

Suppose $\boldsymbol{A}$ is diagonalizable then,
\begin{align}
    d\vec{y} &= \boldsymbol{P}\boldsymbol{D}\boldsymbol{P}^{-1}(\vec{y}-\vec{\mu})dt + \boldsymbol{B}d\xi,   \nonumber \\
\implies d(P^{-1}\vec{y}) &= \boldsymbol{D}(\boldsymbol{P}^{-1}(\vec{y}-\vec{\mu}))dt + \boldsymbol{P}^{-1}\boldsymbol{B}d\xi, \nonumber \\
\implies d\vec{z} &= \boldsymbol{D}(\vec{z}-\tilde{\vec{\mu}})dt + \boldsymbol{\tilde{B}}d\xi
\end{align}
for the latent space, $\vec{z}\equiv \boldsymbol{P}^{-1}\vec{y}$. By inspection, the latent space obeys Ornstein-Uhlenbeck dynamics with $D_{jj}=\lambda_j$ and hence we can approximate each $z_j$,
\begin{align}
    z_j(t+\Delta t) &\approx z_j(t)+D_{jj} ( z_j(t)-\tilde{\mu}_j )\Delta t + \tilde{\epsilon}_j,~~~~\text{where} \nonumber \\
    \tilde{\vec{\epsilon}} \sim \mathcal{N}(0,\boldsymbol{\tilde{B}}\boldsymbol{\tilde{B}}^T|\Delta t|)
\end{align}
which we can map into the observed space using $\boldsymbol{P}$ to get,
\begin{align}
    \Aboxed{ \vec{y}(t+\Delta t) &\approx \vec{y}(t)+\boldsymbol{A}( \vec{y}(t)-\vec{\mu} )\Delta t + \vec{\epsilon} & \vec{\epsilon} \sim \mathcal{N}(0,\boldsymbol{B}\boldsymbol{B}^T |\Delta t|) }
\end{align}
which is equation~(\ref{eq:sf}) with $\boldsymbol{A}\equiv \boldsymbol{W}$. The transformed variance of $\tilde{\epsilon}_i$ is,
\begin{align}
    \langle (\boldsymbol{P}\tilde{\vec{\epsilon}})(\boldsymbol{P}\tilde{\vec{\epsilon}})^T \rangle &= \boldsymbol{P}\langle \tilde{\vec{\epsilon}}\tilde{\vec{\epsilon}}^T\rangle \boldsymbol{P}^T \nonumber \\
    &= \boldsymbol{P} \langle \boldsymbol{\tilde{B}}\boldsymbol{\tilde{B}}^T\rangle \boldsymbol{P}^T |\Delta t| \nonumber \\
    &= \langle \boldsymbol{B}\boldsymbol{B}^T\rangle |\Delta t| \nonumber \\
    &\equiv \boldsymbol{\Sigma} |\Delta t|
\end{align}
\textbf{QED}.

\subsection{Mapping to Sehl and Yates} \label{sec:sehl}
Sehl and Yates performed a meta-analysis of 469 biomarkers across cross-sectional and longitudinal aging studies and observed that the vast majority of biomarkers decay linearly with age \cite{Sehl2001-ld}. In the present section we demonstrate that their linear model describes the steady-state dynamics of our model. In other words, the long-time (old-age) behaviour of our model is consistent with their observations.

In the steady-state our model equation~(7) becomes linear in time,
\begin{align}
    \langle z_{jn} \rangle_{ss} &= \mu_{0j}(\vec{x}) + \mu_{age,j}t_n - \frac{\mu_{age,j}}{|\lambda_j|} \label{eq:zss}
\end{align}
where we have included all time-independent covariates in $\mu_0(\vec{x})$ for convenience. The biomarkers have a one-to-one relationship with the natural variables through the orthogonal transformation $\boldsymbol{P}$ and thus evolve according to
\begin{align}
    \langle y_{ln} \rangle_{ss} &= \sum_j P_{lj}\mu_{age,j}t_n + \sum_jP_{lj}\bigg(\mu_{0j}(\vec{x})- \frac{\mu_{age,j}}{|\lambda_j|}\bigg). \label{eq:yss}
\end{align}
The Sehl and Yates model\cite{Sehl2001-ld} is
\begin{align}
    \frac{y_{ln}}{y_{l,30}} &= 1 - k_l(t_n-30)
\end{align}
for biomarker $y_{l}$ with baseline value of $y_{l,30}$ at age 30; the age is in years and $k_l$ is the rate in \%-change per year. We can rewrite their model as
\begin{align}
    y_{ln} &= -k_ly_{l,30}\text{t}_n+(30k_l+1)y_{l,30},
\end{align}
which is exactly equation~(\ref{eq:yss}) with the substitutions
\begin{align}
    \sum_j P_{lj} \mu_{age,j} &\equiv -k_ly_{l,30},~~~~\text{and} \nonumber \\
    \sum_j P_{lj} \bigg(\mu_{0j}-\frac{\mu_{age,j}}{|\lambda_j|}\bigg) &\equiv (30k_l+1)y_{l,30},
\end{align}
which can be mapped into $\vec{z}$ using $\boldsymbol{P}^{-1}$:
\begin{align}
     \mu_{age,j} &\equiv -\sum_l P^{-1}_{jl}k_ly_{l,30},~~~~\text{and} \nonumber \\
    \mu_{0j}-\frac{\mu_{age,j}}{|\lambda_j|} &\equiv \sum_l P^{-1}_{jl}(30k_l+1)y_{l,30}.
\end{align}
Observe that $\boldsymbol{P}$ permits the drift of only a few $\vec{z}$ to map into many observed biomarkers, $\vec{y}$. Together with our observation that many more biomarkers drift than do natural variables, Figure~\ref{fig:allostasis_vs}, this implies that Sehl and Yates' observation that most biomarkers drift with age may be due to a only few underlying (allostatic) natural variables that are declining with age.

\FloatBarrier

\section{Additional Results} \label{sec:results}
We restricted the main text to only our key results. Here we provide additional information to support our conclusions.

\begin{figure*}[!h]
     \centering
        \includegraphics[width=\textwidth]{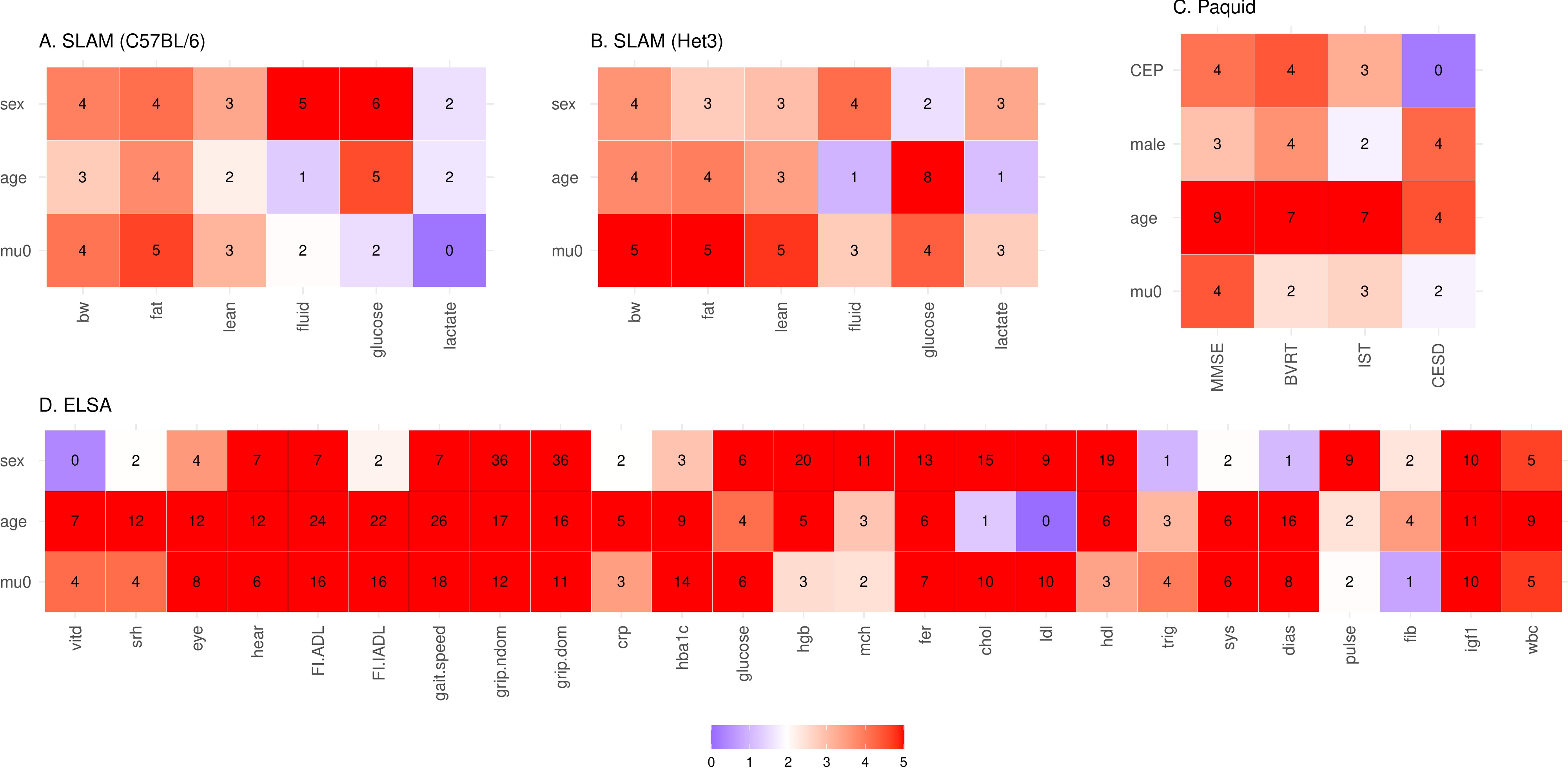}  
    \caption{Covariate significance (z-scores). \textbf{A.} C57BL/6 mice (SLAM). \textbf{B.} Het3 mice (SLAM). \textbf{C.} Paquid (human, dementia). \textbf{D.} ELSA (human). The equilibrium term, $\mu$, was a linear function of these covariates. Most covariates were significant (red or white). Only the blue tiles were not significant at 95\% ($z=1.96$). Tile number is z-score. Colour scale is truncated at $z=5$ ($p=6\cdot 10^{-7}$). See Figures~\ref{fig:cor} and \ref{fig:corcov} for the directions of the covariate effects.}
    \label{fig:zscore}
\end{figure*}

We included covariates, $\vec{x}$, in the equilibrium position, $\mu(\vec{x})$, to reduce confounding effects and to test for the presence of allostasis (which depends on age). Here we tested each parameter for significance using the bootstrap parameter error estimates. The z-score for each covariate is reported in Figure~\ref{fig:zscore}; blue tiles are not significant, white and red are significant at $p\leq 0.05$. Most covariates were significant, particularly age for the human studies. In Section~\ref{sec:select} we found that the effect of covariates on prediction was typically small. This means that the effects of covariates were reliably estimated (small $p$) but did not explain much variation (minor effect on RMSE).

\begin{figure*}
     \centering
         \includegraphics[width=\textwidth]{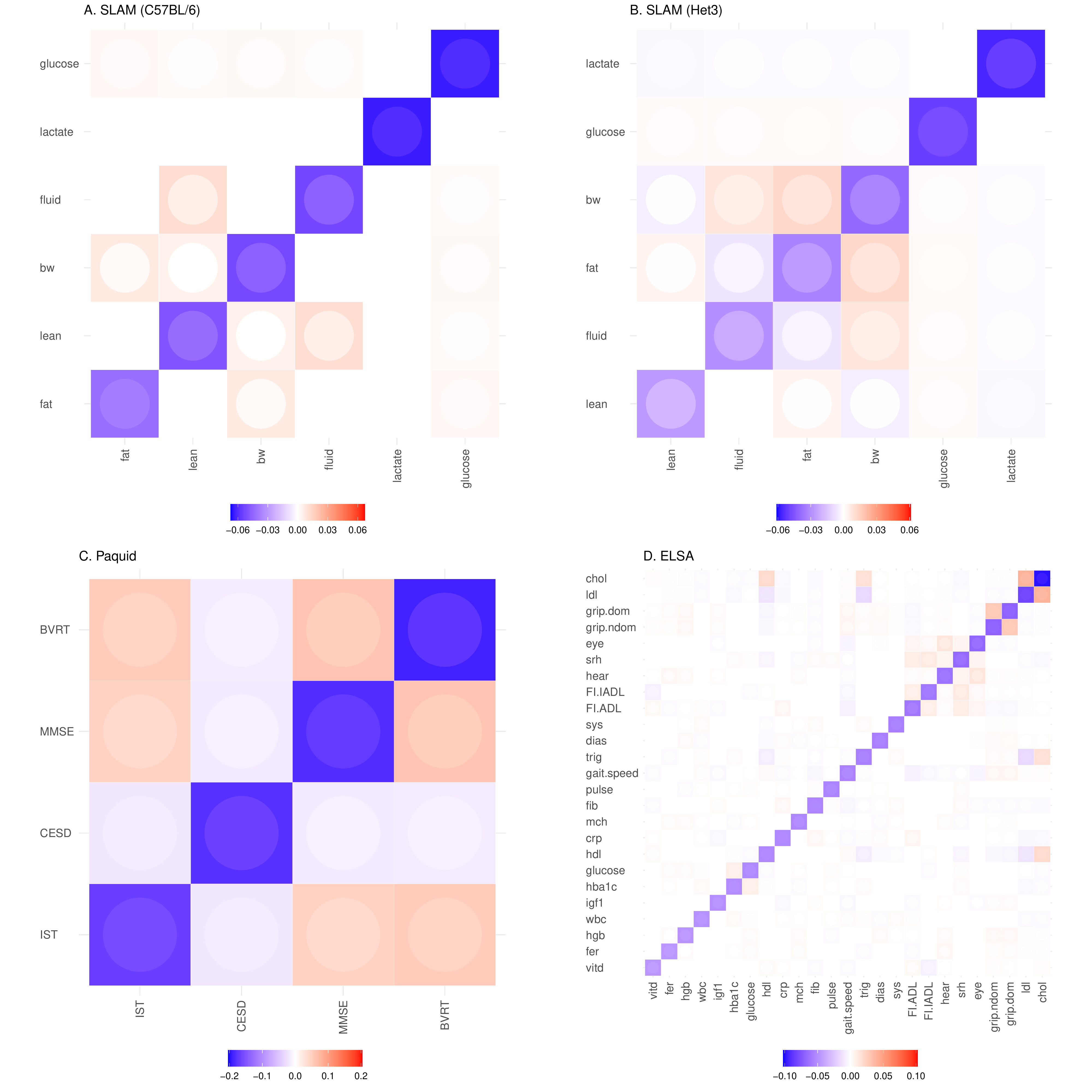}
    \caption{Interaction networks for all datasets. \textbf{A.} C57BL/6 mice (SLAM). \textbf{B.} Het3 mice (SLAM). \textbf{C.} Paquid (human, dementia). \textbf{D.} ELSA (human). Tile colour indicates interaction strength (saturation) and direction (colour) of the interaction from the y-axis variable to the x-axis variable. Inner colour indicates the limit of 68\% confidence interval (CI) closest to zero (i.e.\ standard error). Non-significant interactions, at 68\%, have been whited-out. Variables are sorted by diagonal strength (increasing rate). The matrices are real and symmetric because the data were diagonalized by an orthogonal matrix (PCA).}
    \label{fig:networks}
\end{figure*}

Our model estimates an interaction network, $\boldsymbol{W}$, together with equilibrium positions. In the main text we presented the ELSA network with suppressed diagonal (Figure~2). The complete networks for each dataset are provided in Figure~\ref{fig:networks}. The networks are all symmetrical because we used PCA as a preprocessing step. Relationships indicate how the y-axis variable will affect the x-axis variable during the next timestep.

\begin{figure*}
     \centering
        \includegraphics[width=\textwidth]{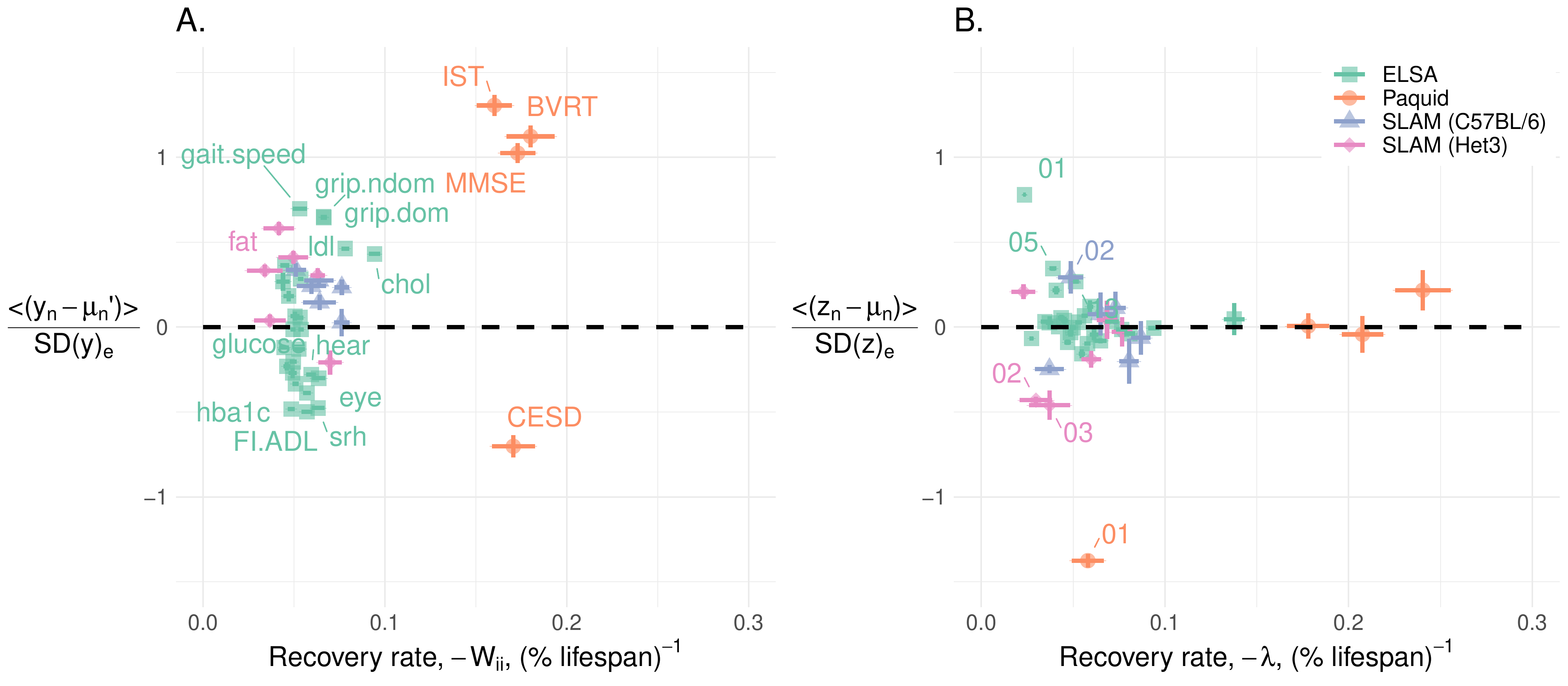}  
    \caption{Homeostasis of biomarkers vs natural variables. The dysruption of homeostasis seems to be diffuse across biomarkers whereas it is concentrated into a few natural variables. \textbf{A.} Observed biomarkers were typically far from equilibrium (dotted line). \textbf{B.} In contrast, most natural variables were close to equilibrium. We inferred that variables close to equilibrium were in homeostasis whereas those far from equilibrium were allostatic. Together these plots suggest that the natural variables were able to condense the effects of allostasis into a few major variables.}
    \label{fig:allostasis_vs}
\end{figure*}

Our model also estimates an equilibrium homeostatic position for each variable, $\mu$. An important question is how strongly do variables adhere to homeostasis in the biomarkers, $\vec{y}$ versus the natural variables, $\vec{z}$?  In the main text we presented the difference between the natural variable mean and the equilibrium position for each variable, $\langle z_j-\mu_j \rangle$. We reproduce that figure beside the observed biomarkers in Figure~\ref{fig:allostasis_vs}. In Figure~\ref{fig:allostasis_vs}B (and Figure~3A) we observed that the natural variables appear to be split into two groups: the majority group was close to $\mu$, indicative of homeostasis, whereas the minority group was far from homeostasis. This latter group had a strong drift term, $\mu_{age}$, which indicated that homeostasis was a moving target i.e.\ allostasis. In Figure~\ref{fig:allostasis_vs}A we show that the observed biomarkers were much more likely to be far from homeostasis than the natural variables (B), implying that the natural variables are able to condense the effects of age-related drift (allostasis) into a few variables (see also Section~\ref{sec:sehl}). 

\begin{figure*}
     \centering
         \includegraphics[width=\textwidth]{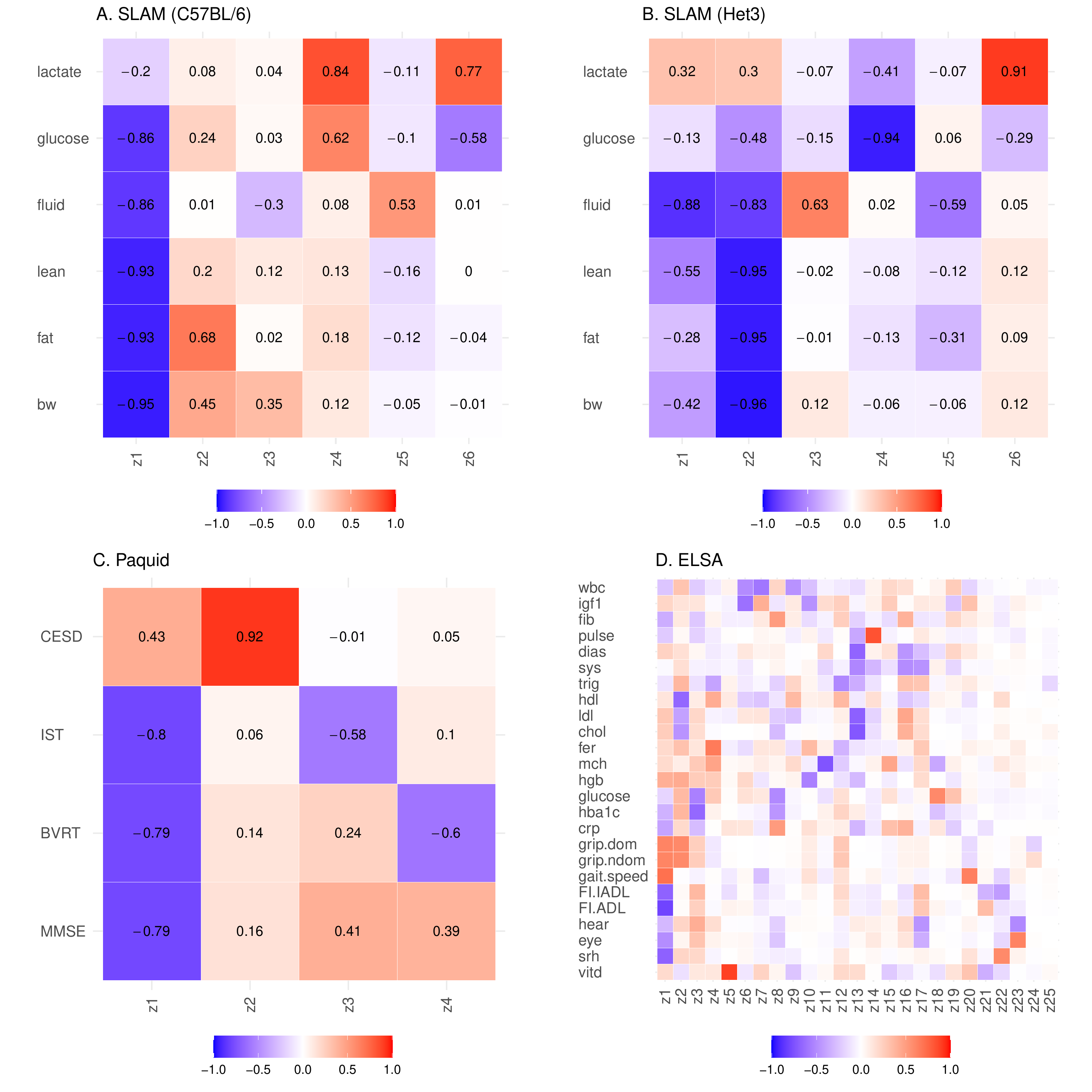}
    \caption{Natural variable correlates --- biomarkers (predictors). \textbf{A.} C57BL/6 mice (SLAM). \textbf{B.} Het3 mice (SLAM). \textbf{C.} Paquid (human, dementia). \textbf{D.} ELSA (human). This helps to describe what information is in each natural variable, $z$, and therefore what each natural variable is capable of controlling. The sign of each $z$ is arbitrary due to idiosyncrasies of the eigendecomposition.}
    \label{fig:cor}
\end{figure*}

\begin{figure*}
     \centering
         \includegraphics[width=\textwidth]{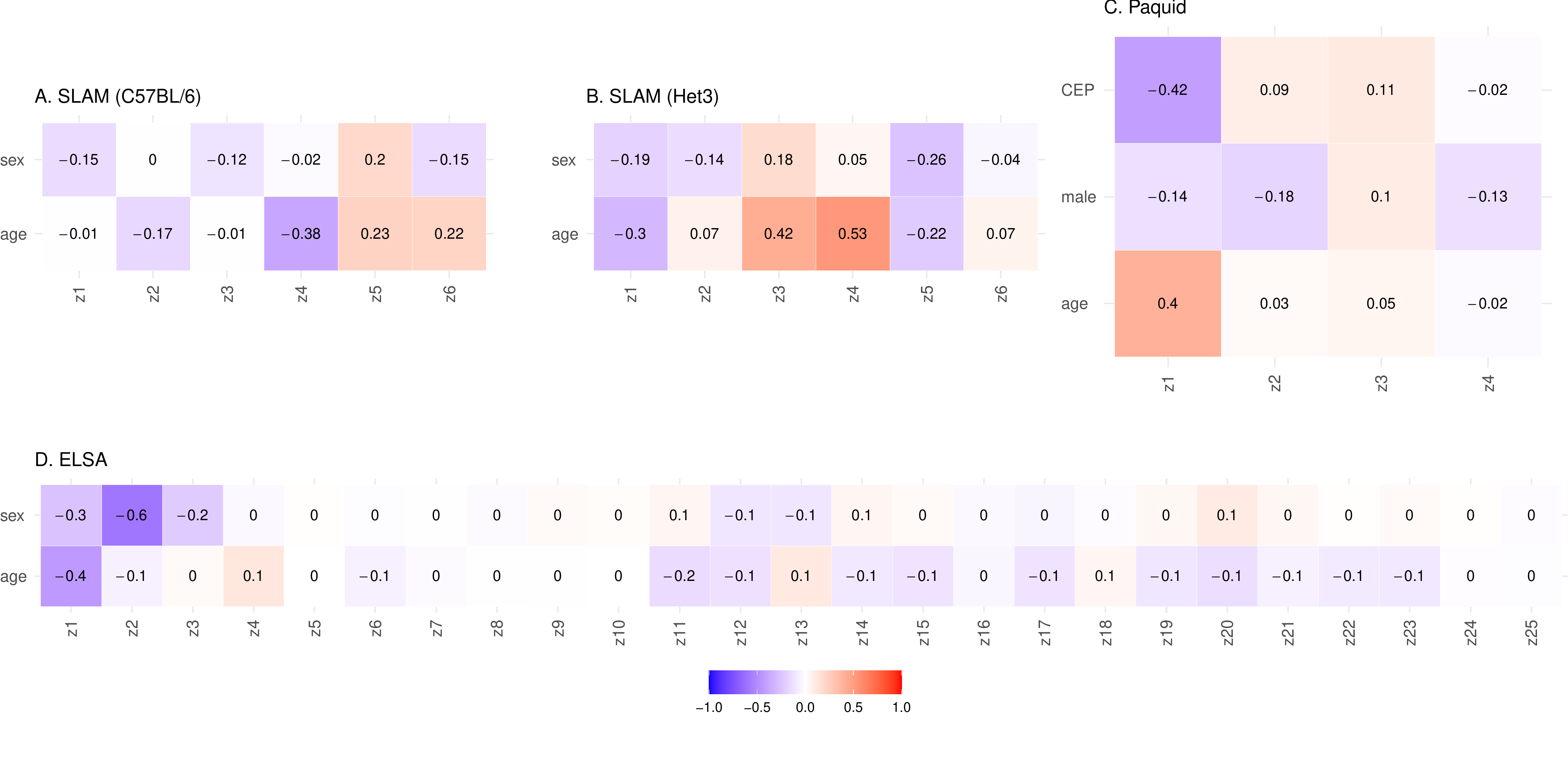}
    \caption{Natural variable correlates --- covariates. \textbf{A.} C57BL/6 mice (SLAM). \textbf{B.} Het3 mice (SLAM). \textbf{C.} Paquid (human, dementia). \textbf{D.} ELSA (human). This provides further information about what information each natural variable, $z$, contains. We expect the strongly drifting variables to exhibit correlations with age, though the sign of each $z$ is arbitrary. Male is a binary sex indicator (1: male, 0: female); sex is the converse (0: male, 1: female). CEP is educational attainment level (1: attained primary, 0: did not).}
    \label{fig:corcov}
\end{figure*}

The natural variables appear to be efficient for representing age-related changes. What do the natural variables mean in terms of observable outcomes? In Figure~\ref{fig:cor} we report the correlations between the accumulating/drifting natural variables and biomarkers. In Figure~\ref{fig:corcov} we report correlations with covariates. Together these give us an idea of what each natural variable represents and, by model implication, is controlling. For example, $z_1$ of Paquid is strongly correlated with the mental acuity scores: MMSE, BVRT and IST, implying it represents overall mental acuity. This may explain why $z_1$ was such a strong predictor of dementia (Figure~\ref{fig:allo_survival}). In this way the age-related decline of mental acuity can be represented by changes to just one variable, $z_1$, but also observed across several biomarkers, MMSE, BVRT and IST.

\begin{figure*}
     \centering
         \includegraphics[width=.8\textwidth]{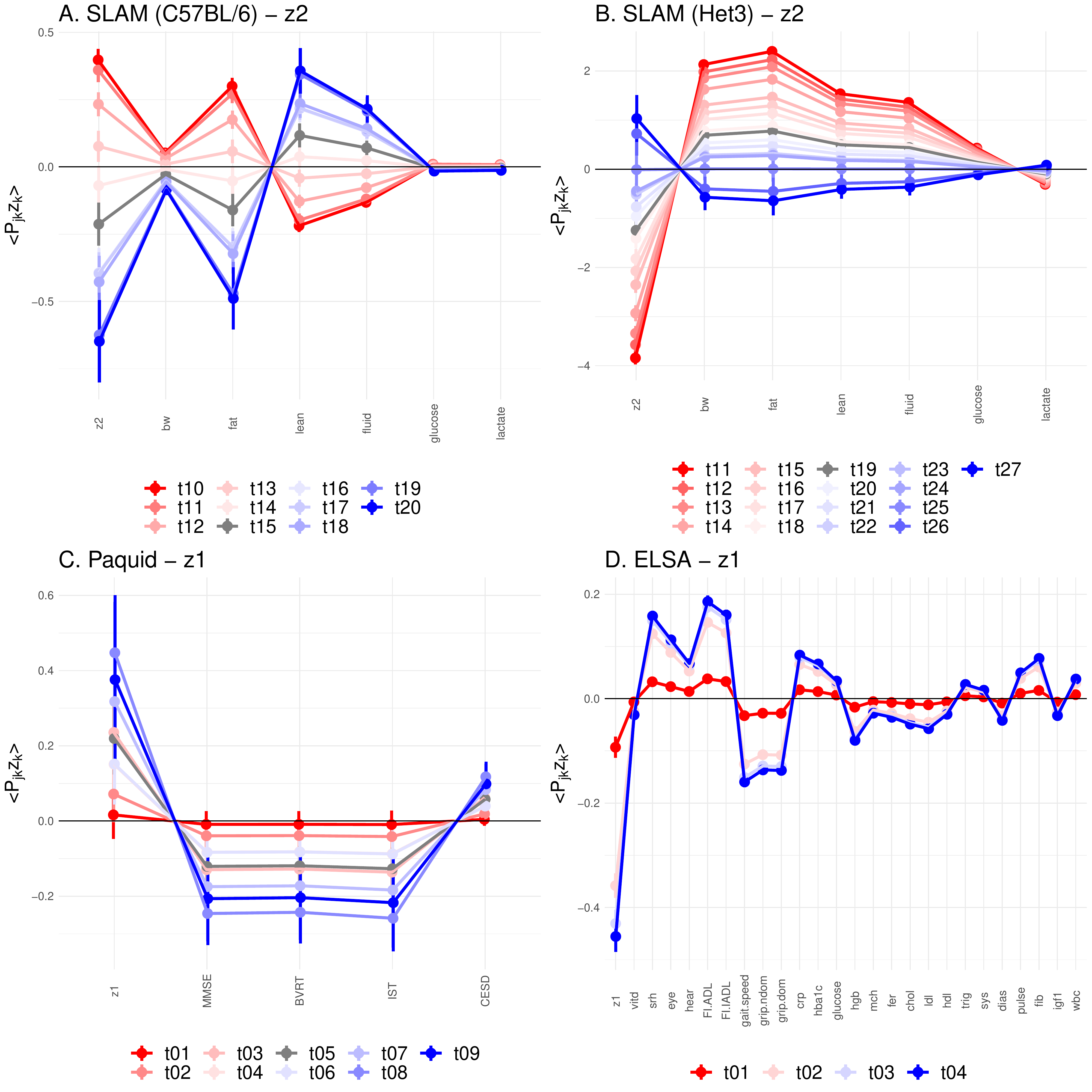}
    \caption{Natural variable drift drives biomarker drift. \textbf{A.} C57BL/6 mice (SLAM). \textbf{B.} Het3 mice (SLAM). \textbf{C.} Paquid (human, dementia). \textbf{D.} ELSA (human). We consider the drift of the primary risk natural variables: $z_1$ for ELSA and Paquid and $z_2$ for SLAM. We observe a continuous drift in the natural variables. We also plot the drift of the biomarkers which is directly caused by each $z$ via $\boldsymbol{P}$. In this manner, a few natural variables can drive drift across several biomarkers. Since $\boldsymbol{P}$ is orthogonal (length-preserving) the drift of each natural variable must be diluted across biomarkers (at most a single biomarker can drift at the same rate). See also the correlation matrices, Figures~\ref{fig:cor} and \ref{fig:corcov}. For the SLAM datasets we've included only timepoints where the average age was over 80~weeks.}
    \label{fig:zdriftall}
\end{figure*}

\begin{figure*}
     \centering
        \includegraphics[width=.55\textwidth]{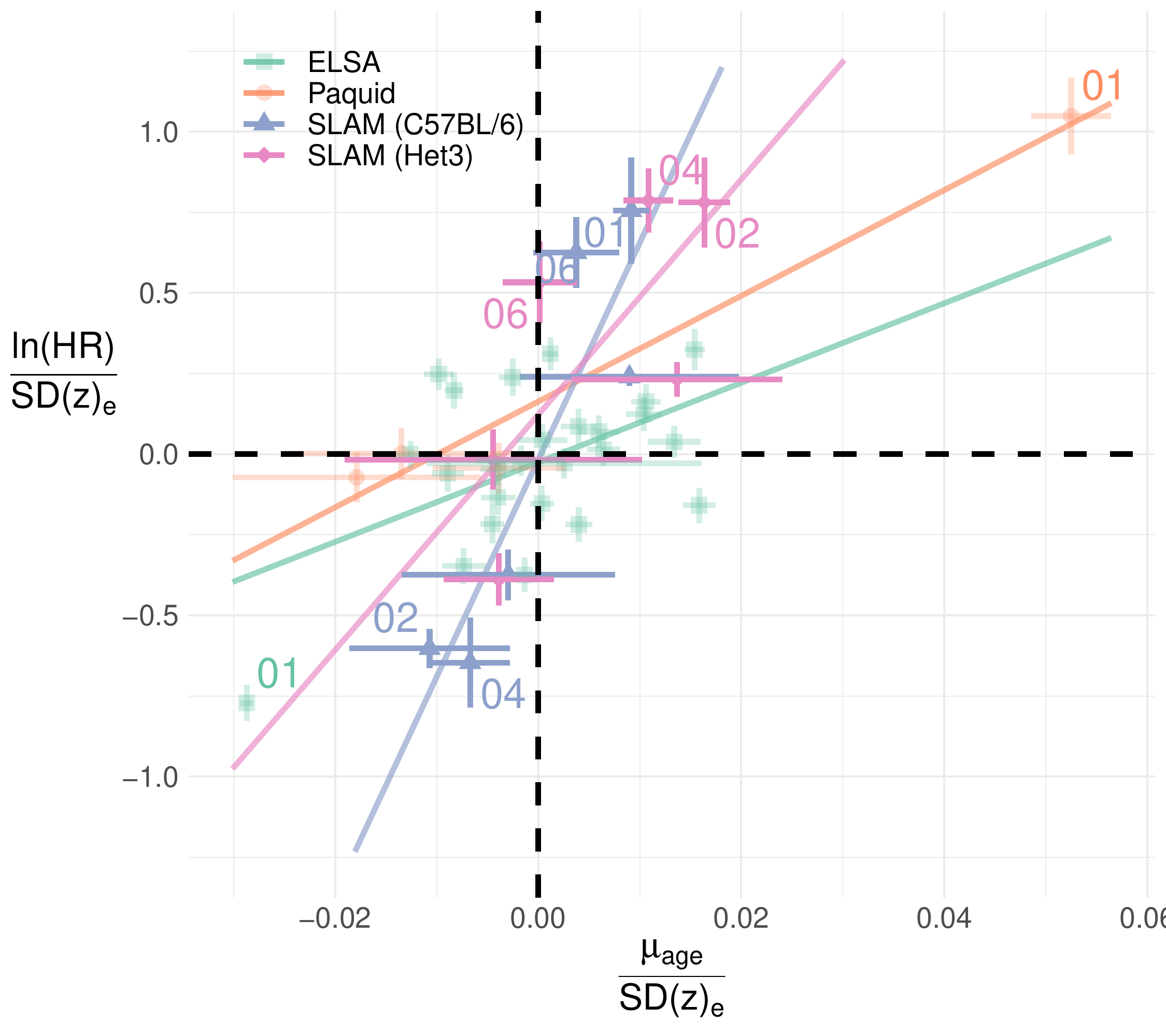}
    \caption{Allostasis drifts towards the risk direction. We fit a Cox model for each natural variable including age and sex as covariates. The Cox coefficient --- i.e. log--hazard ratio (HR) per unit increase --- correlates with the steady-state drift rate, $\mu_{age}$. The dominant risk direction for each dataset has been labelled by eigenvalue rank (e.g.\ $z_1$ is 01). The equilibrium standard deviation provides a native scale for each variable. }
    \label{fig:allo_survival}
\end{figure*}

The linear map, $\boldsymbol{P}$, allows a few natural variables to cause several biomarkers to drift. The effects of allostasis on observed biomarkers via the primary risk natural variables are illustrated in Figure~\ref{fig:zdriftall}. The sign of each natural variable is arbitrary, due to idiosyncrasies in eigendecompositions \cite{Pridham2023-yj}. The dominant survival dimension for the Het3 mouse data was $z_2$, which appears to capture a loss of body fat and muscle, and relative gain of fluid. The dominant $z_2$ dimension for the C57BL/6 was more specific to loss of fat (the $z_1$ signal for C57BL/6 was very similar to the $z_2$ signal for Het3, Figure~\ref{fig:cor}). $z_1$ was the dominant dementia-free-survival dimension for Paquid, and captured a system-wide drop in mental acuity scores (MMSE, BVRT and IST), which likely captures cognitive decline associated with dementia. $z_1$ for ELSA appears to be related to frailty \cite{Pridham2023-yj}, having its effects spread across many variables, especially those related to disability (eye, hear, FI ADL and FI IADL), physical condition (grip strength and gait speed), and self-reported health (SRH); note that higher is better for physical condition variables and worse for the other variables (eye, hear, SRH, FI, etc).  In all cases the effects are strongest in the natural variables, which is ensured by the orthogonality of $\boldsymbol{P}$. This means that the effects of the natural variable drift must be diluted across the observed biomarkers (e.g. Figure~\ref{fig:allostasis_vs}), potentially hiding them within healthy variation. 

The drift rate of the natural variables, $\mu_{age}$, was correlated with the risk of adverse outcome (Figure~4A). We named this phenomenon ``mallostasis'': the tendency of an aging system towards an ever-worsening equilibrium. Here we consider the role of confounding variables by constructing complete survival models for each natural variable and adverse outcome (mortality or dementia onset). We constructed a (time-dependent) survival model for each natural variable, with age, sex and the natural variable as predictors. We then recorded the Cox proportional hazards coefficients, which represent the (conditional) log--hazard ratios per unit increase for each natural variable. We observed that the Cox coefficients correlated with the drift rate, $\mu_{age}$: Figure~\ref{fig:allo_survival}. This provides more robust support for mallostasis: that the steady-state behaviour of aging mice and humans is declining natural variables and commensurately declining health.

\begin{figure*}
     \centering
        \includegraphics[width=\textwidth]{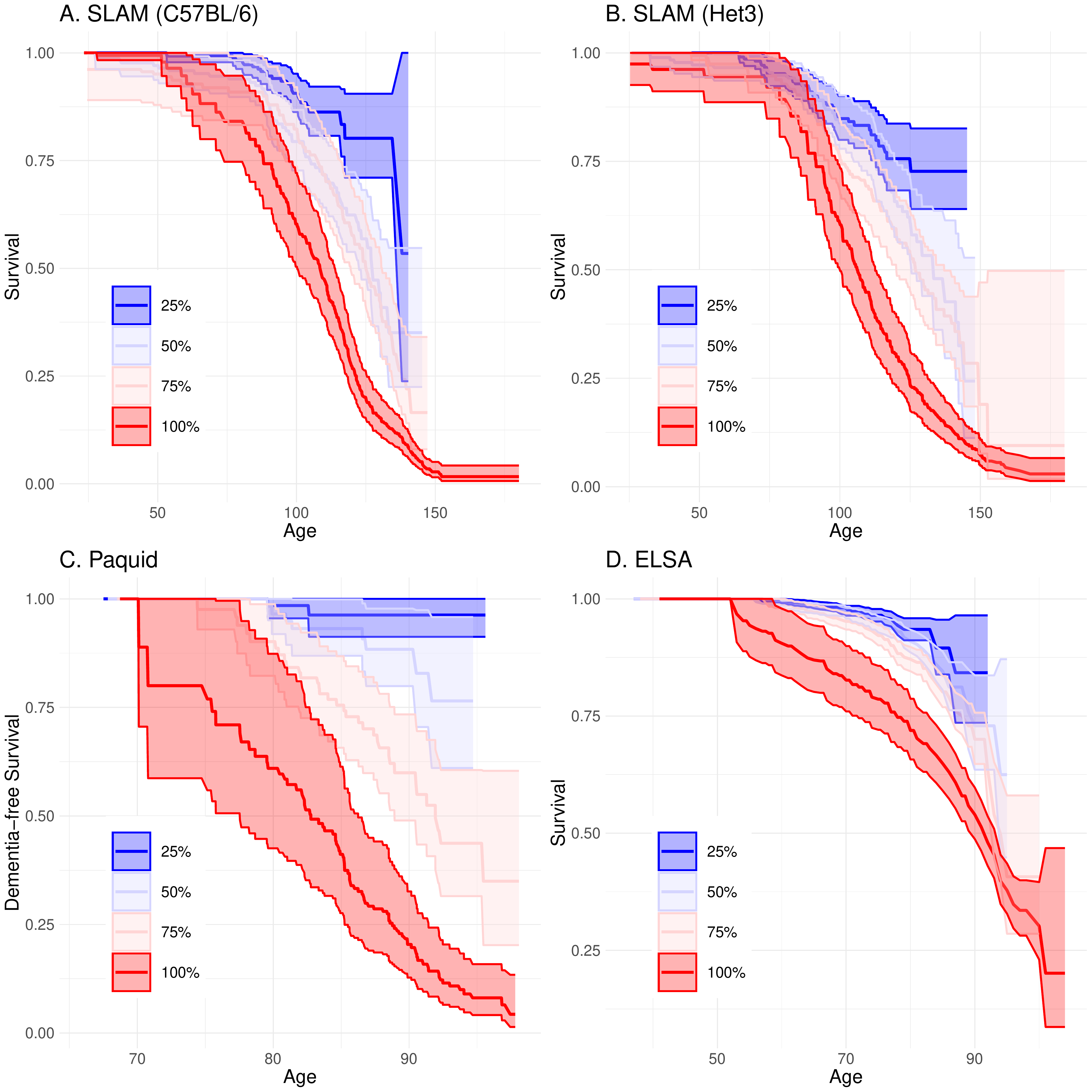}
    \caption{Composite health measure performance. \textbf{A.} C57BL/6 mice (SLAM). \textbf{B.} Het3 mice (SLAM). \textbf{C.} Paquid (human, dementia). \textbf{D.} ELSA (human). A simple estimator of health is $\vec{\mu}_{age}^T\vec{z}$. This leverages mallostasis to infer individual health. Large separation between quartiles (colours) indicates a strong predictor of adverse outcome.}
    \label{fig:ba_survival}
\end{figure*}

As an illustration of mallostasis, we consider a simple composite health measure, $b\equiv \vec{\mu}_{age}^T\vec{z}$. Figure~\ref{fig:allo_survival} demonstrates that Cox coefficient is proportional to $\mu_{age}$ therefore $b$ is proportional to the hazard. This is confirmed in Figure~\ref{fig:ba_survival} which demonstrates that $b$ for each dataset is a good predictor of survival (or dementia onset).

\begin{figure*}
     \centering
         \includegraphics[width=0.7\textwidth]{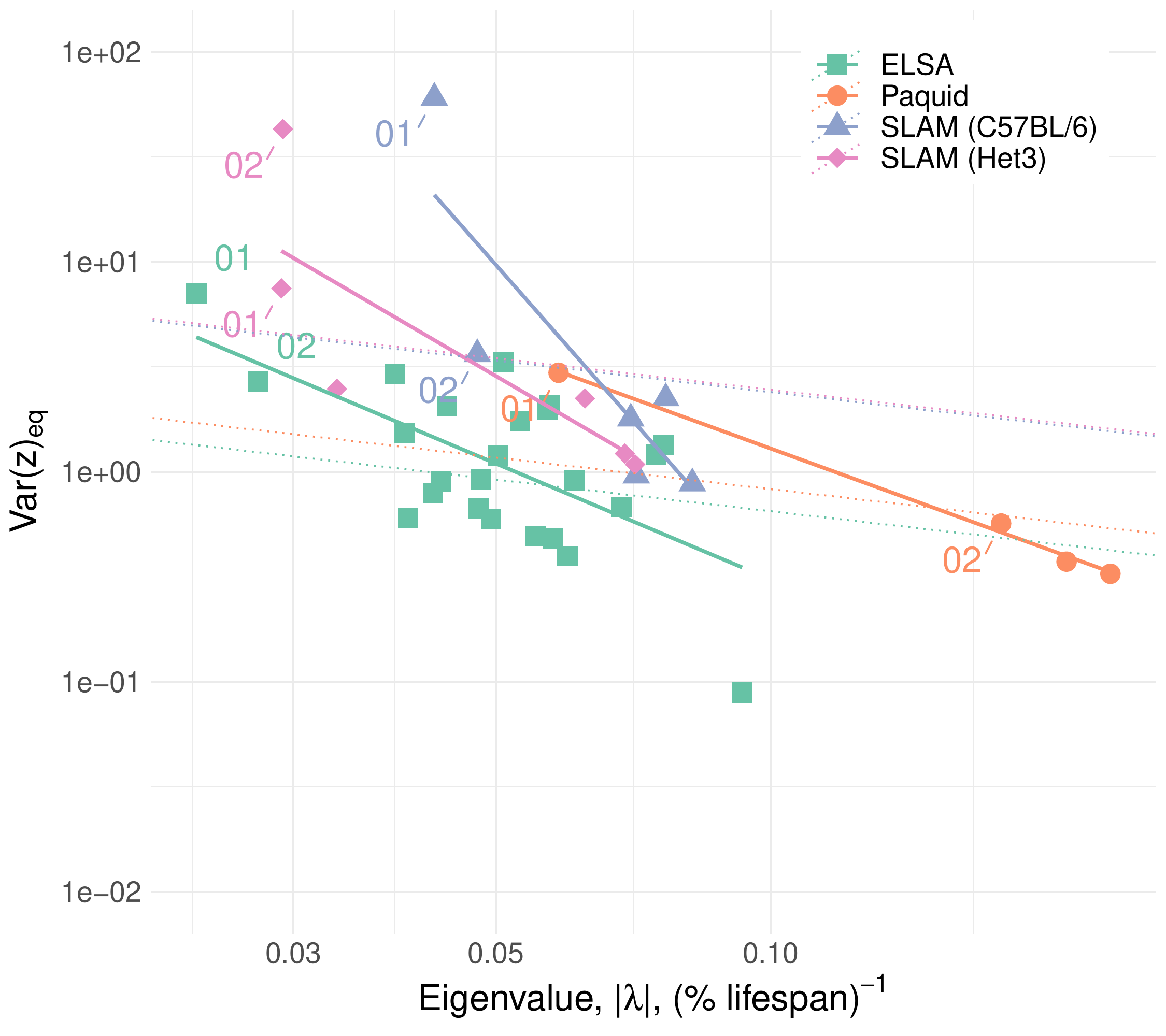}
    \caption{Equilibrium dispersion is primarily determined by eigenvalue strength, $|\lambda|$ equation~(\ref{eq:eqvar}). Smaller eigenvalues are predicted to have larger equilibrium variances. The range of equilibrium variances spans 3 orders of magnitude. The largest variance will drive the observed variation in biomarkers in the steady-state e.g.\ rank 1 will become principal component 1 (equation~(\ref{eq:ypca})). Dotted lines illustrate what the equilibrium variance would be if each dimension had the same noise strength, $\sigma^2$. The fitted solid lines indicate that the noise makes the smaller eigenvalues even more dominant than expected.} 
    \label{fig:eqvar}
\end{figure*}

The natural variables are connected to survival via mallostasis, but how do they relate to the observed phenotype? That is, how do the changes in the natural variables with age affect the observed biomarkers? The total mean and variance are conserved between the biomarkers and the natural variables by Parseval's theorem. This means that natural variables with large means and variances will dominate the means and variances of the observed biomarkers, thus controlling the major changes we see. The steady-state mean grows indefinitely proportional to $\mu_{age}$, what about the variance? The equilibrium (steady-state) dispersion, equation~(\ref{eq:eqvar}), for the natural variables are plotted in Figure~\ref{fig:eqvar}. Smaller eigenvalues are associated with higher variance. Some dimensions (e.g.\ $z_1$ for the C57BL/6) can contain as much as 10x more variance than the next highest dimension. These dimensions will dominate the observed variance in the steady-state. The model predicts that these dimensions will eventually become the dominant principal components (PCs), equation~(\ref{eq:ypca}), implying they would dominate the observed phenotype in the steady-state. Hence what we observe will be dominated by natural variables with small $\lambda$ and large $\mu_{age}$, such as $z_1$ of ELSA, which appears to be closely related to frailty.

\begin{figure*}
     \centering
        \includegraphics[width=\textwidth]{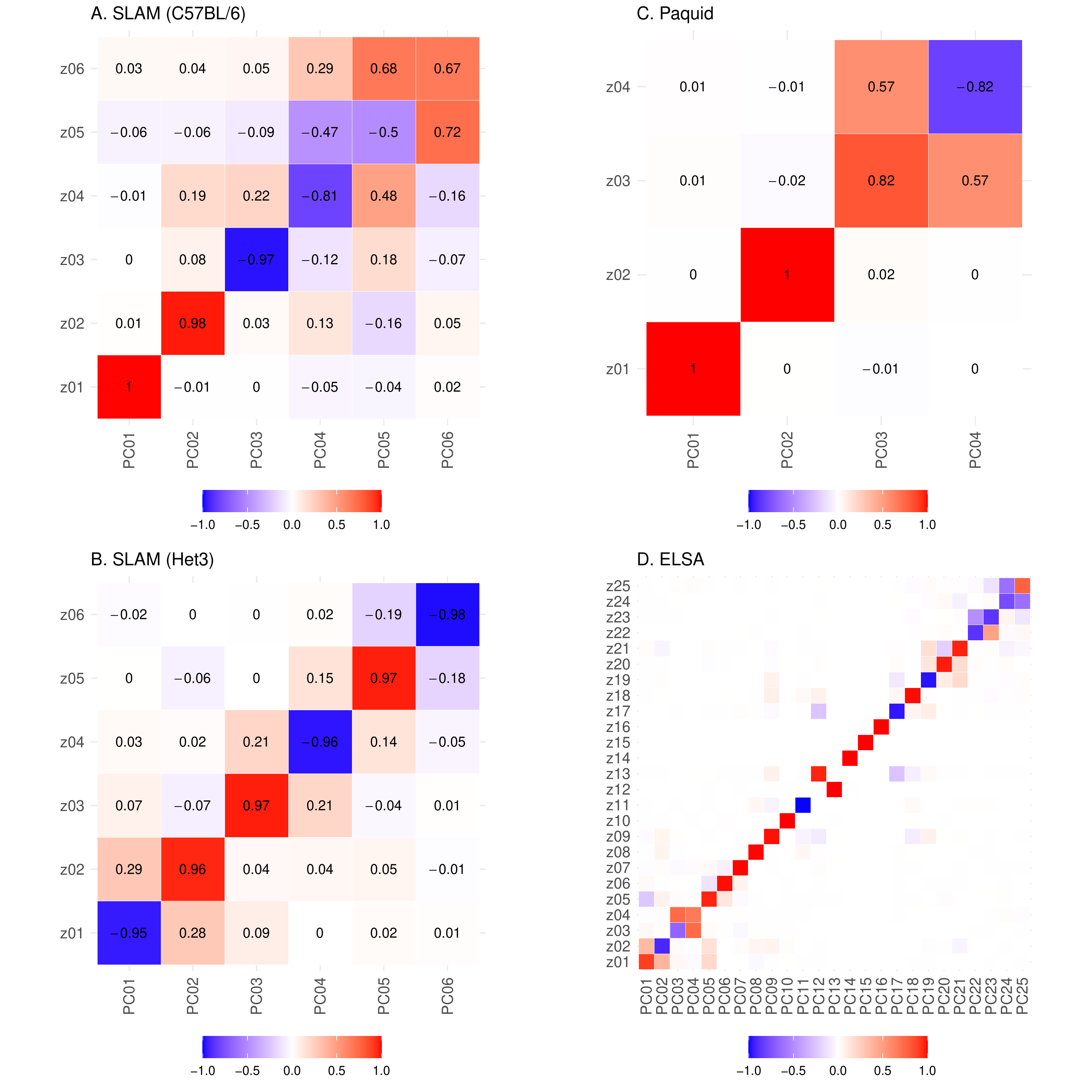}  
    \caption{Principal components are very similar to the natural variables. \textbf{A.} C57BL/6 mice (SLAM). \textbf{B.} Het3 mice (SLAM). \textbf{C.} Paquid (human, dementia). \textbf{D.} ELSA (human). Shown are the dot products between the principal component rotation and $\boldsymbol{P}$. The dot product assesses similarity between the transformations ranging from 1: identical, 0: orthogonal, and $-1$: identical with opposing sign. Identical transformations will generate identical natural variables. If the transformations are identical then all values on the diagonal should be $\pm 1$ (sign is arbitrary\cite{Pridham2023-yj}). We see that the dot products are often close to $\pm 1$, indicating that the transformations are very close, although they do not perfectly coincide.}
    \label{fig:pcdp}
\end{figure*}

We used PCA (principal component analysis) as a preprocessing step, which allowed us to fit a diagonal model, equation~\ref{eq:z}. This simplified analysis and yielded equivalent performance to the full model (Section~\ref{sec:select}). Here we test the self-consistency of the approach. By using PCA, at each bootstrap the eigenvectors of $\boldsymbol{W}$ are principal components, possibly reordered (because we fit a diagonal model for $\boldsymbol{W}$). Averaging over multiple bootstrap replicates removes this equivalence --- although in the steady-state the model predicts that the principal components and eigenvectors of $\boldsymbol{W}$ will coincide, equation~(\ref{eq:ypca}). Here we test the similarity of the PCA rotation and the eigenvector rotation: if they coincide then the principal components are eigenvectors. In Figure~\ref{fig:pcdp} we present the inner product between these matrices, which varies from $-1$ to $1$, with $\pm 1$ representing perfect similarity. We observed strong similarities between the transformations, indicating that the principal components and natural variables will be strongly correlated. This suggests that PCA may be a useful shortcut for approximating the eigenvectors of $\boldsymbol{W}$.

\begin{figure*}
     \centering
        \includegraphics[width=\textwidth]{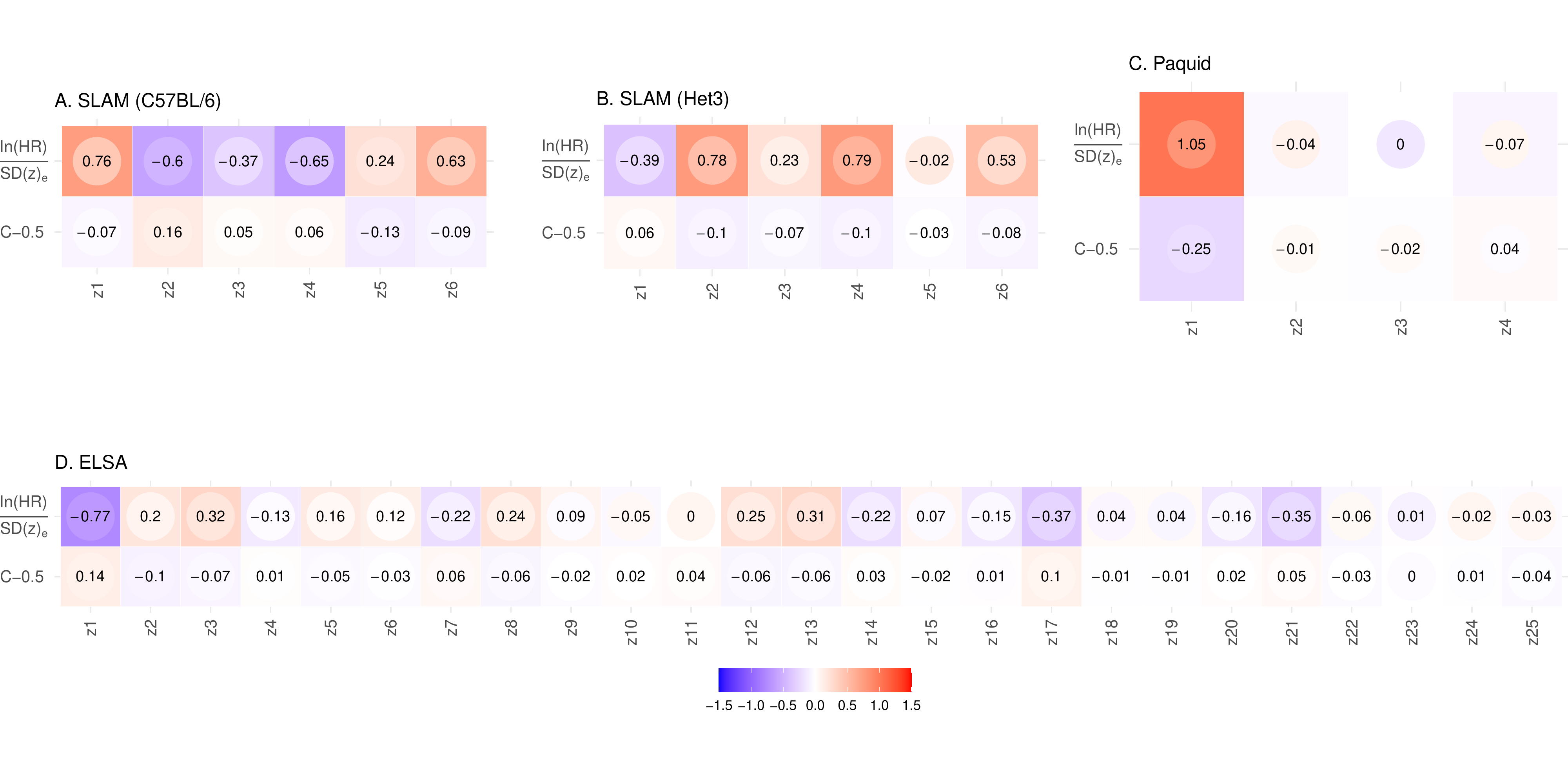}  
    \caption{Survival summary. \textbf{A.} C57BL/6 mice (SLAM). \textbf{B.} Het3 mice (SLAM). \textbf{C.} Paquid (human, dementia). \textbf{D.} ELSA (human). For each dataset, the top row corresponds to the Cox coefficient standardized by the equilibrium dispersion ($\ln{(HR)}/SD(z)_e$) while the bottom is the C-index centered to 0 ($C-0.5$). A Cox coefficient greater than $0$ indicates that higher values are at increase risk and vice versa. A centered C-index greater than $0$ indicates that higher values are at reduced risk and vice versa (opposite of the Cox coefficient). The Cox model is conditioned on age and sex (the same as Figure~\ref{fig:allo_survival}); the C-index is unconditioned. We see that in humans, the first dimension is the dominant determinant in risk of death (ELSA) or dementia (Paquid). It is less clear in mice, where allostatic drift is a better way to identify important survival dimensions (Figure~\ref{fig:allo_survival} or Figure~4A). Inner colour indicates the limit of 95\% confidence interval (CI) closest to zero (non-significant are red on blue or blue on red). }
    \label{fig:survival_summary}
\end{figure*}

Finally, we include a survival summary for each dimension in terms of conditional Cox regression and the C-index in Figure~\ref{fig:survival_summary}. The values are identical to those used in Figures~\ref{fig:allo_survival} (Cox coefficient) and 4 (C-index). This permits the reader to investigate the relative importance of each dimension. Comparing to the correlates of each dimension, Figure~\ref{fig:cor}, one can infer potential mechanisms. For example, $z_2$ of C57BL/6 has a strong survival effect (low is bad) and shows increasing glucose and fat, which could indicate metabolic dysfunction, which C57BL/6 are prone to \cite{Mitchell2015-pg}.



\FloatBarrier

\bibliography{bib}

\end{document}